\documentclass[structabstract]{aa}  
\usepackage{natbib} \bibpunct{(}{)}{;}{a}{}{,} % to follow the A&A style
\usepackage{graphicx}
\usepackage{lscape}
\usepackage{txfonts}
\usepackage{longtable}
\usepackage{amssymb}

\def \HCO+{HCO$^+$} 
\def \H13CO+{H$^{13}$CO$^+$} 
\def\kms{\ifmmode {{\rm \;km\;s^{-1}}}		    	      % km s-1
       \else {\hbox{$\,${\rm km$\;$s$^{\rm -1}$}}}\fi}
\def\solar{\ifmmode_{\mathord\odot\;} \else $_{\mathord\odot}\;$\fi} % _solar
\def\mo{\ifmmode {\,{\it M}\solar\;} \else $\,M$\solar$\;$\fi}	      % M solar		
\def\cm#1{\ifmmode {\,{\rm cm^{-#1}}\;} 		              % cm-1, 
cm-2, cm-3, ...
	\else \hbox{$\,${\rm cm$^{\rm -#1}\;$}}\fi}
\def\am{\ifmmode{^{\scriptscriptstyle\prime}}			       % arcmin
	\else $^{\scriptscriptstyle\prime}$\fi}
\def\deg{\ifmmode{^\circ}\else$^{\circ}$\fi}			       % degree

\def\x {\ifmmode\times\else$\times$\fi}			       	       % times 

\begin{document}
   \title{Kinetic temperatures toward X1/X2 orbit interceptions regions and Giant Molecular Loops in the Galactic center region}
%   \title{On the kinetic temperatures of the molecular gas in the disk and halo {\bf towards} the Galactic center}

    \author{D. Riquelme 
     \inst{1} 
     \thanks{Current address: Max-Planck-Institut f\"ur Radioastronomie, Auf dem H\"ugel 69, 53121 Bonn, Germany}
    \and M.A. Amo-Baladr\'on \inst{2} 
    \and J. Mart\'{i}n-Pintado \inst{2}
    \and R. Mauersberger \inst{3} 
    \and S. Mart\'{i}n \inst{4} 
    \and L. Bronfman \inst{5}}
    \institute{Instituto de Radioastronom\'{i}a Milim\'etrica (IRAM),
              Av. Divina Pastora 7, Local 20, E-18012 Granada, Spain\\ 
     \email{riquelme@mpifr-bonn.mpg.de} 
    \and Centro de Astrobiolog\'ia (CSIC/INTA), Ctra. de Torrej\'on a Ajalvir km 4, E-28850, Torrej\'on de Ardoz, Madrid, Spain
    \and Joint ALMA Observatory, Alonso de C\'ordova 3107, Vitacura, Santiago, Chile  
   \and European Southern Observatory, Alonso de C\'ordova 3107, Vitacura, Casilla 19001, Santiago, Chile
    \and Departamento de Astronom\'{i}a, Universidad de Chile, Casilla 36-D, Santiago, Chile}
   \date{}

% \abstract{}{}{}{}{} 
% 5 {} token are mandatory
 
  \abstract
   {It is well known that the kinetic temperatures, $T_{\rm kin}$, of the molecular clouds in the Galactic center region are higher than in typical disk clouds. However, the $T_{\rm kin}$ of the molecular complexes found at higher latitudes towards the giant molecular loops in the central region of the Galaxy is so far unknown. The gas of these high latitude molecular clouds (hereafter referred to as ``halo clouds'') is located in a region where the  gas in the disk may interact with the gas in the halo in the Galactic center region.}
   {To derive $T_{\rm kin}$ in the molecular clouds at high latitude  and understand the physical process responsible for the heating of the molecular gas both in the Central Molecular Zone (the concentration of molecular gas in the inner $\sim 500$ pc) and in the giant molecular loops.}
   {We measured the metastable inversion transitions of NH$_3$ from $(J,K)=(1,1)$ to $(6,6)$ toward  six positions selected throughout the Galactic central disk and halo. We used rotational diagrams and large velocity gradient (LVG) modeling to estimate the kinetic temperatures toward all the sources. We also observed other molecules like SiO, HNCO, CS, C$^{34}$S, C$^{18}$O, and $^{13}$CO, to derive the densities and to trace different physical processes (shocks, photodissociation, dense gas) expected to dominate the heating of the molecular gas}
   {We derive for the first time $T_{\rm kin}$ of the high latitude clouds interacting with the disk in the Galactic center region. We find high rotational temperatures in all the observed positions. We derive two kinetic temperature components ($\sim 150$ K and $\sim 40$ K) for the positions in the Central Molecular Zone, and only the warm kinetic temperature component for the clouds toward the giant molecular loops. The fractional abundances derived from the different molecules suggest that shocks provide the main heating mechanism throughout the Galactic center, also at high latitudes.}
   {}


   \titlerunning{$T_{\rm kin}$ of X1/X2 orbits interceptions and Giant Magnetic Loops}
   \maketitle
%
%________________________________________________________________
\section{Introduction}
\indent The interstellar molecular gas in the Galactic center
(GC) region (i.e., in the inner $\sim1$ kpc of the Galaxy) shows higher kinetic temperatures,
 $T_{\rm kin}$, than typical disk clouds.  Using metastable inversion transitions of
para-NH$_3$, \citet{Guesten_et_al_1981} derived kinetic temperatures in
the range of 50-120 K towards Sgr A.  Mapping the (1,1), (2,2), and (3,3) inversion transitions of NH$_3$, \citet{Morris_et_al_1983} found
high kinetic temperatures (30-60 K) towards the denser portions of the
GC region. Observing more highly
excited NH$_3$ inversion lines, \citet{Mauersberger_et_al_1986a} and 
  \citet{Huettemeister_et_al_1993a} obtained kinetic temperatures
$T_{\rm kin}\geq 100$ K in all clouds in the GC including  Sgr\,B2 region. 
Similarly high temperatures were also found in the central regions of nearby galaxies, \citep[e.g., ][]{Mauersberger_et_al_2003}.
 From metastable, i.e. $J=K$, inversion transitions of NH$_3$ toward 36 clouds throughout
the GC region, \citet{Huettemeister_et_al_1993b} suggested that in addition to a warm component there is also a ``cool gas component'' with  $T_{\rm kin}\sim 20-30$ K. The extended warm
component in the GC of $\sim 200$ K is not coupled with the dust
(\citealp[$T_{\rm dust}<40$ K, ][]{Rodriguez-Fernandez_et_al_2002, Odenwald_Fazio_1984,
Cox_Laureijs_1989}). High dust temperatures ($T_{\rm dust}\sim 80$ K) are only seen
toward the Sgr\,B2 molecular cloud, which is claimed to be an anomalous region, with recent massive star formation \citep{Bally_et_al_2010}.
So far, to our knowledge, the $T_{\rm kin}$ of molecular clouds has never been determined either  at higher latitudes towards the Giant Molecular Loops \citep[GMLs, ][]{Fukui_et_al_2006}, or in the forbidden and/or high velocity components, explained by the barred potential model as X1 orbits. \\
\indent The kinetic temperature of the molecular gas results from the balance of heating and cooling. 
Molecular clouds cool down by the collisional excitation of molecules and atoms followed by the radiative emission of this energy from the cloud \citep{Hollenbach_1988}. For the physical conditions present in the GC, the cooling is dominated by H$_2$ and CO, while \citet{Huettemeister_et_al_1993b} propose that the dust in the GC region is also an important cooling agent.\\
\indent Dust heated via stars cannot heat gas sufficiently, just because the gas is warmer than the dust. Several heating mechanisms for the GC region have been proposed, as e.g., heating by cosmic rays \citep{Guesten_et_al_1981,Morris_et_al_1983}; heating by X-rays \citep{Watson_et_al_1981,Nagayama_et_al_2007}; magnetic ion-slipping \citep{Scalo_1977}.
The dissipation of mechanical supersonic turbulence through shocks has been proposed for the GC \citep{Fleck_1981, Wilson_et_al_1982, Mauersberger_et_al_1986a}. The shocks can be produced by several phenomena: supernova or hypernova explosions \citep{Tanaka_et_al_2007}; response of the gas in a barred potential model \citep{Binney_et_al_1991}; and when the gas in the GMLs flows down their sides along the magnetic field lines, and joins with the gas layer of the Galactic plane generating shock front at the ``foot points'' of the loops \citep{Fukui_et_al_2006}.

\indent NH$_3$ is one of the best thermometers for measuring the gas kinetic temperatures in molecular clouds \citep[see, ][]{Ho_Townes_1983}. Observing several metastable inversion transitions, one can determinate the kinetic temperature of the molecular clouds.

\indent In this paper, we derive for the first time the kinetic
temperatures of the molecular clouds in the disk-halo interaction regions (foot points of the GMLs and positions where the X1 orbits intercept X2 orbits in a barred potential). We use metastable inversion transitions of NH$_3$ and other molecular tracers (SiO, HNCO, CS) to estimate the kinetic temperatures and densities, and discuss the heating mechanisms of the molecular gas in the GC.
%__________________________________________________________________
\section{Observations}

\subsection{Effelsberg observations}
\indent We observed the metastable inversion transition of NH$_3$
$(J,K)= (1,1)$, $(2,2)$, $(3,3)$, $(4,4)$, $(5,5)$, and $(6,6)$ using the Effelsberg 100m
telescope\footnote{Based on observations with the 100-m telescope of
the Max-Planck-Institut f\"ur Radioastronomie at Effelsberg}
in April 2010, and April 2011. We used the primary focus $\lambda= 1.3$ cm ($18-26$ GHz)
receiver, which has 2 linear polarizations, and a Fast Fourier
Transform Spectrometer (FFTS) in the ``broad IF  band'' mode with a
bandwidth of 500 MHz, providing an effective spectral resolution of
49.133 kHz or 0.386 \kms. We observed the $(1,1)$, $(2,2)$ and $(3,3)$ line
simultaneously, with a band centered at 23.783 GHz, and the $(4,4)$
and $(5,5)$ line in a second setup (centered at 24.336 GHz). The $(6,6)$
was observed in the third setup, centered at 25.056 GHz, using
the 100 MHz bandwidth FFTS spectrometer, which provides an effective
spectral resolution of 9.827 kHz or 0.073 \kms. The beam width of the telescope at 23.7 GHz is 42.2''. The spectra were observed
in position switching mode, with the emission-free reference positions from
\citet{Riquelme_et_al_2010b}, which were checked in the first setup, where the most intense lines are detected. Each position was observed for 12
min in the first setup, 24 min in the second setup, and 32 min in the third setup. The calibration in Effelsberg was done by the
periodic injection of a constant signal (noise cal). To convert the
data to $T_{\rm A}^*$ we corrected for the noise-cal (in K), opacity
and elevation dependent antenna gain\footnote{http://www.mpifr-bonn.mpg.de/div/effelsberg/calibration/1.3cmpf-.html}.
The uncertainty in the calibration is of the order of $5-10\%$. The
main beam temperatures, T$_{\rm MB}$, were obtained by using $T_{\rm
MB}=T^*_{\rm A}\cdot \frac{1}{{\rm Beff}}$, where the beam efficiency, ${\rm Beff}$, is
$0.52$ at $24$ GHz. 
The pointing was checked every two hours against the source 1833-212,
providing an accuracy of better than $10''$. \\ 
\indent In this work, we observe the positions selected in \citet{Riquelme_et_al_2010a}. To avoid confusion, we use the notation of that work. 
The ``Central Molecular Zone'' \citep[CMZ, ][]{Morris_Serabyn_1996} corresponds to the region about $-0\deg.5<l<1\deg.5$.  Since the clouds of 
the CMZ are aligned along the Galactic plane within b$\sim$ 0\deg, this can be viewed as an extension of the Galactic disk, towards galactocentric 
radii $<1$ kpc and will therefore be called ``disk''. When one observed position (from those called ``disk'') have kinematical components 
associated to both, the X1 and the X2 orbits in the barred potential model, we called them explicitly as `Disk\,X1'' and ``Disk\,X2''. The 
 source ``Disk\,2'' which corresponds to Sgr\,B2, is located toward the X2 orbits. Since this source does not have the velocity components 
associated to the X1 orbits, we just call this source as ``Disk''. Gas toward the GMLs regions is labeled as ``halo'' in this paper, to 
differentiate them from the molecular clouds in the Galactic plane. This does not imply that the findings in this paper can be applied to the disk 
or the halo of the Galaxy as a whole, because all of the positions included in this work belong to the GC region.

\begin{figure*}
\includegraphics[width=0.95\textwidth, angle=0]{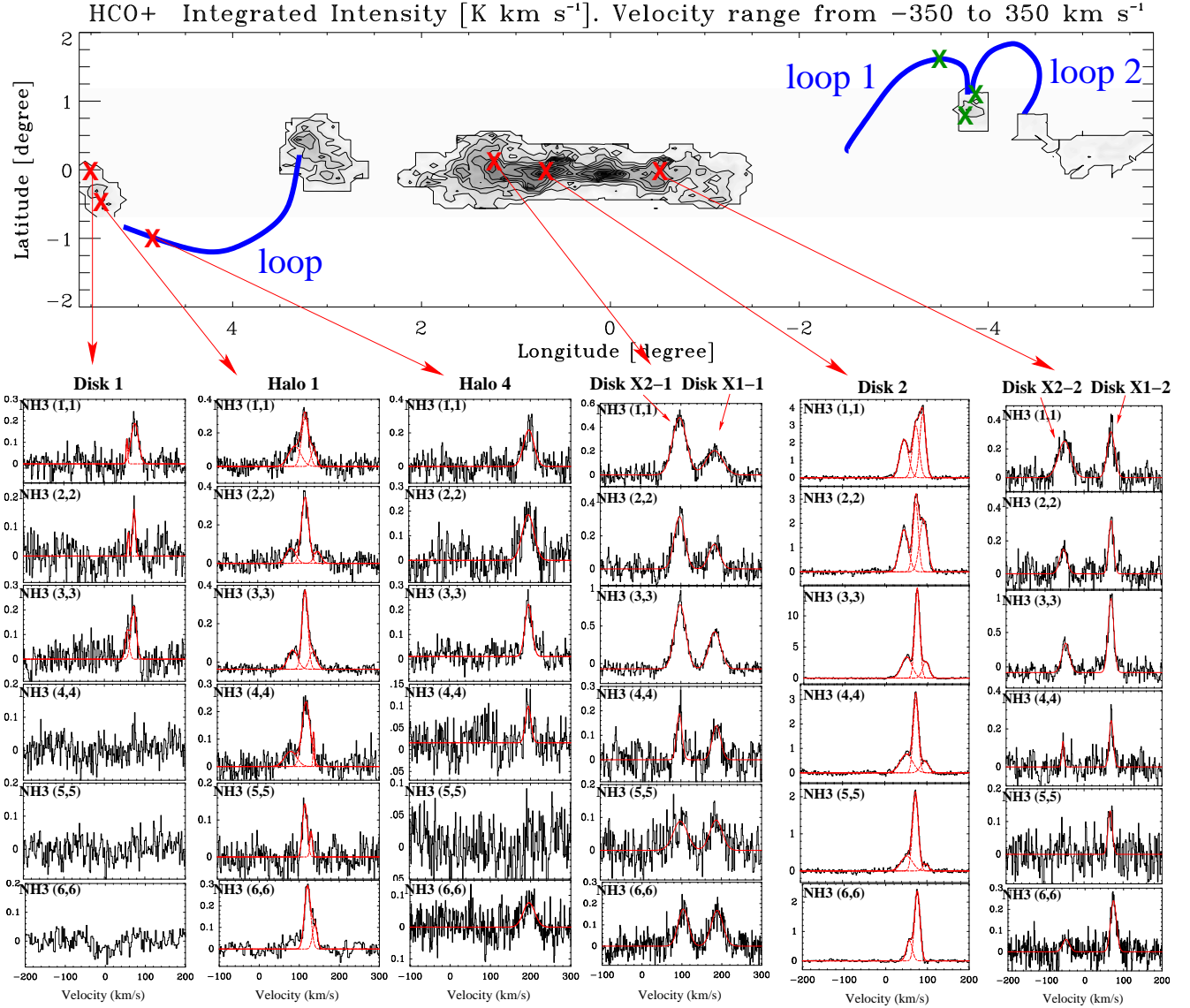}
\caption{Spectra toward selected positions in the GC in the
  metastable inversion transitions from (1,1) to (6,6) of NH$_3$. The positions are indicated in the HCO$^+$ integrated intensity map from 
\citet{Riquelme_et_al_2010b}. The GMLs found by \citet{Fukui_et_al_2006} are indicated in blue. The positions which could not seen from Effelsberg are indicated with green crosses. As indicated in Table \ref{pos}, our Disk\,2 position corresponds to Sgr\,B2.}
\label{espectros}
\end{figure*}

\indent We observed six out of nine positions from \citet{Riquelme_et_al_2010a} visible from Effelsberg shown in Fig. \ref{espectros}, one
in the footpoint of the GMLs (Halo\,1), one in the top of the loop (Halo\,4), two in the
disk toward the location of the expected interactions between the X1
and X2 orbits (Disk\,X1-1, Disk\,X1-2, Disk\,X2-1, Disk\,X2-2,) in the barred potential model \citep{Binney_et_al_1991}
and a pair of positions toward the GC plane (Disk\,1, Disk\,2) as reference (Table \ref{pos}). 

\subsection{Observations with the IRAM 30m telescope}
\indent In order to constrain the physical properties of the gas, we
also observed the $J=2-1, v=0$ rotational transitions of SiO, $^{29}$SiO, and
$^{30}$SiO, the $J=2-1$, $3-2$ rotational transitions of CS and the $J=2-1$
of C$^{34}$S, the $J=10-9$ transition of HNCO, and the $J=1-0$ rotational
transition of $^{13}$CO and C$^{18}$O. The observations were carried
out with the IRAM-30m telescope\footnote{Based on observations carried
out with the IRAM 30m telescope. IRAM is supported by INSU/CNRS
(France), MPG (Germany), and IGN (Spain).} at Pico Veleta (Spain) in
several periods from June 2009 to October 2010. For the 3mm lines, we
used the E090 band of the Eight Mixer Receiver (EMIR)\footnote{http://www.iram.es/IRAMES/mainWiki/EmirforAstronomers}, which provides a bandwidth
of $\sim 8$ GHz simultaneously in both polarizations per sideband, and for CS $(3-2)$
emission, we use the E150 band of EMIR receiver, which provide a
bandwidth of $\sim 4$ GHz simultaneously in both polarizations. As the
backend, we used the Wideband Line Multiple Autocorrelator (WILMA), providing a resolution of 2
MHz or 6.6 km/s at 91 GHz and 4.1 km/s at 146 GHz. We observed the nine selected positions from
\citet{Riquelme_et_al_2010a} that were all observable with the 30m telescope. In this work, we use the
antenna temperature scale $T^*_{\rm A}$, which can be converted to
main beam temperature $T_{\rm MB}=T^*_{\rm A}\cdot \frac{{\rm Feff}}{{\rm Beff}}$, where the forward efficiency ${\rm Feff}$ is $95\%$ and the
main-beam efficiency is ${\rm Beff} = 81\%$ at $86$ GHz, and ${\rm Feff}=
93\%$ and ${\rm Beff} = 74\%$ at $145$ GHz. The beam width of the
telescope is $29''$ at $86$ GHz, and $16''$ at 145 GHz.\\

\begin{table*}{\small
\begin{center}
\begin{tabular}{lccccl}
\hline\hline
%\\ 
Source name$^a$       & Associated  & \multicolumn{2}{c}{Galactic coordinates}& \multicolumn{2}{c}{Equatorial coordinates}          \\
                      & object      &  $l$ [\deg]      & $b$ [\deg]           & $\alpha_{J2000}$ & $\delta_{J2000}$                 \\ \hline    
Halo\,1               & M$+5.3-0.3$ & $5.45$           &  $-0.324$            & $17^{\rm h}59^{\rm m}17.8^{\rm s}$ &$-24\deg24'38''$\\
Halo\,4               & Top Loop    & $4.75$           &  $-0.8$              & $17^{\rm h}59^{\rm m}34.9^{\rm s}$ &$-25\deg15'16''$\\
Disk\,X1-1, Disk\,X2-1& 1.3 complex & $1.28$           &  $ +0.07$            & $17^{\rm h}48^{\rm m}21.9^{\rm s}$ &$-27\deg48'19''$\\
Disk\,X1-2, Disk\,X2-2& Sgr C       &$359.446$         &  $-0.124$            & $17^{\rm h}44^{\rm m}46.9^{\rm s}$ &$-29\deg28'25''$\\
Disk\,1               & Galactic plane at $l\sim 5\deg.7$&$5.75$& $0.25$      & $17^{\rm h}57^{\rm m}46.5^{\rm s}$ &$-23\deg51'51''$\\
Disk\,2               & Sgr B2      & $0.6932$         &  $ -0.026$           & $17^{\rm h}47^{\rm m}21.9^{\rm s}$ &$-28\deg21'27''$\\ \hline
\end{tabular}
\caption{Observed positions in NH$_3$ lines. \label{pos}}
\end{center}
$^a$ following the notation of \citet{Riquelme_et_al_2010a}}
\end{table*}

%_
\section{Results}
Fig. \ref{espectros} shows the  ammonia spectra taken toward 
each position in all the metastable inversion transitions observed in
this work. Most of the metastable inversion transitions of NH$_3$ were detected, except the (4,4), (5,5), and (6,6) of ``Disk\,1'' and the (5,5) 
of ``Halo\,4''. The criteria used to define if a emission line is detected or not, was to have a line peak temperature $> 3 \sigma_{\mathrm{rms}}$, 
where $\sigma_{\mathrm{rms}}$ is the root mean square per spectral channel. If the intensity of the line does not reach this value, we still assume that a 
line is actually detected if the line has an integrated intensity in the velocity width (as defined by the (3,3) line which presents the highest 
signal-to-noise ratio) $> 3 \sigma$.
%_________________________________________________________________
\subsection{Optical depth of NH$_3$}
\indent Each NH$_3$ inversion transition is split into five components:
a ``main component'' and four symmetrically placed ``satellites'' (the quadrupole hyperfine (HF) structure). Due to the large linewidth of the molecular 
clouds in the GC, the magnetic splitting ($<0.2 \kms$) cannot be resolved. Under the assumption of local thermodynamical equilibrium (LTE), 
the relative intensities of the four satellite HF components  can be used to estimate the optical depth $\tau$ of the main component of the 
metastable inversion transitions. Knowing $\tau$, we can estimate the NH$_3$ column density, and the rotational temperature from the ratios of the peak
or integrated intensities.\\
\indent We use the ``NH$_3$ method'' from CLASS\footnote{http://www.iram.fr/IRAMFR/GILDAS} to
determine the optical depth for the $(1,1)$, $(2,2)$ and $(3,3)$ lines.
To define the linewidth (which was used as a fixed parameter in the NH$_3$ method), we use the $(3,3)$ transition, because these spectra have the
best signal-to-noise ratio in our observations and the HF components are much weaker than those of the $(1,1)$ and $(2,2)$ lines. As we can see in 
Table \ref{observation}, all the NH$_3$ lines observed in this work are optically thin toward all sources, except the $(1,1)$ transition toward
 Sgr\,B2. Following the criteria of \citet{Huettemeister_et_al_1993b} based on the lower peak
intensities in these lines, with respect to the $(1,1)$, $(2,2)$ and $(3,3)$ ones, we assume that the $(4,4)$ and $(5,5)$ are also optically thin. 
Table \ref{observation} presents the results from simple Gaussian fits for all the observed positions, allowing all the parameters to be free. 
\subsection{Physical conditions of the gas from CS and NH$_3$}
\indent To derive the $n(\mathrm{H_2})$ and $T_{\mathrm{kin}}$ of the gas, we combine the CS and NH$_3$ molecular emission, in an iterative way. 
First, we used MASSA software\footnote{http://damir.iem.csic.es/mediawiki$-$1.12.0/index.php/MASSA\\\_Users\_Manual} to derive the rotational 
temperatures and column densities using Boltzmann diagrams \citep[see, ][for a detailed explanation and equations of the method]{Goldsmith_Langer_1999} 
(Table \ref{results}). The rotational temperature, which is a lower limit of the actual kinetic temperature, $T_{\mathrm{kin}}$, was used as a fixed 
parameter in RADEX \citep[see, ][for detailed explanation of the formalism adopted in this statistical equilibrium radiative transfer code]{Vandertak_et_al_2007} 
to derive the $n(\mathrm{H_2})$ and CS column densities. Then, using the $n(\mathrm{H_2})$ obtained from CS, we used RADEX to derive the kinetic temperature 
from the para-NH$_3$ transitions (see \S \ref{radex_NH3}). With the kinetic temperature, we derived then 
the final $n(\mathrm{H_2})$ and CS column densities (Table \ref{resultsCSLTE}).

\begin{table*}
\caption{NH$_3$ physical parameters (rotational temperatures and column densities) derived for each source using MASSA software. 
Bold faced values indicated the most likely result consistent with the non-LTE analysis. \label{results}}
\small
\begin{tabular}{lccccccccc}
\hline\hline
%\\ 
Source& $n^a$ &$T_{\rm rot}$&$T_{\rm rot}$&$T_{\rm rot}$&$T_{\rm rot}$&$N({\rm NH_3})$& $N({\rm NH_3})$ & $N({\rm NH_3})$&  $N($o-NH$_3)$ \\
      &   & \begin{scriptsize}(11-22)\end{scriptsize} &\begin{scriptsize}(22-44-55)\end{scriptsize}& \begin{scriptsize}(11-22-44-55)\end{scriptsize}&\begin{scriptsize}(33-66)\end{scriptsize}&\begin{scriptsize}(11-22)\end{scriptsize} &\begin{scriptsize}(22-44-55)\end{scriptsize}& \begin{scriptsize}(11-22-44-55)\end{scriptsize}& \begin{scriptsize}(33-66) \end{scriptsize}      \\
      &   &       [K]      &        [K]        &      [K]         &     [K]      &10$^{14}$ cm$^{-2}$&10$^{14}$ cm$^{-2}$&10$^{14}$ cm$^{-2}$&10$^{14}$ cm$^{-2}$ \\ \hline    
Halo1 &1  & $46.7 \pm 0.9$ & $117.5 \pm  1.7$  &${\bf 92.7\pm0.8}$& $81.0\pm0.8$ &  $2.14 \pm 0.07$  & $3.53 \pm 0.10$   &${\bf 3.72\pm0.06}$&  $5.37\pm 1.1$ \\
      &2  & $43.2 \pm 0.3$ & $131.8 \pm  0.9$  &${\bf 96.6\pm0.3}$&$156.4\pm9.6$ &  $4.93 \pm 0.06$  & $9.09 \pm 0.11 $  &${\bf 9.49\pm0.06}$&  $11.1 \pm 1.2$ \\
      &3  &                &                   & $80.9 \pm 1.1$   &$169  \pm31$  &                   &                   & $2.06 \pm 0.06$   &  $3.3 \pm 1.0$\\\hline
Halo4 &1  &                &                   & $56.5 \pm 2.3$   &$138  \pm10$  &                   &                   & $5.48 \pm 0.45$   &   $5.39\pm 0.76$ \\\hline
DiskX1-1&1& $27.2 \pm 1.5$ & $156   \pm 12  $  &                  &$164.1\pm6.6$ &  $4.85 \pm 0.53$  & $7.7  \pm 1.2 $   &                   &  $12.07\pm 0.91$ \\\hline
DiskX2-1&1& $27.4 \pm 0.8$ &  $91.7 \pm  3.6$  &                  &$122.8\pm3.8$ &  $9.75 \pm 0.52$  & $9.87 \pm 0.78$   &                   &  $16.56\pm 0.98$ \\\hline
DiskX1-2&1& $33.1 \pm 2.2$ & $112.7 \pm  6.5$  &                  &$138.2\pm5.1$ &  $3.99 \pm 0.48$  & $5.75 \pm 0.71$   &                   &  $10.93\pm 0.80$ \\\hline
DiskX2-2&1& $22.0 \pm 1.4$ &  $72.5 \pm  6.1$  &                  &$106.9\pm9.7$ &  $4.44 \pm 0.55$  & $2.85 \pm 0.58$   &                   &   $5.09\pm 0.90$\\\hline
Disk1 &1  & $19.1 \pm 2.9$ & $154   \pm 48  $  &                  &$139  \pm26 $ &  $0.74 \pm 0.20$  & $0.66 \pm 0.41$   &                   &   $0.93\pm 0.39$ \\
      &2  & $26.9 \pm 2.4$ &  $82.9 \pm  9.8$  &                  & $100 \pm12 $ &  $1.41 \pm 0.20$  & $1.29 \pm 0.34$   &                   &  $2.02\pm 0.46$\\\hline
Disk2 &1  &${\bf38.1\pm0.6}$&${\bf110.3\pm1.0}$& $90.6 \pm 0.5$   & $95.7\pm1.5$ &${\bf31.6\pm1.0}$&${\bf48.1 \pm1.1}$   & $60.3 \pm 1.0$    &  $61.0 \pm 2.0$ \\
      &2  & $50.8 \pm 1.6$ & $145.6 \pm  2.3$  &${\bf113.8\pm0.5}$&$112.9\pm0.6$ & $42.1  \pm 2.1 $  & $84.9  \pm 3.1$   &${\bf101.6 \pm1.0}$&  $133.2 \pm 1.4$  \\
      &3  &${\bf33.0\pm0.9}$&${\bf63.1\pm1.2}$ & $51.7\pm 0.5$    & $52.4\pm3.2$ &${\bf33.1 \pm1.3 }$&${\bf32.0 \pm1.7}$ & $47.2 \pm 1.0$    &  $37.6 \pm6.4$ \\\hline
\end{tabular}
\begin{list}{}{}
\item $^a$ Cloud number defined by the different velocity components (see Table \ref{observation})
\item $N(para-{\rm NH_3}$) correspond to the sum of all observed para-NH$_3$ column densities. When the data is consistent with a two temperature model, 
$N(para-{\rm NH_3})$ correspond to the sum of column 7 and 8. If only one temperature regime is present, the $N(para-{\rm NH_3})$ corresponds to column 9.
\end{list}
\end{table*}  

\subsubsection{$n(\mathrm{H_2})$ derived from the CS data}
\indent We used the non-LTE excitation radiative transfer code
RADEX to derive the $n(\mathrm{H_2})$ and CS column densities from line intensities of the observed CS lines. The modeling suggests that the CS emission 
is optically thin with opacities ranging from 0.05 to 0.96. The results are showed in Table \ref{resultsCSLTE}. The error were estimated assuming a 10\% 
calibration error as the typical flux calibration uncertainty at the 30-m telescope, and we give an upper and lower value based on the minimum and maximum 
value from the LVG diagrams (see from Fig. \ref{modeloCSHalo1} to Fig. \ref{modeloCSSgrB2}). It is important to note that $n(\mathrm{H_2})$ in some sources 
is poorly constrained, which translates in large errors or upper limits shown in the Table \ref{Tkin_nH2}. If we derive the $n(\mathrm{H_2})$ using the 
rotational temperature (which is a lower limit to the kinetic temperature), the $n(\mathrm{H_2})$ differ on average by $\sim 27$\%. 

\begin{table*}
\caption{Physical parameters derived from CS using non-LTE (RADEX) model. \label{resultsCSLTE} The kinetic temperature used in RADEX is indicated. 
When more than one $T_{{\rm kin}}$ regime was present in one position, we derived the physical parameters from each $T_{{\rm kin}}$. We assumed a 10\% calibration error.}
\begin{tabular}{lc|c|ccccc}
\hline\hline
%\\  
Source&Cloud& $T_{kin}$ & Tex CS(2-1) & Tex CS(3-2) & $\tau_{CS (2-1)}$&$\tau_{CS (3-2)}$ \\
     &number& [K] &[K] & [K] & &    \\\hline
Halo1 &1    &$115$  &$<3.8$            &$<3.6$             &$>0.25$               &$>0.16$                  \\  
      &2    &$ 90$  &$12.6^{9.3}_{4.3}$ & $8.2^{2.5}_{1.8}$ & $0.25^{0.35}_{0.15}$ & $0.54^{0.34}_{0.18}$    \\
      &3    &$135$  &$ 4.6^{1.7}_{1.1}$ &$4.4^{1.0}_{0.9}$ & $0.31^{1.06}_{0.18}$ & $0.35^{0.77}_{0.14}$    \\\hline
Halo2 &1    &$113$  &$4.1^{1.3}_{0.8}$ & $4.0^{0.9}_{0.8}$ & $0.22^{0.75}_{0.13}$ & $0.14^{0.31}_{0.06}$    \\
      &2    &$113$  &$3.8^{1.3}_{0.8}$ & $3.8^{0.9}_{0.9}$ & $0.28^{15.3}_{0.18}$ & $0.15^{5.81}_{0.07}$    \\\hline
Halo3 &1    &$113$  &$4.7^{1.8}_{0.0}$ & $4.4^{1.1}_{0.0}$ & $0.61^{0.10}_{0.38}$ & $0.75^{0.12}_{0.35}$    \\
      &2    &$113$  &$6.1^{3.1}_{1.7}$ & $5.2^{1.2}_{1.0}$ & $0.07^{0.11}_{0.04}$ & $0.05^{0.04}_{0.02}$    \\
Halo4 &1    &$ 95$  &$5.4^{2.3}_{1.2}$ & $4.7^{1.1}_{0.8}$ & $0.21^{0.28}_{0.12}$ & $0.17^{0.12}_{0.06}$    \\\hline
Halo5 &1    &$ 95$  &$10.4^{8.4}_{3.4}$& $7.0^{2.2}_{1.4}$ & $0.11^{0.13}_{0.07}$ & $0.21^{0.10}_{0.07}$    \\\hline
Disk-X1-1&1 &$ 38$  &$10.9^{5.6}_{3.1}$& $7.1^{1.9}_{1.2}$ & $0.12^{0.10}_{0.06}$ & $0.24^{0.09}_{0.07}$    \\
         &  &$300$  &$9.3^{8.3}_{3.0}$ & $6.8^{2.2}_{1.4}$ & $0.15^{0.20}_{0.09}$ & $0.25^{0.14}_{0.09}$    \\\hline
Disk-X2-1&1 &$ 38$  &$6.7^{2.4}_{1.6}$ & $5.3^{1.1}_{0.9}$ & $0.37^{0.51}_{0.19}$ & $0.48^{0.37}_{0.15}$    \\
         &  &$100$  &$5.9^{2.5}_{1.4}$ & $5.0^{1.3}_{1.1}$ & $0.50^{1.23}_{0.29}$ & $0.55^{0.85}_{0.22}$    \\\hline
Disk-X1-2&1 &$ 52$  &$13.7^{8.4}_{4.4}$& $8.5^{2.4}_{1.7}$ & $0.16^{0.16}_{0.08}$ & $0.42^{0.18}_{0.13}$    \\
         &  &$215$  &$12.9^{15.5}_{4.7}$&$8.4^{2.9}_{1.8}$ & $0.18^{0.24}_{0.12}$ & $0.43^{0.23}_{0.15}$    \\\hline
Disk-X2-2&1 &$ 28$  &$10.0^{3.8}_{2.5}$& $6.7^{1.6}_{1.2}$ & $0.20^{0.18}_{0.09}$ & $0.30^{0.13}_{0.08}$    \\
         &  &$ 95$  &$9.3^{7.1}_{2.8}$ & $6.6^{2.1}_{1.3}$ & $0.22^{0.28}_{0.14}$ & $0.31^{0.17}_{0.11}$    \\\hline
Disk1  &1   &$ 23$  &$<3.4$            & $<3.3$            & $>0.22$              & $>0.14$                 \\
       &    &$154$  &$<3.0$            & $<3.0$            & $>0.57$              & $>0.30$                 \\
       &2   &$ 38$  &$5.5^{2.1}_{1.2}$ & $4.7^{1.0}_{0.7}$ & $0.17^{0.20}_{0.09}$ & $0.17^{0.10}_{0.05}$    \\
       &    &$ 82$  &$5.0^{1.9}_{1.1}$ & $4.5^{1.0}_{0.8}$ & $0.21^{0.37}_{0.12}$ & $0.19^{0.18}_{0.07}$    \\\hline
Disk2  &1   &$ 68$  &$6.8^{3.5}_{1.7}$ & $5.4^{1.4}_{0.9}$ & $0.16^{0.19}_{0.09}$ & $0.21^{0.12}_{0.07}$    \\
       &    &$200$  &$6.1^{2.9}_{1.6}$ & $5.3^{1.3}_{1.0}$ & $0.20^{0.28}_{0.11}$ & $0.23^{0.17}_{0.08}$    \\
       &2   &$145$  &$6.7^{2.9}_{1.7}$ & $5.7^{1.4}_{1.2}$ & $0.70^{14.8}_{0.41}$ & $0.96^{15.2}_{0.40}$    \\
       &3   &$ 50$  &$16.1^{9.2}_{5.6}$& $9.4^{2.7}_{2.1}$ & $0.19^{0.22}_{0.09}$ & $0.45^{0.23}_{0.13}$    \\
       &    &$ 80$  &$16.1^{16.2}_{6.2}$&$9.3^{3.5}_{2.2}$ & $0.19^{0.27}_{0.12}$ & $0.45^{0.26}_{0.16}$    \\ \hline
\end{tabular}				     		   
\end{table*}

%__________________________________________________________________
\subsubsection{LVG analysis from NH$_3$ \label{radex_NH3}}
\begin{figure*}
\vbox{
\hbox{
\includegraphics[width=0.38\textwidth, angle=90]{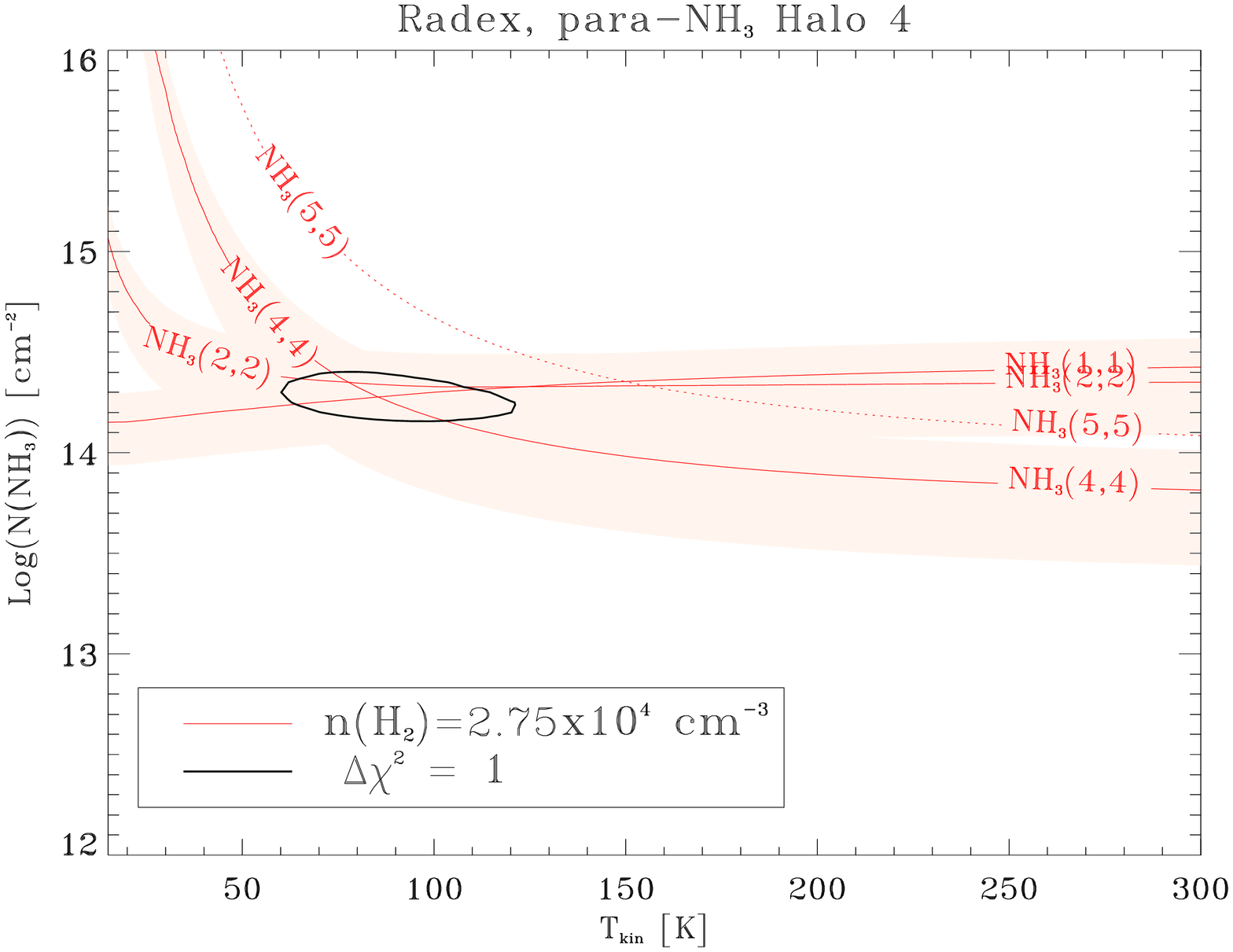}
%\hspace{-1 cm}
\includegraphics[width=0.38\textwidth, angle=90]{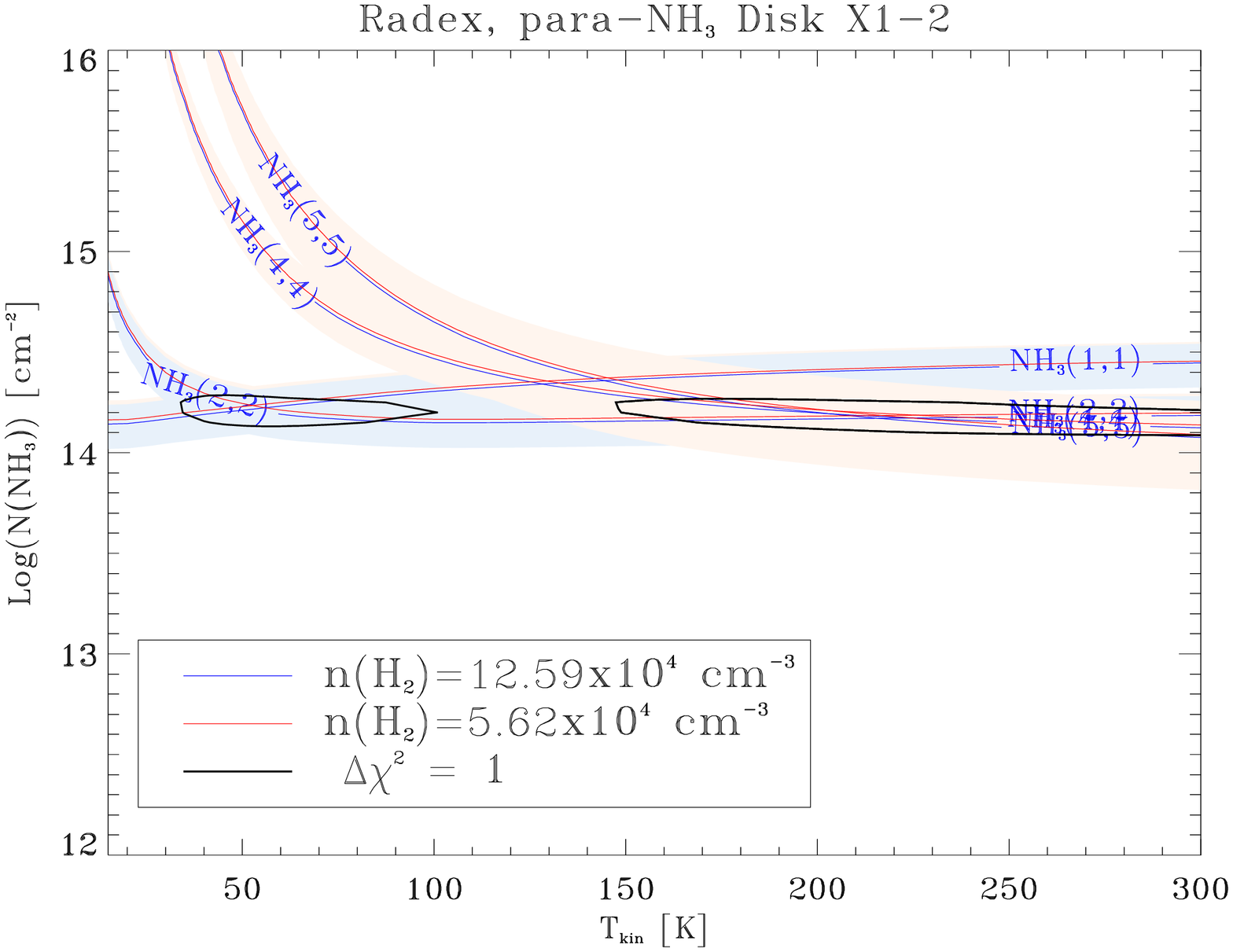}
}
\hbox{
\includegraphics[width=0.5\textwidth, angle=0]{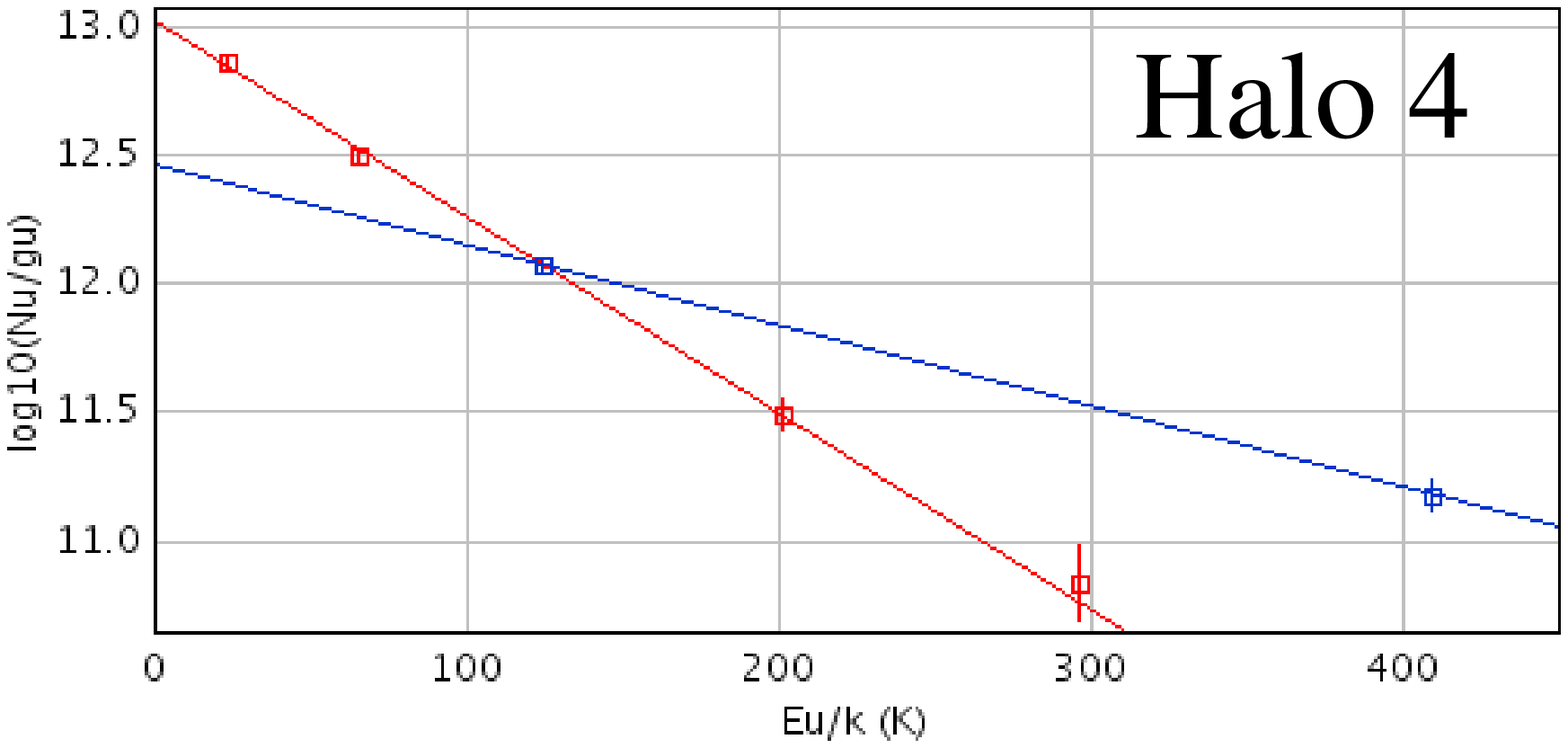}
\includegraphics[width=0.5\textwidth, angle=0]{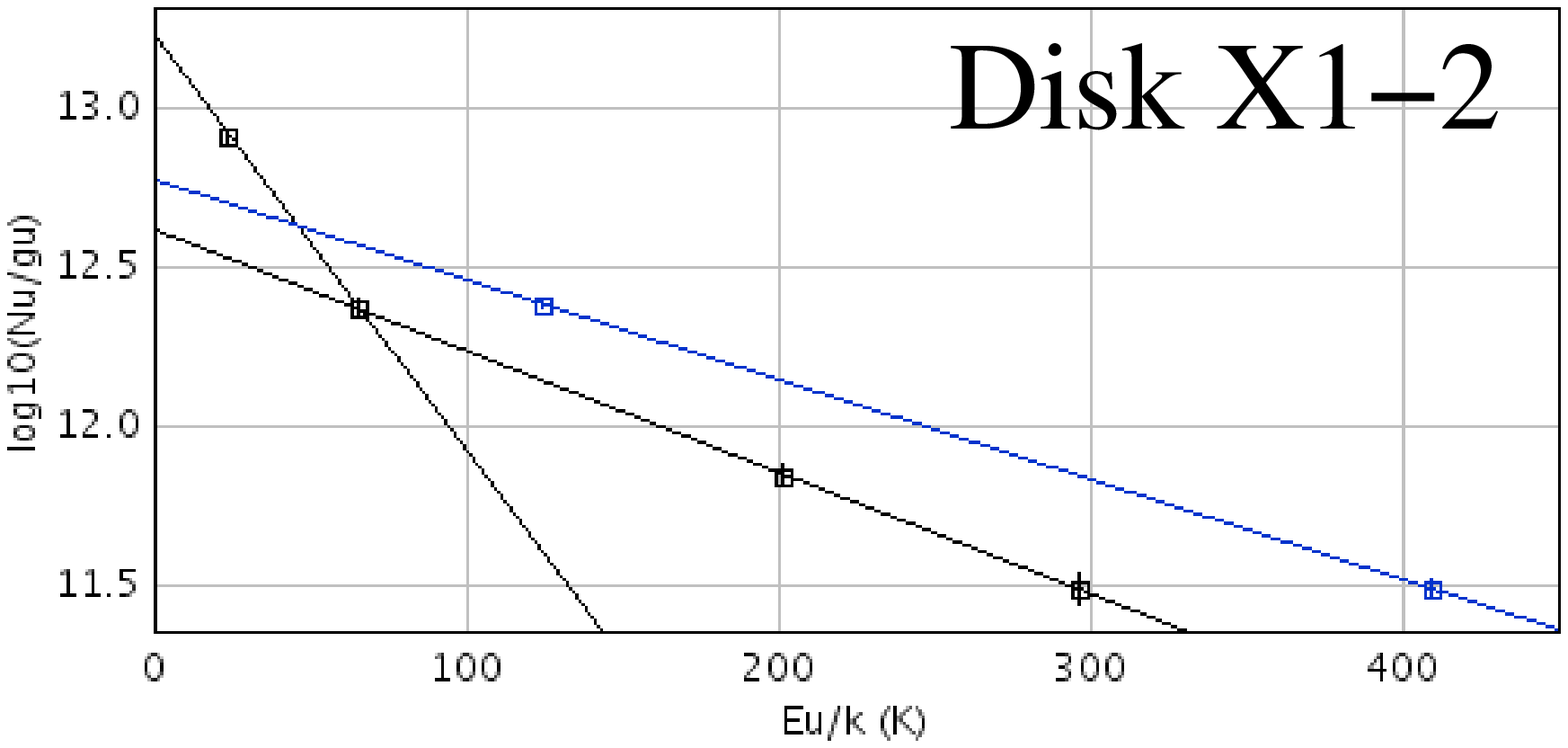}
}
}
\caption{Example of rotational (bottom) and LVG (top) diagrams of NH$_3$. Left: Halo\,4. Right: Disk\,X1-2. In the Boltzmann diagram (bottom), 
para-NH$_3$ species are fitted with the red line, and the ortho-NH$_3$ species are fitted with the blue line. In the LVG diagrams (top), we plot the peak intensity of the metastable inversion transitions of para-NH$_3$. For the source Halo\,4, we derive only one kinetic temperature component (warm),  which is plotted in red lines, and the error in red shaded region. The source Disk\,X1-2 was modeled using two kinetic temperature component model, one warm and one cold, that were plotted in red and blue lines respectively with the correspondly associated error shown as shaded regions. We show the $n(\mathrm{H_2})$ used in the LVG model for the warm and cold components. To estimate the uncertainty of the derived parameters, we computed the $\chi^2$ of the line intensities over the grid used for the LVG model. We impose $\Delta \chi^2=\chi^2-\chi^2_{\mathrm{min}}=1$, which translates in the $68.3$\% (1 $\sigma$) confidence level projected for each parameter axis}  (see text for details). 
\label{modelo}
\end{figure*}

\indent To estimate the kinetic temperatures of the gas, we also used a non-LTE excitation and radiative transfer code RADEX. 
 Using the value of $n(\mathrm{H_2})$ derived
from the CS LVG analysis (Table \ref{Tkin_nH2}), and the velocity widths  (see Table \ref{observation}), we can derive the
$T_{\rm kin}$ and $N_{{\rm NH_3}}$.  Fig. \ref{modelo} shows an example of this
procedure, and Table \ref{Tkin_nH2} shows the results. In Fig. \ref{modelo} and from Fig. \ref{modeloHalo1} to Fig. \ref{modeloSgrB2}, we show in blue the results
corresponding to the metastable inversion transitions $(1,1)$-$(2,2)$ (low
rotational temperature), and in red, the results corresponding to the
metastable inversion transitions $(2,2)$-$(4,4)$-$(5,5)$ (high rotational
temperature). For the cases where only one temperature regimen was a
possible solution, we plotted the result in red in the LVG plot. 
LVG models indicate that the results from LTE are reliable. Additionally, for every observed position, we checked the two temperature component 
assumption by comparison to synthetic spectra  under LTE approach using MASSA software. We found that for the positions where we derived two 
kinetic temperature components, the modeled line profile fit better the observed
emission, while a single warm component was not enough to reproduce the observed profile. When the $(4,4)$ or $(5,5)$ inversion transitions 
were not detected, the upper limits to their emission were plotted in dashed lines. This upper limit to the emission
was obtained as 3$\sigma_{\mathrm{rms}}$ level. The individual fits to all sources are shown in the Online Appendix (Fig.\ref{modeloHalo1} to \ref{modeloSgrB2}). 

As a result of our analysis, we derive two kinetic temperature (one cool and one warm) components in the CMZ and only one warm component in the halo positions. 
In the CMZ the cool component range from $23$ K to $68$ K with an average value of $40$ K; and the warm component range from $80$ K to $>300$ K with an average 
value of $150$ K. This reference values should be taken with caution due to the large uncertainty of the kinetic temperature derived from the LVG (see below). 
To estimate the uncertainty of the derived parameters, we computed the $\chi^2$ of the line intensities over the grid used for the LVG model. We impose 
$\Delta \chi^2=\chi^2-\chi^2_{\mathrm{min}}=1$, which translates in the $68.3$\% confidence level projected for each parameter axis 
\citep[see, e.g., ][Section \S 15.6]{NumericalRecipes}. The black ellipses shows the error in the model (Fig. \ref{modelo}).

\begin{table*}
\caption{Kinetic temperatures and densities derived from LVG calculations from NH$_3$ and CS data \label{Tkin_nH2}}
\begin{tabular}{lccccccl}
\hline\hline
%\\ 
Source &Cloud  &\multicolumn{2}{c}{low temperature}&\multicolumn{2}{c}{high temperature}&\multicolumn{2}{c}{single temperature from p$-$NH$_3$}\\
       &number  &  $T_{{\rm kin}}$  & $n({\rm H_2})$   &$T_{{\rm kin}}$  & $n({\rm H_2})$ &$T_{{\rm kin}}$  & $n({\rm H_2})$   \\ 
       &  &    [K]  & 10$^4$ cm$^{-3}$   & [K]  & 10$^4$ cm$^{-3}$&    [K]  &10$^4$ cm$^{-3}$     \\ \hline    
Halo1  &1  & & & & &$>115$ &$<1.00$              \\
       &2  & & & & & $>90$ &$8.49^{5.63}_{4.08}$ \\
       &3  & & & & &$>135$ &$1.58^{1.72}_{1.23}$ \\\hline
Halo2  &1  & & & & &$113^c$&$1.26^{1.56}_{0.98}$ \\
       &2  & & & & &$113^c$&$<2.51$              \\\hline
Halo3  &1  & & & & &$113^c$&$1.50^{2.05}_{0.09}$ \\
       &2  & & & & &$113^c$&$3.55^{2.76}_{1.96}$ \\\hline
Halo4  &1  & & & & &$ 95 $ &$2.75^{2.54}_{1.49}$  \\\hline
Halo5  &1  & & & & &$ 95^c$      &$7.47^{5.11}_{3.18}$  \\\hline
DiskX1-1&1 &$38$ &$13.0^{8.21}_{5.05}$ &$>300^a$ &$3.55^{2.56}_{1.66}$ & & \\\hline
DiskX2-1&1 &$38$ &$5.62^{4.38}_{3.11}$ &$100$    &$2.70^{2.73}_{1.90}$ & & \\\hline
DiskX1-2&1 &$52$ &$13.3^{8.09}_{5.34}$ &$215$    &$5.62^{3.88}_{2.46}$ & & \\\hline
DiskX2-2&1 &$28$ &$14.1^{8.88}_{5.98}$ &$95$     &$6.20^{5.02}_{2.95}$ & & \\\hline
Disk1  &1  &$23$ &$<1.12$ &$>154^b$ &$<0.16$ & & \\
       &2  &$38$ &$4.57^{3.84}_{2.33}$ &$>82^b$  &$2.51^{2.40}_{1.55}$ & & \\\hline
Disk2  &1  &$68$ &$4.97^{3.94}_{2.40}$ &$200$    &$2.47^{2.00}_{1.35}$ & & \\
       &2  &     &                   &         & &$>145$ &$2.51^{2.50}_{2.41}$ \\
       &3  &$50$ &$15.8^{9.27}_{6.94}$ &$80$ &$11.6^{8.32}_{5.32}$ & & \\\hline
\end{tabular}
\begin{list}{}{}
\item $^a$ LVG gives a $T_{{\rm kin}}$ greater than 300 K which is the value allowed by the collisional rates given by \citet{Danby_et_al_1988}.
\item $^b$ Due that the values for the (4,4)-(5,5) metastable inversion transitions are upper limits to the actual $T_{\rm MB}$,  the modeled curve in the LVG plot is outside the allowed range, and we give a lower limit to the kinetic temperature using the rotational temperature form the LTE plot.
\item $^c$ The assumed value for the kinetic temperature for positions were there are no NH$_3$ observations is the average of the $T_{{\rm kin}}$ from similar positions.
\end{list}
\end{table*}  
%\end{landscape}

%__________________________________________________________________
\subsection{Column densities and relative abundances from other molecules}
\indent To shed light on the physical processes that are heating the
molecular gas, we derived also relative abundances of NH$_3$ with respect to the following molecules: 
SiO, which is a well-known shock tracer
\citep{Martin-Pintado_et_al_1992}, CS which is a high-density gas tracer
\citep[$n> 10^4 {\rm cm}^{-3}$][]{Mauersberger_Henkel_1989, Mauersberger_et_al_1989}, and HNCO which is a tracer of shocks, and very high densities, 
\citep[$n_{{\rm H_2}}\geq 10^6 {\rm cm}^{-3}$, ][]{Jackson_et_al_1984,Martin_et_al_2008,Zinchenko_et_al_2000} with a high photodissociation rate 
(Table \ref{fractional_abundances_molecules}). We also derive the fractional abundances of these molecules with respect to H$_2$ as traced by 
$^{13}$CO (Table \ref{fractional_abundances_H2}). We assumed that they arise from the same volume. This assumption may not be fulfilled for all of our
observed position due to the  observed differences in the velocity center and
linewidth (see Table \ref{SiO-CS-HNCO}).  In all the calculations we assume that the GC sources are extended, therefore we take $T_{\rm B}\sim T_{\rm MB}$.

\subsubsection{Column densities of SiO, HNCO, $^{13}$CO, and C$^{18}$O}
\indent We used the non-LTE excitation radiative transfer code
RADEX to derive the column densities. For the
species with only one observed transition (SiO, HNCO, and CO isotopes), we are forced to make some assumptions about the physical properties 
of the gas ($T_{\mathrm{kin}}$ and $n(\mathrm{H_2})$). We used the kinetic temperatures derived by NH$_3$, and the $n(\mathrm{H_2})$ from the CS 
data  (Table \ref{Tkin_nH2}). The results are shown in Table \ref{resultsmolecules}. The column densities agree within a factor of 3-4 if we 
use the LTE approach (Table \ref{SiO-CS-HNCO}) for a $T_{\mathrm{ex}}=10$ K (which is consistent with the value derived for SiO by \citet{Huettemeister_et_al_1998}, 
and with our $T_{\mathrm{ex}}$ value derived from CS shown in Table \ref{resultsCSLTE}).

Using $n(\mathrm{H_2})\sim 10^3$ cm$^{-3}$ (because the critical density of CO is lower than for CS) the column densities for the CO isotopes 
are a factor 2-4 lower than using the $n(\mathrm{H_2})\sim 10^4$ cm$^{-3}$.

\begin{table*}
\caption{Column densities from different molecules derived from RADEX using as a fixed parameter the kinetic temperatures and the $n(\mathrm{H_2})$  
(Table \ref{Tkin_nH2})} \label{resultsmolecules}
\begin{tabular}{lcccccccccc}
\hline\hline
%\\ 
Source&Cloud& T$_{\rm {kin}}$&  N(p-NH$_3$)& N(CS)&  N(SiO)   & N(HNCO)&N(C$^{34}$S)&  N($^{13}$CO) & N(C$^{18}$O)& N(H$_2$)$^a$\\
     &number&[K]&[$10^{14}\rm {cm}^{-2}$]& [$10^{13}\rm {cm}^{-2}$]&[$10^{13}\rm {cm}^{-2}$]   &[$10^{13}\rm {cm}^{-2}$]&[$10^{13}\rm {cm}^{-2}$] &[$10^{16}\rm {cm}^{-2}$]  &[$10^{16}\rm {cm}^{-2}$] &[$10^{21}\rm {cm}^{-2}$] \\\hline
Halo1 &1    & $115$ &$1.07^{0.68}_{0.27}$&  $>6.73$              & $>2.24$             &$>0.50$               & $-$                   &$4.47^{0.52}_{3.47}$  &$0.52^{0.07}_{0.41}$    &$23.53^{2.74}_{18.26}$ \\  	 
      &2    & $ 90$ &$2.24^{0.11}_{0.26}$& $15.99^{6.50}_{2.56}$ & $4.35^{1.96}_{0.87}$&$2.89^{0.79}_{0.46}$  &$1.20^{0.44}_{0.17}$   &$4.03^{0.67}_{0.75}$  &$0.97^{0.17}_{0.18}$    &$21.21^{3.53}_{3.95}$  \\	 
      &3    & $135$ &$0.70^{0.15}_{0.34}$& $6.53^{17.17}_{2.74}$ & $1.88^{5.20}_{0.86}$&$-$                   & $-$                   &$2.00^{0.75}_{1.06}$  &$0.16^{0.08}_{0.09}$    &$10.53^{3.95}_{5.58}$  \\\hline	 
Halo2 &1    & $113$ &  $-$               & $5.75^{16.65}_{2.73}$ & $0.76^{2.45}_{0.38}$&$1.64^{4.83}_{0.75}$  &$0.33^{1.67}_{0.21} $  & $-$                  & $-$                    &  $-$   \\	 
      &2    & $113$ &  $-$               & $>2.91$               & $>0.34$             &$-$                   &$>0.13$                &$-$                   & $-$                    &  $-$   \\\hline	 
Halo3 &1    & $113$ &  $-$               & $22.52^{3.55}_{10.64}$& $3.98^{0.59}_{1.99}$&$4.47^{0.55}_{1.74}$  &$1.43^{0.36}_{0.70} $  &$-$                   &  $-$                   &  $-$   \\	 
      &2    & $113$ &  $-$               & $3.71^{3.05}_{1.07}$  & $-$                 &$1.27^{0.72}_{0.19}$  &$0.29^{0.34}_{0.13} $  &$-$                   &  $-$                   &  $-$   \\\hline	 
Halo4 &1    & $ 95$ &$1.99^{0.53}_{0.56}$& $6.18^{5.32}_{2.17}$  & $1.00^{1.00}_{0.38}$&$3.98^{2.69}_{0.80}$  &$0.40^{0.42}_{0.18} $  &$7.35^{1.83}_{2.07}$  &$0.25^{0.08}_{0.08}$    &$38.68^{9.63}_{10.89}$ \\\hline	 
Halo5 &1    & $ 95$ &  $-$               & $5.29^{1.83}_{0.93}$  & $0.99^{0.42}_{0.22}$&$2.58^{0.76}_{0.32}$  &$0.47^{0.14}_{0.08} $  &$-$                   &  $-$                   &  $-$   \\\hline
Disk-X1-1&1 & $ 38$ &$1.99^{0.67}_{0.56}$& $12.64^{3.33}_{1.91}$ & $3.91^{1.30}_{0.76}$&$3.05^{0.74}_{0.48}$  &$0.64^{0.20}_{0.12} $  &$8.84^{1.12}_{1.15}$  &$0.51^{0.19}_{0.18}$    &$46.53^{5.89}_{6.05}$ \\ 	 
         &  & $300$ &$1.58^{0.58}_{0.48}$& $13.78^{6.31}_{2.82}$ & $3.98^{2.00}_{0.90}$&$2.88^{0.67}_{0.33}$  &$0.71^{0.35}_{0.18} $  &$32.8^{9.39}_{8.73}$  &$1.87^{1.08}_{0.87}$    &$172.63^{49.42}_{45.95}$ \\\hline	 
Disk-X2-1&1 & $ 38$ &$4.47^{0.39}_{0.69}$& $19.07^{15.10}_{5.44}$& $3.36^{2.95}_{1.12}$&$9.59^{4.00}_{1.02}$  &$0.80^{0.64}_{0.24} $  &$16.9^{2.45}_{2.94}$  &$1.37^{0.23}_{0.27}$    &$33.80^{4.90}_{5.88}$ \\	 
         &  & $100$ &$3.30^{0.43}_{0.98}$& $22.56^{37.65}_{8.27}$& $3.60^{6.76}_{1.36}$&$10.6^{13.0}_{2.01}$  &$0.98^{1.60}_{0.37} $  &$29.2^{7.55}_{11.1}$  &$2.51^{0.77}_{0.98}$    &$58.40^{15.10}_{22.20}$ \\\hline	 
Disk-X1-2&1 & $ 52$ &$1.66^{0.27}_{0.31}$& $10.60^{2.63}_{1.28}$ & $2.09^{0.62}_{0.35}$&$3.19^{0.89}_{0.56}$  &$0.71^{0.15}_{0.09} $  &$3.10^{0.40}_{0.41}$  &$0.32^{0.05}_{0.05}$    &$16.32^{2.11}_{2.16}$ \\	 
         &  & $215$ &$1.51^{0.30}_{0.33}$& $11.03^{3.71}_{1.70}$ & $2.10^{0.78}_{0.37}$&$2.96^{0.96}_{0.39}$  &$0.74^{0.22}_{0.11} $  &$8.09^{1.81}_{1.78}$  &$0.82^{0.23}_{0.21}$    &$42.58^{9.53}_{9.37}$ \\\hline	 
Disk-X2-2&1 & $ 28$ &$1.90^{0.51}_{0.55}$& $10.67^{3.49}_{1.73}$ & $0.22^{0.07}_{0.04}$&$2.08^{0.42}_{0.32}$  &$0.93^{0.29}_{0.15} $  &$12.8^{1.55}_{1.58}$  &$1.14^{0.13}_{0.14}$    &$25.60^{3.10}_{3.16}$ \\	 
         &  & $ 95$ &$1.00^{0.60}_{0.52}$& $11.07^{5.26}_{2.48}$ & $0.27^{0.19}_{0.08}$&$2.00^{0.57}_{0.25}$  &$0.96^{0.48}_{0.19} $  &$28.7^{5.26}_{5.83}$  &$2.62^{0.52}_{0.53}$    &$57.40^{10.52}_{11.66}$ \\\hline	 
Disk1  &1   & $ 23$ &$0.40^{1.84}_{0.26}$& $>2.90$               & $>0.25$             &$>1.12 $              &$>0.03$                &$1.26^{0.52}_{0.55}$  &$0.13^{0.05}_{0.07}$    &$2.52^{1.04}_{1.10}$ \\	 
       &    & $154$ &$>0.16$             & $>7.20$               & $>1.43$             &$>2.82 $              &$>0.09$                &$1.32^{0.15}_{0.61}$  &$0.14^{0.02}_{0.08}$    &$2.64^{0.30}_{1.22}$ \\        	                   
       &2   & $ 38$ &$0.63^{0.20}_{0.19}$& $3.89^{2.80}_{1.25}$  & $0.45^{0.37}_{0.17}$&$1.72^{0.79}_{0.27}$  &$0.16^{0.13}_{0.06} $  &$2.26^{0.35}_{0.40}$  &$0.19^{0.04}_{0.04}$    &$4.52^{0.70}_{0.80}$ \\	 
       &    & $ 82$ &$>0.45$             & $4.54^{5.48}_{1.70}$  & $0.50^{0.62}_{0.22}$&$2.00^{1.99}_{0.52}$  &$0.19^{0.26}_{0.08} $  &$3.46^{0.85}_{1.08}$  &$0.28^{0.09}_{0.10}$    &$6.92^{1.70}_{2.16}$ \\\hline	 
Disk2  &1   & $ 68$ &$14.1^{0.19}_{0.15}$& $8.98^{5.18}_{2.46}$  & $2.98^{2.12}_{0.87}$&$-$                   &$3.05^{1.83}_{0.76} $  & $-$                  &$-$                     &$-$   \\	 
       &    & $200$ &$14.1^{0.02}_{0.16}$& $10.03^{8.07}_{3.04}$ & $3.16^{2.71}_{1.06}$&$-$                   &$3.45^{2.86}_{1.04} $  & $-$                  & $-$                    &$-$   \\	 
       &2   & $145$ &$25.1^{0.07}_{0.07}$& $>16.51$              & $>5.63$             &$>264.4$              &$>5.67$                &$101.28^{28.47}_{78.30}$&$19.63^{6.48}_{16.39}$&$202.56^{56.94}_{156.60}$ \\	 
       &3   & $ 50$ &$14.1^{0.09}_{0.05}$& $16.89^{4.31}_{1.86}$ & $-$                 &$-$                   &$-$                    &  $-$                 &  $-$                   &  $-$   \\	 
       &    & $ 80$ &$12.1^{0.07}_{0.07}$& $17.05^{5.18}_{2.08}$ & $-$                 &$-$                   &$-$                    &  $-$                 &   $-$                  &  $-$   \\\hline
\end{tabular}				     		         
\begin{list}{}{}
\item $^a$ Column densities of H$_2$ derived from $^{13}$CO using a conversion factor of $5.0\times 10^{-6}$ \citep{Rodriguez-Fernandez_et_al_2001} for the normal GC gas, and a factor of $1.9\times 10^{-6}$ for the gas in the disk-halo interaction regions (see text for details).
\end{list}
\end{table*}

The total column density of H$_2$ can be estimated from the CO isotopologues, that have lower optical depths than the main isotope, and therefore a more reliable estimation of the column density  ($N({\rm H_2})=N({\rm ^xC^yO})\times [{\rm ^xC^yO/H_2}]$, where $x$ and $y$ correspond to the isotopic substitution used for the carbon and oxygen atoms). For our calculations, we assume an abundance ratio CO/H$_2$ of 10$^{-4}$ \citep[see, e.g., ][ and references therein]{Rodriguez-Fernandez_et_al_2001}, and we use the $^{13}$CO emission. Therefore, we also need the $^{12}$C/$^{13}$C isotopic ratio.  Recently, \citet{Riquelme_et_al_2010a}
derived a high $^{12}$C/$^{13}$C isotopic value ($>40$) in some of the sources
studied in this work. Such a high isotopic
value was found mainly toward the disk-halo interaction regions, therefore, we
still use the standard value of 20 \citep[see, e.g.,
][]{Wilson_Matteucci_1992} in the typical GC gas for the
sources ``Disk1'', ``Disk2'' and for the sources with kinematic of X2
orbits (Disk\,X2-1, Disk\,X2-2). For the sources which are in the
disk-halo interaction regions, we used the value of 53 corresponding to
the typical value found in the 4 kpc molecular ring
\citep{Wilson_Rood_1994}, which also was used by
\citet{Torii_et_al_2010b} and \citet{Kudo_et_al_2011} in the GMLs regions. This translates in using a $\mathrm{[^{13}CO/H_2]}$ conversion 
factor of $5.0\times 10^{-6}$  for the normal GC gas, and  $1.9\times 10^{-6}$ for the gas in the disk-halo interaction regions (Table \ref{resultsmolecules}).
We decided not to use the C$^{18}$O emission to trace the total column density of
H$_2$, because of the uncertainties in the  $^{16}$O/$^{18}$O isotopic ratio in
the disk-halo interaction regions, which could be affected by the unprocessed
gas that is being accreted toward the GC region \citep{Riquelme_et_al_2010a}.

\subsubsection{Fractional abundances}
\indent We derived beam averaged fractional abundances with respect to H$_2$ for all the observed molecules (Table \ref{fractional_abundances_H2}), and
in Table \ref{fractional_abundances_molecules} we show the results of the
fractional abundances of NH$_3$ with respect to the other molecules,
and the fractional abundances of SiO and HNCO with respect to CS and
C$^{34}$S.
\begin{table*}
\caption{Fractional abundances of SiO, HNCO, NH$_3$, CS, and C$^{34}$S with respect to H$_2$ \label{fractional_abundances_H2}}
\begin{tabular}{lccccccl}
\hline\hline
%\\ 
Source & $n^a$&$X({\rm SiO})$&$X({\rm HNCO})$&$X({\rm NH_3})^b$& $X({\rm CS})$&$X({\rm C^{34}S})$ \\
       &      & $\times10^{-9}$&$\times10^{-9}$ &$\times10^{-8}$&$\times10^{-9}$&$\times10^{-10}$\\\hline
Halo1  &1     &$1.42 \pm 0.95$ &$0.32 \pm 0.21$&$1.70 \pm 1.25$  &$4.27 \pm 2.84$ &                \\ 
       &2     &$2.33 \pm 0.79$ &$1.45 \pm 0.39$&$2.12 \pm 0.40$  &$9.16 \pm 2.25$ & $6.36 \pm 1.84$\\ 
       &3     &$4.17 \pm 3.73$ &               &$1.42 \pm 0.77$  &$14.15\pm 12.38$&                \\\hline
Halo2  &1     &                &               &                 &                &                \\       
       &2     &                &               &                 &                &                \\\hline
Halo3  &1     &                &               &                 &                &                \\
       &2     &                &               &                 &                &             \\\hline
Halo4  &1     &$0.34 \pm 0.20$ &$1.29 \pm 0.58$&$1.11 \pm 0.37$  &$2.04 \pm 1.13$ & $1.37 \pm 0.87$\\\hline
Halo5  &1     &                &               &                 &                &                \\\hline
DiskX1-1&1    &$0.39 \pm 0.12$ &$0.28 \pm 0.07$&$0.35 \pm 0.09$  &$1.31 \pm 0.37$ & $0.67 \pm 0.20$\\\hline
DiskX2-1&1    &$1.20 \pm 0.58$ &$3.08 \pm 1.12$&$1.72 \pm 0.41$  &$6.94 \pm 3.24$ & $2.94 \pm 1.39$\\\hline
DiskX1-2&1    &$0.77 \pm 0.18$ &$1.12 \pm 0.25$&$1.10 \pm 0.21$  &$3.95 \pm 0.86$ & $2.60 \pm 0.54$\\\hline
DiskX2-2&1    &$0.07 \pm 0.02$ &$0.52 \pm 0.10$&$0.76 \pm 0.18$  &$2.91 \pm 0.70$ & $2.55 \pm 0.60$\\\hline
Disk1  &1     &$3.60 \pm 1.01$ &$8.44 \pm 2.37$&$2.40 \pm 0.67$  &$21.63\pm 6.08$ & $2.57 \pm 0.72$\\\hline
       &2     &$1.12 \pm 0.49$ &$4.22 \pm 1.45$&$1.94 \pm 0.36$  &$9.94 \pm 4.13$ & $4.21 \pm 1.87$\\\hline
Disk2  &1     &                &               &                 &                &                \\
       &2     &$0.37 \pm 0.26$ &$17.31\pm12.10$&$3.29 \pm 2.30$  &$1.08 \pm 0.76$ & $3.71 \pm 2.60$\\
       &3     &                &               &                 &                &                \\\hline
\end{tabular}
\begin{list}{}{}
\item $^a$ Cloud number defined by the different velocity components.
\item $^b$ Column density of NH$_3$ correspond to the $N(p-NH_3)+N(o-NH_3)=2\times N(p-NH_3)$ assuming a ortho-to-para ratio of 1.
\end{list}
\end{table*} 

The results for Disk\,1, will be disscused below. $X({\rm SiO})$ varies from $0.07-4.17 \times 10^{-9}$ with the largest value toward the 
foot points of the GMLs.  $X({\rm HNCO})$ shows less variation in all the observed sources ranging from $0.28$ to $3.08\times 10^{-9}$, 
except for the Disk\,2, which has a large abundance of $17.31\times 10^{-9}$. The fractional abundances with respect to H$_2$ depend on 
a reliable estimation of the H$_2$ column density, which depends in a number of assumption (H$_2$ to CO conversion factor, physical parameters 
used to derive the column densities of $^{13}$CO). Therefore, we also compare the column density of the different molecular tracer 
(shock, photodissociation) with respect to CS (dense tracer) and its  C$^{34}$S isotope (Table \ref{fractional_abundances_molecules}), 
which is almost certainly optically thin because in the GC region $^{34}$S is $\sim 10$ times rarer than the main isotope \citep{Chin_et_al_1996}.

Although there are no big differences from source to source in the $N(\mathrm{SiO})/N(\mathrm{CS})$ and $N(\mathrm{SiO})/N(\mathrm{C^{34}S})$ 
ratios, we can see that the highest values are found toward Halo\,1 and Disk\,X1 sources, with a difference up to 1 order of magnitude if we 
compare the Disk\,X1-2 with the Disk\,X2-2 sources. 
The relative abundance $N(\mathrm{HNCO})/N(\mathrm{CS})$ and $N(\mathrm{HNCO})/N(\mathrm{C^{34}S})$ toward the Disk\,2 source is by far highest. The source Disk\,2
corresponds to Sgr\,B2M (20'',100'') from 
\citet{Martin_et_al_2008}, which is classified as a ``typical
Galactic center cloud'' in their work. They found a large
HNCO/$^{13}$CS abundance ratio of $68\pm13$ in that source. 

We exclude  the source Disk\,1 from the previous analysis, because the determination of their physical parameters (kinetic temperature) 
would be overestimated. The metastable inversion transitions (4,4) and (5,5) of NH$_3$ were not detected, therefore the kinetic temperature 
should not be high. In our radiative transfer calculations we use an upper limit to the rotational temperature (154 K), which was taken as 
the kinetic temperature of the gas. It is probable that the actual kinetic temperature (if there is a high temperature regimen in this source) 
could be much lower (similar to the value for the disk X2 positions, which correspond to typical gas in the CMZ). The LVG column density of SiO 
on this source (Table \ref{resultsmolecules}) is a factor of $\sim 30$ larger than the value from the LTE (Table \ref{SiO-CS-HNCO}), while the 
differences for the other positions are only a factor 2-4.

%%______________________________________________________________
\section{Discussion}
\subsection{Kinetic temperatures toward the Galactic center region}
\indent The derived kinetic temperatures for the halo positions are consistent with one high temperature regime ($>90$ K), 
while the clouds in the CMZ are consistent with two temperature regimes ($\sim 40$ and $\sim 150$ K).
\subsubsection{Single temperature regime in the loop region}
\indent We derived a single high kinetic temperature regime ($>95$
K) for the halo sources (Halo\,1 and Halo\, 4), towards the GML discovered by \citet{Fukui_et_al_2006}. Surprisingly, both the Halo\,1 
position in the footpoint of the GMLs and Halo\,4 in the top of the loop do not show any trace of low kinetic temperatures, which otherwise 
are present throughout the CMZ as discussed in previous works \citep{Huettemeister_et_al_1993b, Huettemeister_et_al_1998}. 
\citet{Torii_et_al_2010b} derived kinetic temperature of 30-100 K or higher, and densities of $10^{3}$ cm$^{-3}$ using multitransitional 
CO observations toward the foot point of the GML (loop 1 and 2). This foot point corresponds to our Halo\,2, and Halo\,3, which could not 
be observed with the Effelsberg telescope due to their low declination. Furthermore, \citet{Torii_et_al_2010a} made a comparison between the 
foot points 1 and 2 with the two broad velocity features, the Clump 2 and $l=5\deg.5$ complex, finding that they share common properties such 
as the vertical elongation to the plane and large velocity spans of $50-150$ km\,s$^{-1}$ suggesting that they have a similar origin. Therefore, 
the physical processes that are occurring in the Halo\,1 position should be similar to those in the well studied foot point of the loop1 and 2. 

\indent Is our Halo\,1 position really located at the foot point of the GML?, or it is just along the line of sight given that we see the  GC region edge-on? 
This source is at $l=5\deg.5$ and at $\sim 790$ pc in projection of the GC $(l,b)=(0,0)$ position assuming a GC distance of $8.23\pm 0.2$ kpc 
\citep{Genzel_et_al_2010}. It was previously observed by several authors 
\citep{Bitran_et_al_1997,Fukui_et_al_2006,Sawada_et_al_2001, Lee_Lee_2003, Riquelme_et_al_2010b}; its main velocity component is at 
$98$ \kms \citep[from the HCO$^+$ data from ][]{Riquelme_et_al_2010b}, and  a large velocity width in all the species is observed in 
this and previous works, which indicates that this source indeed belongs to the GC region. Furthermore, we cannot rule out that the gas seen in 
the Halo\,1 position has some velocity component belonging to the GC region, but at smaller or larger distances. High resolution maps of the 
foot point region as well as maps of the complete loop are needed to reveal the morphology and kinematic of the complete loop to confirm the 
association of this position to the GMLs scenario, because this loop is tentatively detected by \citet{Fukui_et_al_2006}. Therefore, the high 
$^{12}$/$^{13}$C isotopic ratio found by \citet{Riquelme_et_al_2010a} toward this position, and towards the well studied foot point of the 
loop 1 and 2 would provide evidence for the GMLs scenario.

\indent Additional support for the GMLs scenario comes from the kinetic temperature gradient and large NH$_3$ abundances of the high 
metastable inversion transitions. The sense of the temperature gradient can help to establish whether the shocks are due to the GMLs 
scenario or to the ejection of gas from the disk due to star formation. Temperature that decreases from the disk (low latitudes) to 
the halo in the GC region would indicate that the material is falling from the halo to the galactic disk supporting the GML scenario, 
because the post shocked gas which has cooled down is at larger latitude than the recently heated material at the shock front 
\citep[see e.g., Fig. 16 of ][]{Genzel_1992}. The gradient will be in the opposite way if the material is being ejected. 
We observed that the kinetic temperature is slightly larger in the foot point than in the top of the loop, which tentatively supports the loops scenario.\\

\subsubsection{The two temperature components model in the CMZ clouds}
\indent  In the CMZ (Disk\,X1-1, Disk\,X1-2, Disk\,X2-1, Disk\,X2-2, and Disk\,2),
our results are consistent with a two components model, with  
at least at two different temperatures, one cool and one warm (see Table \ref{results} and \ref{Tkin_nH2}). 
This result is in agreement with \citet{Huettemeister_et_al_1993b, Huettemeister_et_al_1998}, who found that 
the data were consistent with two rotational temperature components: one cool ($\sim 25$ K) and other warm ($>100$ K). 
Furthermore, they found that for a typical GC molecular cloud, 25\% of the gas has high temperatures, and this gas has 
low H$_2$ density; while the remaining 75\% of the total gas mass is cooler at densities of $\geq 10^4$ cm$^{-3}$. Both gas
components are in pressure equilibrium. Our result, on the other hand, indicates that there is as much gas in the low 
temperature component ($\sim$ 50\%) as in the high temperature component ($\sim$ 50\%). 
A possible explanation for such a discrepancy may be a selection effect: The positions from
\citet{Huettemeister_et_al_1993b} are selected as intensity peaks in the
CS maps from \citet{Bally_et_al_1987}, therefore these correspond
to high density regions, as CS is a dense tracer. Our sources, on the contrary,
correspond to shock positions, therefore expecting that the amount of
warm molecular gas will be larger than at the positions from
\citet{Huettemeister_et_al_1993b}.   

\subsection{On the heating and cooling of the molecular gas in the GMLs and in  the CMZ}
\indent Although the number of positions 
observed in the halo is rather limited, a remarkable result obtained in this work, is the single high kinetic temperature regime 
($>95$ K) for the halo sources, in contrast with the two temperature regimens (cool and warm) present throughout the CMZ.  
 
Therefore, the question that naturally arises is whether the cooling in the CMZ is more effective than in the halo positions, 
or if the heating mechanism in the GMLs is so efficient that the molecular gas has no time to be cooled down.
\subsubsection{The cooling of the gas in the GC region}
In the following, we describe the cooling rates for H$_2$ and CO emissions, and for the gas-dust coupling.

\citet{Goldsmith_Langer_1978} derived the temperature dependence of the total cooling rate for a variety of molecular 
hydrogen densities and at velocity gradient of 1 km s$^{-1}$ pc$^{-1}$ for the temperature range of 10 to 60 K. According 
to them, for $n(\mathrm{H_2})\leq10^3\, \mathrm{cm}^{-3}$ the cooling is dominated by CO; and for 
$10^3\leq n(\mathrm{H_2})\leq 10^5\, \mathrm{cm}^{3}$, CI, O$_2$, and the isotopic species of CO,  contribute with the 30\% to 70\% of the total cooling. 
We use the expressions of the total cooling rates ($\Lambda_{\mathrm{total}}$) derived by them for the different physical conditions of the molecular gas 
in the positions where this formulation is valid (see Table \ref{c-h-rate}). \citet{Neufeld_et_al_1995} and \citet{Neufeld_Kaufman_1993} derived the 
radiative cooling rates in a wider range of temperatures ($10 \le T \le 2500$ K), and stated that the dominant coolants for the molecular interstellar
 gas are CO, H$_2$, O, and H$_2$O. The efficiency of the cooling depends of the gas temperature, the H$_2$ particle density, and the optical depth 
parameter $\tilde{N}(\mathrm{H_2})=n(\mathrm{H_2})/|dv/dz|$, where $dv/dz$ is the velocity gradient along the line of sight. The latter depends on 
the geometry and velocity structure 
assumed \citep[see Table 1 in][]{Neufeld_Kaufman_1993}. Here we assume 
$\tilde{N}(\mathrm{\mathrm{H_2}})=N(\mathrm{H_2})/\Delta v$; where $N(\mathrm{H_2})$ is the $\mathrm{H_2}$ column density estimated in 
column 11 of Table \ref{Tkin_nH2}, and $\Delta v$ is the measured line width of CS(2-1) (Table \ref{SiO-CS-HNCO}).
We compute the total cooling rates according to their model in Table \ref{c-h-rate} by spline interpolation of the values for their 
molecular cooling function \citep{Neufeld_et_al_1995}.

Later, \citet{LeBourlot_et_al_1999} derived the cooling rate by H$_2$ ($\Lambda_{\mathrm{H_2}}$) for a wider range of physical parameters 
($100 \leq T_{\mathrm{kin}}\leq 10^4$ K; $1\leq n(\mathrm{H_2}) \leq 10^8$ cm$^{-3}$). We used the FORTRAN  subroutine provided by them only 
towards the sources allowed by the parameters range, i.e., for the warm gas (Table \ref{c-h-rate}).  

The cooling rate for the coupling of the dust and gas is given by \citep{Goldsmith_Langer_1978} 
\begin{equation}
\label{cooling_gas_dust}
\Lambda_{\mathrm{gd}}=2.4\times 10^{-33} T_g^{1/2}(T_g-T_d)n^2(\mathrm{H_2}) \mathrm{\,erg\,s^{-1} cm^{-3}}~.
\end{equation}
where the grains parameters (grain size and accommodation coefficient) where taken from \citet{Leung_1975}.
\citet{Huettemeister_et_al_1993b} argued that the cold 
gas component is coupled to the dust temperature at high densities,
therefore, the dust in the GC region would be a cooling agent. They proposed that the density of the cold gas must be at least an order of
magnitude higher than that of the hot gas; otherwise the hot gas would also cooled down. We find that the gas density of the hot component is only slightly
lower than in the cold component (a factor $\sim 2-4$ from the CS data in Table \ref{resultsCSLTE}). Although the gas-dust coupling becomes 
significant at $n(H_2)\sim10^5$ cm$^{-3}$ \citep{Juvela_Ysard_2011}, which is reached only in the cool component regime in few positions in 
the CMZ (Table \ref{resultsCSLTE}), we estimated this cooling rate for all the positions (Table \ref{c-h-rate}). 

Estimating total cooling rates, e.g., following \citet{Goldsmith_Langer_1978}, one should account for two factors: the depletion of 
coolant species, and the lack of processed gas in the halo. 
The depletion of the coolant species can increase the gas temperature at low and moderate densities of $n(\mathrm{H_2})\leq 10^4$ cm$^{-3}$. 
This effect was studied by \citet{Goldsmith_2001} in dark clouds. In the physical conditions of the GC, the depletion of the coolant species 
is unlikely due to the high densities and low temperatures which are needed to have this effect. 
Second, as we noted before, in the density range of the GC clouds, CI, O$_2$, and the isotopes of CO,  contribute 30\% to 70\% of the total cooling. 
\citet{Riquelme_et_al_2010a} found a high $^{12}$C/$^{13}$C isotopic ratio towards the disk-halo connection regions (halo, disk X1) which they 
interpreted as non-processed gas being accreted towards the GC. If this interpretation is correct (since the gas in the loop is less processed 
than the gas in the CMZ), one would expect a lower metallicity and, hence, a less efficient cooling by molecular or atomic lines. On the other 
hand, a high isotopic ratio was also found towards the disk X1 positions, and those positions have indeed a cool temperature regime.

The low temperature regime toward the X1 orbit positions can be explained by the large H$_2$ cooling rate derived from \citet{LeBourlot_et_al_1999} 
and the large total cooling rates derived from \citet{Neufeld_et_al_1995} (see Table \ref{c-h-rate}). However, the cooling rate in the Halo positions 
are also large, and those positions do not present the cool temperature regime.

In the following, we estimate the heating rates affecting the GC region to study if a rise of the heating mechanism should be the responsible of 
the lack of low temperature regime in the halo sources.
\subsubsection{The heating of the gas in the GC}
\indent We find high gas kinetic temperatures for basically all the observed
positions of our sample. Therefore the heating mechanism responsible for the widespread high temperatures should apply for the gas in the entire GC region, 
with little effect on the dust. The heating mechanism in the GC should be different from that the heating of warm clouds (T$_{\mathrm{kin}}>100$ K) in the disk, 
where the molecules are heated by collisions with hot dust, heated by embedded stars. 

\indent Several heating mechanisms have been proposed to explain the
high kinetic temperatures in the GC region. 
\begin{enumerate}
\item Cosmic rays heating: Heating by a large flux of low energy cosmic rays 
\citep{Guesten_et_al_1981,Morris_et_al_1983, Huettemeister_et_al_1993b, Yusef-Zadeh_et_al_2007a}. This mechanism requires a cosmic ray 
ionization rate ($\zeta_{\rm CR}$) of one or two orders of magnitude larger than the Galactic value of 10$^{-17}{\rm s}^{-1}$, which will 
influence also the gas-phase chemistry, increasing the atomic hydrogen due to the increased cosmic ray dissociation rate of H$_2$, and 
also molecular ion emission like HCO$^+$.  \citet{Yusef-Zadeh_et_al_2007a,Yusef-Zadeh_et_al_2007b} argue that such a high ionization rate 
is found in the GC region. From absorption lines of H$_3^+$ originating from a diffuse, hot molecular component in the CMZ, 
\citet{Goto_et_al_2008} estimated an ionization rate of $\sim 10^{-15}\, s^{-1}$. We assume a similar value in the gas observed by us, 
although it is presumably denser and cooler.
The heating rate depends on $n$ and $\zeta_{\rm CR}$. It is difficult to derive it for each observed position, and this mechanism cannot be ruled out.
\item X-ray heating: Heating by an extended diffuse source of soft X-ray emission \citep{Watson_et_al_1981} is  unlikely since the X-ray emission 
is less  extended than the NH$_3$ emission, and there is no obvious source for the required luminosity in the extended soft X-ray emission 
\citep{Morris_et_al_1983}. Heating by an extremely hot plasma emitting X-rays.  \citet{Nagayama_et_al_2007} found that the NH$_3$ emitting 
region in the CMZ, with $T_{\rm kin}= 20-80$ K and $T_{\rm kin} > 80$ K, is surrounded, in the longitude-velocity space, by a high-pressure 
region \citep{Sawada_et_al_2001}, where the gas is less dense and hotter ($n(H_2)<10^3\, \mathrm{cm^{-3}}$, $T_{\rm kin}>100$ K). Because the 
high pressure region is  found to be coincident with the hot emitting X-rays, they argued that the thermal energy radiated from the hot plasma 
emitting X-ray plasma can heat the gas in the high pressure region. The heating rate due to this mechanism is very uncertain, but 
Ao et al., (2012, in prep.) argue that this mechanisms is not able to account for the high kinetic temperatures in the GC region.
\item Ion-Slip heating: The GC is pervaded by a magnetic field of few mG \citep[see e.g., ][]{Ferriere_et_al_2007}, and their presence can also 
influence the heating of the GC. The ion-slip heating has been proposed for the molecular clouds in the GC region, where the heating rate depends 
on  the magnetic field $B$, ionization rate, neutral number density $n_n$,  ion number density $n_i$, number of collision per second, the reduced 
mass of the ions and neutrals, and the scale of the cloud $R$ (in pc) \citep[see, ][]{Scalo_1977}. Because of the uncertainty of some on these 
values for each observed position, it is difficult to estimate the heating rate. 
\item Ultraviolet heating:  The high NH$_3$ abundances in the GC region require effective shielding from UV radiation because ammonia is easily 
photodisociated by ultraviolet radiation \citep{Rodriguez-Fernandez_et_al_2001}. This is also confirmated by the large abundance of the HNCO molecule, 
which is also photodissociated by UV radiation. We discard UV heating in the GC region.\\
HNCO could be formed via gas phase reactions, but formation in grains seems to be more efficient \citep[see e.g.,][and references
therein]{Martin_et_al_2008}. Presumably this molecule is
released to the ISM by grain erosion and/or disruption by shocks
\citep{Zinchenko_et_al_2000}, and is easily photodissociated by UV
radiation. The shocks that release the molecule from the grain
mantles should be slow enough to not dissociated the molecule. We find
that the X(HNCO) is low in the Disk\,X1-1 source, where we expect
shocks, but no UV emission. Also we can see an enhancement
toward the Halo\,1 source (cloud number 2), and the Disk\,X2-1 source, which could be due to the shocks present in these regions, 
are strong enough to evaporate the
molecule from the grain mantles but too slow to  not dissociate the HNCO.
This molecule can be used to trace the shocks
properties throughout the GC region.
\item Shocks: The dissipation of mechanical turbulence through shocks would offer a compelling answer, because the GC region shows 
an ubiquitous presence of shocks, as traced by the SiO emission (\citealp{Martin-Pintado_et_al_2000, Riquelme_et_al_2010b}).
The heating rate due to the dissipation of turbulence (of velocity $v_t$ on a spatial scale of $R_c$) is given by \citep{Black_1987}
\begin{equation}
\label{heating_turbulence}
\Gamma_{\mathrm{turb}}\approx 3.5 \times 10^{-28} v_t^3 n_H \left( \frac{1 \mathrm{pc}}{R_c}\right) \mathrm{\,erg\,s^{-1} cm^{-3}}
\end{equation}

We derive the heating rate of the dissipation of turbulence for each position in Table \ref{c-h-rate}. We use $R_c=5$ pc 
\citep{Rodriguez-Fernandez_et_al_2001}. It is important to note that this equation is highly dependent of $v_t$, therefore 
if the molecular clouds are not resolved in the beam size, the heating rate would be overestimated.\\
The origin of this turbulence in the positions studied in this work, can be the following:  the large scale dynamics in a barred 
potential model \citep{Binney_et_al_1991} can produce the shocks found towards the high velocity clouds associated with the X1 orbits 
(in the 1.3 complex, and Sgr\,C).  This is supported by the higher kinetic temperatures found in the disk\,X1 sources than in the 
disk\,X2 sources (Table \ref{Tkin_nH2}). In  our ``halo'' positions the shocks can be produced by the GMLs scenario, which is supported
 by the broad velocity width at the foot point of the loops. 
However, high kinetic temperatures and the large SiO emission are widespread throughout all the GC region. Supernova or hypernova 
explosions could cause for the high temperature found in the lower velocity components (in our notation, disk\,X2) of the 1\deg.3 complex 
\citep{Tanaka_et_al_2007}.  For the lower longitudes in the CMZ, the heating could be explained by interactions with SNRs close to Sgr\,A 
and also with non-thermal filaments in the radio arc, and cloud-cloud collisions, and expanding bubbles in the vicinity of Sgr\,B \citet{Martin-Pintado_et_al_1997}. 
Cloud-cloud collisions were also proposed in the GC region \citep{Wilson_et_al_1982}, which is favored by the large linewidth typical 
in the GC clouds. \citet{Guesten_et_al_1985} argue that if this mechanism is acting in the molecular gas, the linewidth and the 
temperatures should be correlated.  \citet{Mauersberger_et_al_1986a} found such a correlation for the clouds observed in the GC, 
which would support this mechanism. Like \citet{Huettemeister_et_al_1993b}, we do not confirm such a clear correlation between the 
rotational temperature associated with the inversion transitions (2,2), (4,4), and (5,5) (Table \ref{results}) and the Doppler width 
of our sources (see Fig \ref{trotational_ancho}). For the few observed positions we can neither confirm nor reject the cloud-cloud 
collisions as primary heating mechanism in our sources.
\end{enumerate}
The heating rates derived from Eq. \ref{heating_turbulence} for the halo and disk positions are similar.  Therefore, it is probably that this 
mechanism by itself is not causing the lack of the low temperature regimen observed in the halo positions. An extra heating input would be 
required for the halo positions. \citet{Torii_et_al_2010b} propose that the gas in the foot point of the GML is heating by C-shocks, and 
that the warmest region of the foot point (the ``U shape'') would also be heated by magnetic reconnection or by upward flowing gas bounced 
by the narrow neck in the foot point. They estimated that the total available energy (considering the magnetic and gravitational energy) 
injected to the U shape is $1.8-2.6\times 10^{37} \mathrm{erg s^{-1}}$ in $\sim 10^6$ years. Consider a cloud size of 3 pc radius, as estimated 
by \citet{Torii_et_al_2010b}, the heating rate is $\Gamma \sim 3.9-7.8 \times 10^{-21}$ erg cm$^{-3}$ s$^{-1}$. This heating rate is negligible 
in comparison with the values obtained  from the dissipation of turbulence alone. Alternatively, the CMZ can be understood as a highly turbulent 
medium, where many phenomena are taking place (shocks produced by Galactic potential, SN explosions, star formation, interaction with non-thermal filaments, 
cosmic rays, the presence of a supermassive black hole, etc) and coexist in the central few hundred parsec of the Galaxy.  All of these phenomena modify 
the physical parameter of the region. The two kinetic temperature regimes present in this region, which are in pressure equilibrium 
\citep{Huettemeister_et_al_1998}, could be the results of the interplay of the different phenomena mentioned above. On contrary, in the 
GMLs scenario, the gas goes down towards the Galactic plane following the magnetic field lines, which lead a tidy movement of the gas producing 
shocks front at the foot point of the loops. The continuous shock at the foot point is not affected by other phenomena (like, e.g., star formation), 
therefore  the gas is continuously heated. High resolution maps of the foot point of the GMLs are needed not only to confirm this hypothesis
 but also to resolve the linewidth of the molecular clouds to better estimate the heating rate for dissipation of turbulence (Eq. \ref{heating_turbulence}). 
Summarizing, in spite of the limited number of positions studied in this work, we propose that the high kinetic temperature found in all 
of the sources are produced by shocks. We discarded UV heating due to the large abundance of HNCO and NH$_3$ molecule. From our data it is 
not possible to confirm or rule out x-ray and ion-slip heating.

\begin{figure}
\includegraphics[width=0.35\textwidth, angle=90]{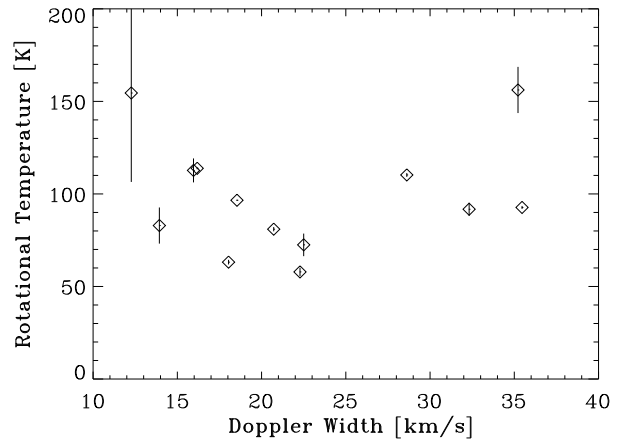}
\caption{$T_{rot}$ is plotted versus the average velocity linewidth from all the metastable inversion transitions. The $T_{rot}$ correspond 
to the values in bold from the Table \ref{results}}
\label{trotational_ancho}
\end{figure}

\section{The ammonia ortho-to-para ratio and its implications on kinetic temperatures now and there}
\indent Radiative and collisional transitions between ortho-NH$_3$ $(K=0, 3, 6, 9...)$ and para-NH$_3$ $(K=1, 2, 4, 5...)$ are 
forbidden because of their different nuclear spins. The time scale of conversion between ortho and para species is $10^6$ yrs in 
the gas phase \citep{Cheung_et_al_1969}. Therefore ortho-NH$_3$ and para-NH$_3$ can be almost treated as different species. 
The spin temperatures between ortho and para-NH$_3$, or the ortho/para ratio, may reflect the conditions at the time of the formation of 
NH$_3$, while rotational temperature within the same species reflect present conditions \citep{Ho_Townes_1983,Umemoto_et_al_1999}.

A ratio of 1.0 would be expected if NH$_3$ is formed in the gas phase reactions at high temperatures, and a larger value could be explained 
by a formation on cold dust grains and a subsequent release into the gas phase. For temperatures of about 10 K, the typical temperature of 
cold dust, the ortho-to-para ratio will be larger than 2.

\citet{Dulieu_2011}, however, points out that there might be a differential desorption on ortho and para molecules from dust grains and that this, 
and not the formation process, determines the gas phase ortho/para ratios.

\indent From the LVG analysis of the ammonia molecule, we derive an ortho-to-para ratio average for all of our sources of $0.70$ with a 
standard deviation of $0.20$ (see Table \ref{ortho_para_ratio}). In this calculation, we take into account only the warm component of the 
para-NH$_3$ with the ortho-NH$_3$. While the ortho-NH$_3$  transitions are probably tracing higher kinetic temperatures than the para-species 
(see Table \ref{parameters_ortho}), the ortho-to-para ratio derived using RADEX for the warm kinetic temperature component approaches to the 
statistical equilibrium value. 
This result is consistent with \citet{Huettemeister_et_al_1993b} estimation of the $(3,3)$ column density. 
Non thermal emission has been predicted for the  NH$_3$ (3,3) line by \citet{Walmsley_Ungerecht_1983}, and has been observed e.g. by 
\citet{Mauersberger_et_al_1986a,Mauersberger_et_al_1986b}. Also the $(6,6)$ line might be a maser source in some case \citep{Lebron_et_al_2011}. 
Our RADEX calculations were made assuming a background of 2.7 K. We have verified that in our models the ortho lines tend to be weak masers 
for a large part of our parameter space. 
\begin{table}
\caption{Ortho-to-para NH$_3$ ratio from LVG model \label{ortho_para_ratio}}
\begin{tabular}{lcc}
\hline\hline
%\\ 
Source   &$^a$ &$N(o-NH_3)/N(p-NH_3)^b$ \\\hline
Halo1    &1     &$0.83$\\ 
         &2     &$0.63$\\ 
         &3     &$0.68$\\\hline
Halo4    &1     &$0.48$\\\hline
DiskX1-1 &1     &$1.0$\\\hline
DiskX2-1 &1     &$0.76$\\\hline
DiskX1-2 &1     &$0.93$\\\hline
DiskX2-2 &1     &$0.76$\\\hline
Disk1    &1     &$2.02^c$\\
         &2     &$0.72^c$\\\hline
Disk2    &1     &$0.59$\\
         &2     &$0.63$\\
         &3     &$0.34$\\\hline
\end{tabular}
\begin{list}{}{}
\item $^a$ Cloud number defined by the different velocity components.
\item $^a$ We take into account only the warm component of the para-NH$_3$ with the ortho-NH$_3$ (see text for details).
\item $^c$ This value was not considered in the average of the ortho-to-para ratio, because it is possible that this position does not have the high kinetic temperature regimen.
\end{list}
\end{table}

%______________________________________________________________

\section{Conclusions}

   \begin{enumerate}
      \item We have used the metastable inversion transitions of NH$_3$ from $(J,K)=(1,1)$ to $(6,6)$ to derive the gas kinetic temperature toward 
six positions selected in the Galactic central disk and for higher latitude molecular gas.  We also observed other molecules like SiO, HNCO, CS, 
C$^{34}$S, C$^{18}$O, and $^{13}$CO, to derive the densities and to trace different  physical processes (shocks, photodissociation, dense gas) which 
was  used to reveal the heating mechanisms affecting the molecular gas in these regions.
      \item The GC molecular gas consists of roughly two kinetic temperature components in the CMZ. Only the warm kinetic temperature regime is 
found in the ``halo'' positions, and in the Disk\,2 position, which corresponds to Sgr\,B2. The results obtained in this paper apply for the disk 
and halo positions defined in this work within the GC region, and do not represent the conditions for the Galaxy as a whole.
      \item The kinetic temperatures are high, not only in the typical GC clouds, but also in the high latitude and high velocity clouds observed 
in this paper.
      \item Shocks are a compelling heating mechanism of the molecular clouds in the GC region. This is supported by the high gas kinetic temperature 
and by the increased SiO abundance in the location where shock are expected. Due to the fragile nature of the HNCO (enhanced by shock but easily 
photodissociated by UV radiation and strong shocks), this molecule could be used to reveal the characteristic of the shock. Other heating mechanisms 
previously proposed for the GC, however, cannot been ruled out.
      \item The high kinetic temperatures found in both, the X1 orbits and in the foot point of the GMLs, seem to support to the large scale dynamics 
induced by the bar potential, and the GMLs scenario as origins for the shocks respectively.
     
   \end{enumerate}

\begin{acknowledgements}
D.R. and M.A.A-B. were supported by Radionet during the observations.
D.R. was supported by DGI grant AYA 2008-06181-C02-02. J.M-P. and S.M. have been partially supported by the Spanish MICINN under grant numbers 
ESP2007-65812-C02-01 and AYA2010-21697-C05-01. L.B. acknowledges support from CONICYT projects FONDAP 15010003 and Basal PFB-06. SM acknowledge 
the co-funding of this work under the Marie Curie Actions of the European Commission (FP7-COFUND).
D.R. thanks the Joint ALMA Observatory for its hospitality. We thank to David Neufeld for kindly providing the tabulated values of their molecular cooling function.
\end{acknowledgements}

\bibliographystyle{aa} % style aa.bst
\bibliography{referencias} % your references Yourfile.bib

%______________________________________________________________
%online appendix
%\Online
\clearpage
\begin{appendix}
\section{Complementary Tables}
\onecolumn

\begin{landscape}
\begin{table}
\caption{Fractional abundances of NH$_3$ with respect to SiO, HNCO, CS, C$^{34}$S, and of SiO and HNCO with respect to CS and C$^{34}$S \label{fractional_abundances_molecules}}
{
\begin{tabular}{lcccccccccl}
\hline\hline
%\\ 
Source &$n^a$ &{\large $\frac{N(NH_3)}{N(SiO)}$}&{\large $\frac{N(NH_3)}{N(HNCO)}$}&{\large $\frac{N(NH_3)}{N(CS)}$}&{\large $\frac{N(NH_3)}{N(C^{34}S)}$}& {\large $\frac{N(SiO)}{N(CS)}$}&{\large $\frac{N(SiO)}{N(C^{34}S)}$}&{\large $\frac{N(HNCO)}{N(CS)}$}&{\large $\frac{N(HNCO)}{N(C^{34}S)}$}&{\large $\frac{N(SiO)}{N(HNCO)}$} \\\hline
Halo1  &1 &$11.97\pm 3.64$ &$53.64 \pm 16.28$&$3.99 \pm 1.21$  &                  &$>0.33$         &                &$>0.07$         &                &$>4.48$          \\ 
       &2 &$9.11 \pm 2.68$ &$14.59  \pm 3.09$&$2.32 \pm 0.41$  &$33.40 \pm 7.84$  &$0.25 \pm 0.09$ &$3.67 \pm 1.35$ &$0.16 \pm 0.04$ &$2.29 \pm 0.70$ &$1.60 \pm 0.57$  \\ 
       &3 &$3.41 \pm 2.67$ &                 &$1.01 \pm 0.76$  &                  &$0.29 \pm 0.31$ &                &                &                &                 \\\hline
Halo2  &1 &                &                 &                 &                  &$0.14 \pm 0.15$ &$1.69 \pm 2.01$ &$0.29 \pm 0.31$ &$3.47 \pm 4.05$ &$0.49 \pm 0.53$  \\
       &2 &                &                 &                 &                  &$>0.12$         &                &                &                &               	 \\\hline
Halo3  &1 &                &                 &                 &                  &$0.17 \pm 0.09$ &$2.60 \pm 1.50$ &$0.20 \pm 0.10$ &$3.08 \pm 1.58$ &$0.85 \pm 0.42$  \\
       &2 &                &                 &                 &                  &                &                &$0.33 \pm 0.17$ &$3.89 \pm 2.58$ &                 \\\hline
Halo4  &1 &$32.32\pm 18.12$&$8.60  \pm 3.46$ &$5.46 \pm 2.84$  &$81.43\pm 49.50$  &$0.17 \pm 0.12$ &$2.52 \pm 1.97$ &$0.64 \pm 0.38$ &$9.47 \pm 6.41$ &$0.27 \pm 0.17$  \\\hline
Halo5  &1 &                &                 &                 &                  &$0.19 \pm 0.07$ &$2.18 \pm 0.80$ &$0.49 \pm 0.15$ &$5.60 \pm 1.64$ &$0.39 \pm 0.14$  \\\hline
DiskX1-1&1&$8.91 \pm 2.30$ &$12.46 \pm 2.51$ &$2.69 \pm 0.65$  &$52.80\pm 13.87$  &$0.30 \pm 0.08$ &$5.93 \pm 1.74$ &$0.22 \pm 0.05$ &$4.24 \pm 1.04$ &$1.40 \pm 0.34$  \\\hline
DiskX2-1&1&$14.33 \pm 6.28$&$5.57  \pm 1.70$ &$2.48 \pm 1.04$  &$58.46\pm 24.92$  &$0.17 \pm 0.10$ &$4.08 \pm 2.44$ &$0.44 \pm 0.22$ &$10.49\pm 5.33$ &$0.39 \pm 0.20$  \\\hline
DiskX1-2&1&$14.32 \pm 2.76$&$9.83  \pm 1.76$ &$2.78 \pm 0.48$  &$42.41 \pm 6.94$  &$0.19 \pm 0.04$ &$2.96 \pm 0.62$ &$0.28 \pm 0.06$ &$4.31 \pm 0.86$ &$0.69 \pm 0.15$  \\\hline
DiskX2-2&1&$111.3 \pm 34.7$&$14.52 \pm 3.28$ &$2.60 \pm 0.70$  &$29.67 \pm 7.91$  &$0.02 \pm 0.01$ &$0.27 \pm 0.08$ &$0.18 \pm 0.04$ &$2.04 \pm 0.47$ &$0.13 \pm 0.04$  \\\hline
Disk1  &1 &$6.67 \pm 0.00$ &$>2.84$          &$>1.11$          &$93.33 \pm 0.00$  &$>0.17$         &$>14.00$        &$>0.39$         &$>32.83$        &$>0.43$          \\
       &2 &$17.28 \pm 6.91$&$4.59  \pm 1.32$ &$1.95 \pm 0.72$  &$45.96\pm 18.58$  &$0.11 \pm 0.06$ &$2.66 \pm 1.51$ &$0.42 \pm 0.20$ &$10.02\pm 4.98$ &$0.27 \pm 0.13$  \\\hline
Disk2  &1 &$74.31\pm 23.50$&                 &$24.65\pm 7.26$  &$71.03\pm 20.94$  &$0.33 \pm 0.14$ &$0.96 \pm 0.41$ &                &                &                 \\
       &2 &$89.23 \pm 0.19$&$>1.90$          &$30.43\pm 0.06$  &$88.60 \pm 0.19$  &$>0.34$         &$>0.99$         &$>16.01$        &$>46.63$        &$>0.02$          \\
       &3 &                &                 &$14.30\pm 1.86$  &                  &                &                &                &                &                 \\\hline
\end{tabular}
\begin{list}{}
\item $^a$ Cloud number defined by the different velocity components.
\end{list}
}
\end{table} 
\end{landscape}

\begin{table}
 \caption{Estimation of heating and cooling rates for each source\label{c-h-rate}}
\begin{tabular}{lc|cc|ccccc}
\hline\hline
Source&$n^a$ &$T_{kin}$& $n(H_2)$&$\Lambda_{total}^{b,e}$ & $\Lambda_{total}^{c,e}$  & $\Lambda_{H_2}^{d,e}$ &$\Lambda_{gd}^{d}$ & $\Gamma_{turb}^{e}$  \\
      & &  [K]& [$\times 10^4$cm$^{-3}$]&&&\\\hline 
Halo1  &1   & 115 &$  1.00$ &       & 14.0 &  0.78  & 0.018 & 27.5 \\       
Halo1  &2   &  90 &$  8.49$ &       & 242  &  4.36  & 0.86  & 58.4 \\       
Halo1  &3   & 135 &$  1.58$ &       & 46.8 & 31.4   & 0.057 & 12.9 \\\hline 
Halo2  &1   & 113 &$  1.26$ &       & 21.6 &  0.015 & 0.028 & 31   \\       
Halo2  &2   & 113 &$  0.97$ &       & 13.3 &  0.82  & 0.016 & 16.5 \\\hline 
Halo3  &1   & 113 &$  1.50$ &       & 30.7 &  1.3   & 0.039 & 62.6 \\       
Halo3  &2   & 113 &$  3.55$ &       & 136  &  3.1   & 0.22  & 178  \\\hline 
Halo4  &1   &  95 &$  2.75$ &       & 44.9 &  1.4   & 0.10  & 41   \\\hline 
Halo5  &1   &  95 &$  7.47$ &       & 210  &  3.9   & 0.74  & 51.4 \\\hline 
Disk-X1&1   &  38 &$ 13.0 $ &   1.5 & 36.7 &        & 0.081 & 849  \\       
       &    & 300 &$  3.55$ &       & 534  & 261    & 0.79  & 232  \\\hline 
Disk-X2&1   &  38 &$  5.62$ &   1.5 & 12.7 &        & 0.015 & 234  \\       
       &    & 100 &$  2.70$ &       & 48.5 &  1.2   & 0.11  & 113  \\\hline 
Disk-X1&2   &  52 &$ 13.3 $ &   3.6 & 101  &        & 0.51  & 76.3 \\       
       &    & 215 &$  5.62$ &       & 746  & 112    & 1.3   & 32.2 \\\hline 
Disk-X2&2   &  28 &$ 14.1 $ &  0.60 & 16.5 &        & 0.57  & 158  \\       
       &    &  95 &$  6.20$ &       & 150  &  3.3   & 0.51  & 69.4 \\\hline 
Disk1  &1   &  23 &$  1.12$ & 0.071 & 1.26 &        & 0.005 & 5.29 \\      
       &    & 154 &$  0.16$ &       & 1.37 &  0.51  & 0.001 & 0.76 \\       
Disk1  &2   &  38 &$  4.57$ &  0.28 & 25.8 &        & 0.010 & 31.4 \\       
       &    &  82 &$  2.51$ &       & 45.1 &        & 0.064 & 17.3 \\\hline 
Disk2  &1   &  68 &$  4.97$ &   1.3 & 48.6 &        & 0.17  & 250  \\       
       &    & 200 &$  2.47$ &       & 179  & 34     & 0.23  & 124  \\       
Disk2  &2   & 145 &$  2.51$ &       & 55.8 &  8.1   & 0.16  & 32.5 \\       
Disk2  &3   &  50 &$ 15.8 $ &   3.2 & 86.8 &        & 0.60  & 129  \\       
       &    &  80 &$ 11.6 $ &       & 224  &        & 1.3   & 94.7 \\\hline 
\end{tabular}
\begin{list}{}
\item $^a$ Cloud number defined by the different velocity components.
\item $^b$ from \citet{Goldsmith_Langer_1978}.
\item $^c$ from \citet{Neufeld_et_al_1995, Neufeld_Kaufman_1993}
\item $^d$ using program provided by \citet{LeBourlot_et_al_1999}.
\item $^e$ [$\times 10^{-21}\mathrm{erg\,s^{-1} cm^{-3}}$]
\end{list}
\end{table}

\begin{table}
\caption{Column density derived for NH3 (3,3) and NH3 (6,6) using RADEX \label{parameters_ortho}}
\begin{tabular}{lc|c|cc}
\hline\hline
Source  &$^a$  &n(H2)              & T(kin) & N(ortho-NH3)\\\hline
Halo1   &1 &$1.00\times 10^4$  & $>300$   & $1   \times 10^{14}$ \\
Halo1   &2 &$8.49\times 10^4$  & $>300$   & $1.26\times 10^{14}$ \\
Halo1   &3 &$1.58\times 10^4$  & $>300$   & $0.45\times 10^{14}$ \\\hline
Halo4   &1 &$2.75\times 10^4$  & $>170$   & $0.95\times 10^{14}$ \\\hline
DiskX1-1&1 &$3.55\times 10^4$  & $>300$   & $1.58\times 10^{14}$ \\\hline
DiskX2-1&1 &$2.70\times 10^4$  & $>200$   & $2.51\times 10^{14}$ \\\hline
DiskX1-2&1 &$5.62\times 10^4$  & $>280$   & $1.41\times 10^{14}$ \\\hline
DiskX2-2&1 &$6.20\times 10^4$  & $265 $   & $0.76\times 10^{14}$ \\\hline
Disk1   &1 &$0.16\times 10^4$  & $>300$   & $0.32\times 10^{14}$ \\
Disk1   &2 &$2.51\times 10^4$  & $>300$   & $0.32\times 10^{14}$ \\\hline
Disk2   &1 &$2.47\times 10^4$  & $200 $   & $8.32\times 10^{14}$ \\
Disk2   &2 &$2.51\times 10^4$  & $>300$   & $12.6\times 10^{14}$ \\
Disk2   &3 &$11.6\times 10^4$  & $130 $   & $4.79\times 10^{14}$ \\\hline
\end{tabular}
\begin{list}{}
\item $^a$ Cloud number defined by the different velocity components.
\end{list}
\end{table}

\onecolumn
{\small
\begin{longtable}{ccccccccc}
\caption{Results from Gaussian fit and optical depth from NH$_3$ method. \label{observation}}\\
\hline\hline
Source  & Cloud     &Transition & $v_{\rm LSR}$& T$_{\rm MB}$ & $\Delta v_{1/2}$& Integrated intensity$^e$ & rms &$\tau$ \\
        &number$^f$  &  (j,k)    &      [\kms]  &       [K]    &      [\kms]     &  [K \kms]     & [mk]        &     \\\hline    
\endfirsthead
\caption{continued.}\\
\hline\hline
Source  & Cloud     &Transition & $v_{\rm LSR}$& T$_{\rm MB}$ & $\Delta v_{1/2}$& Integrated intensity$^e$ & rms &$\tau$ \\
        &number$^f$  &  (j,k)    &      [\kms]  &       [K]    &      [\kms]     &  [K \kms]            &  [mk]  & \\\hline    
\hline
\endhead
\hline
\endfoot
Halo\,1 &1  & (1,1)    & $85.6  \pm  0.4$  & $0.10$ & $29.0 \pm 0.4$ &$3.19  \pm 0.04$ &$52$&$<0.40$\\
        &2  & (1,1)    & $115.8 \pm  0.4$  & $0.32$ & $23.6 \pm 0.4$ &$7.91  \pm 0.04$ &$52$&$0.10\pm 0.09$\\
        &3  & (1,1)    & $139.3 \pm  0.4$  & $0.09$ & $18.8 \pm 0.4$ &$1.77  \pm 0.04$ &$52$&$<0.64$\\
        &1  & (2,2)    & $83.1  \pm  0.4$  & $0.07$ & $37.8 \pm 0.4$ &$2.94  \pm 0.04$ &$51$&$<0.86$\\
        &2  & (2,2)    & $116.7 \pm  0.4$  & $0.35$ & $18.4 \pm 0.4$ &$6.77  \pm 0.04$ &$51$&$0.21\pm 0.01$\\
        &3  & (2,2)    & $147.1 \pm  0.4$  & $0.06$ & $35.5 \pm 0.4$ &$2.29  \pm 0.04$ &$51$&$4.9\pm 2.0^d$\\
        &1  & (3,3)    & $ 87.7 \pm  0.4$  & $0.22$ & $38.8 \pm 0.4$ &$9.23  \pm 0.07$ &$51$&$0.10\pm 0.03$ \\
        &2  & (3,3)    & $117.2 \pm  0.4$  & $0.85$ & $16.8 \pm 0.4$ &$15.16 \pm 0.07$ &$51$&$0.10\pm 0.03$ \\
        &3  & (3,3)    & $138.6 \pm  0.4$  & $0.19$ & $20.8 \pm 0.4$ &$ 4.29 \pm 0.07$ &$51$&$0.16\pm 0.01$ \\
        &1  & (4,4)    & $ 82.0 \pm  0.4$  & $0.06$ & $32.9 \pm 0.4$ &$ 2.03 \pm 0.02$ &$50$&  \\	  	
        &2  & (4,4)    & $118.7 \pm  0.4$  & $0.23$ & $21.5 \pm 0.4$ &$ 5.34 \pm 0.02$ &$50$&  \\		
        &3  & (4,4)    & $137.4 \pm  0.4$  & $0.05$ & $16.6 \pm 0.4$ &$ 0.91 \pm 0.02$ &$50$&  \\ 	
        &1  & (5,5)    & $ 87.7^a$         & $0.12^b$& $38.8^a$      &$0.467^b$        &$41$&  \\
        &2  & (5,5)    & $115.7 \pm  0.4$  & $0.15$ & $12.4 \pm 0.9$ &$ 1.92 \pm 0.13$ &$41$&  \\
        &3  & (5,5)    & $131.0 \pm  0.9$  & $0.06$ & $11.9 \pm 3.2$ &$ 0.74 \pm 0.09$ &$41$&  \\
        &1  & (6,6)    & $87.7^a$          & $0.13^b$&$38.8^a$ &$0.613^b$              &$41$& \\
        &2  & (6,6)    & $120.5 \pm  0.8$  & $0.30$  &$17.3 \pm 2.1$ &$ 5.48 \pm 0.61$ &$41$& \\
        &3  & (6,6)    & $138.5 \pm  1.9$  & $0.12$  &$14.4 \pm 2.9$ &$ 1.78 \pm 0.54$ &$41$& \\\hline
Halo\,4 &1  & (1,1)    & $196.4 \pm  0.9$  & $0.22$ & $30.2 \pm 2.1$ &$6.99 \pm  0.42$ &$83$&$ <0.31$ \\
        &1  & (2,2)    & $194.7 \pm  1.1$  & $0.19$ & $34.1 \pm 2.9$ &$6.74 \pm  0.47$ &$84$&$ 1.72 \pm 0.53$ \\
        &1  & (3,3)    & $195.5 \pm  0.5$  & $0.37$ & $20.3 \pm 1.2$ &$7.93 \pm  0.38$ &$80$&$ <0.31$ \\
        &1  & (4,4)    & $194.5 \pm  1.1$  & $0.09$ & $16.0 \pm 2.9$ &$1.44 \pm  0.20$ &$49$&  \\
        &1  & (5,5)    & $195.5^a$         & $0.15^b$& $20.3^a$ &$0.417$ &$50$&  \\
        &1  & (6,6)    & $199.4 \pm  1.7$  & $0.09$ & $24.3 \pm 4.0$ &$2.26 \pm  0.32$ &$51$&  \\\hline
Disk\,X1-1&1 &(1,1)    & $183.1 \pm 1.0$ & $0.20$  &$48.1 \pm  2.2$ &$10.43 \pm 0.45$ &$74$&$0.86 \pm 0.27$  \\
        & 1  &(2,2)    & $182.9 
\pm 1.3$ & $0.15$  &$32.1 \pm  2.6$ &$ 5.10 \pm 0.38$ &$71$&$0.10 \pm 0.11$  \\
        & 1  &(3,3)    & $184.1 \pm 0.4$ & $0.43$  &$34.8 \pm  1.0$ &$15.88 \pm 0.39$ &$71$&$0.10 \pm 0.02$  \\
        & 1  &(4,4)    & $188.1 \pm 1.1$ & $0.14$  &$24.7 \pm  2.2$ &$ 3.66 \pm 0.33$ &$74$&  \\
        & 1  &(5,5)    & $187.1 \pm 1.7$ & $0.09$  &$36.4 \pm  4.1$ &$ 3.56 \pm 0.34$ &$63$&  \\
        & 1  &(6,6)    & $188.3 \pm 1.1$ & $0.17$  &$35.0 \pm  2.6$ &$ 6.25 \pm 0.41$ &$59$&  \\\hline
Disk\,X2-1&1 &(1,1)    & $96.4 \pm 0.4$  & $0.48$  &$40.8 \pm  1.0$ &$20.91 \pm 0.42$ &$74$&$<0.84$  \\
        & 1  &(2,2)    & $96.5 \pm 0.5$  & $0.32$  &$30.7 \pm  1.4$ &$10.36 \pm 0.38$ &$71$&$0.10 \pm 0.02$  \\
        & 1  &(3,3)    & $98.1 \pm 0.2$  & $0.74$  &$32.9 \pm  0.6$ &$25.95 \pm 0.39$ &$71$&$0.10 \pm 0.01$  \\
        & 1  &(4,4)    & $96.0 \pm 0.6$  & $0.19$  &$15.7 \pm  1.7$ &$ 3.16 \pm 0.28$ &$74$&  \\
        & 1  &(5,5)    & $98.6 \pm 2.2$  & $0.09$  &$41.5 \pm  4.6$ &$ 3.91 \pm 0.37$ &$63$&  \\
        & 1  &(6,6)    &$103.4 \pm 1.1$  & $0.17$  &$30.8 \pm  2.6$ &$ 5.70 \pm 0.40$ &$59$& \\\hline
Disk\,X1-2&1 &(1,1)    & $69.1 \pm 0.5$  & $0.32$  &$22.7 \pm  1.3$ &$ 7.77 \pm 0.37$ &$78$&$0.10 \pm 0.03$ \\
        & 1  &(2,2)    & $69.4 \pm 0.4$  & $0.32$  &$14.5 \pm  1.3$ &$ 4.99 \pm 0.33$ &$80$&$0.10 \pm 0.06$ \\
        & 1  &(3,3)    & $69.1 \pm 0.2$  & $0.87$  &$17.3 \pm  0.6$ &$16.10 \pm 0.42$ &$91$&$0.10 \pm 0.04$ \\
        & 1  &(4,4)    & $69.9 \pm 0.4$  & $0.25$  &$12.2 \pm  1.4$ &$ 3.23 \pm 0.26$ &$71$&  \\
        & 1  &(5,5)    & $67.0 \pm 0.7$  & $0.13$  &$13.1 \pm  1.4$ &$ 1.84 \pm 0.20$ &$62$&  \\
        & 1  &(6,6)    & $74.1 \pm 0.6$  & $0.24$  &$17.8 \pm  1.7$ &$ 4.58 \pm 0.33$ &$61$&  \\\hline
Disk\,X2-2&1 &(1,1)    & $-50.2 \pm 0.8$ & $0.25$  &$37.9 \pm  1.9$ &$10.22 \pm 0.46$ &$78$&$1.00 \pm 0.29$ \\
        & 1  &(2,2)    & $-53.0 \pm 1.2$ & $0.15$  &$22.1 \pm  3.1$ &$ 3.51 \pm 0.39$ &$80$&$<0.28$ \\
        & 1  &(3,3)    & $-47.1 \pm 0.6$ & $0.35$  &$22.9 \pm  1.4$ &$ 8.41 \pm 0.45$ &$91$&$<0.23$ \\
        & 1  &(4,4)    & $-54.1 \pm 0.6$ & $0.14$  &$ 6.7 \pm  1.5$ &$ 0.99 \pm 0.18$ &$71$&  \\
        & 1  &(5,5)    & $-47.1^a$       & $0.19^b$&$22.9^a$        & $0.55^b$        &$62$&  \\
        & 1  &(6,6)    & $-49.3 \pm 2.8$ & $0.07$  &$21.0 \pm  4.9$ &$ 1.31 \pm 0.31$ &$61$&  \\\hline
Disk\,1 & 1  &(1,1)    & $64.5 \pm 0.4 $ & $0.08$ &$20.0\pm 0.4 $  & $1.73\pm 0.03$  &$61$ &$<1.25 $\\
        & 2  &(1,1)    & $78.2 \pm 0.4 $ & $0.16$ &$18.4\pm 0.4 $  & $3.05\pm 0.03$  &$61$ &$<0.34 $\\
        & 1  &(2,2)    & $59.3 \pm 0.8 $ & $0.08$ &$ 5.1\pm 1.6 $  & $0.45\pm 0.15$  &$63$ &$<1.35 $\\
        & 2  &(2,2)    & $73.4 \pm 0.5 $ & $0.16$ &$ 8.5\pm 1.5 $  & $1.47\pm 0.20$  &$63$ &$<0.26 $\\
        & 1  &(3,3)    & $56.5 \pm 1.0 $ & $0.11$ &$11.7\pm 3.1 $  & $1.36\pm 0.27$  &$66$ &$<0.56 $\\
        & 2  &(3,3)    & $72.4 \pm 0.6 $ & $0.21$ &$14.9\pm 0.5 $  & $3.41\pm 0.03$  &$66$ &$<0.32 $\\
        & 1  &(4,4)    & $56^a$  & $0.16^b$& $11.7^a$    &$ 0.33^b$ &$53$ & \\
        & 2  &(4,4)    & $72^a$  & $0.16^b$& $15.0^a$    &$ 0.38^b$ &$53$ & \\
        & 1  &(5,5)    & $56^a$  & $0.14^b$& $11.7^a$    &$ 0.29^b$ &$47$ & \\
        & 2  &(5,5)    & $72^a$  & $0.14^b$& $15.0^a$    &$ 0.33^b$ &$47$ & \\
        & 1  &(6,6)    & $56^a$  & $0.15^b$& $11.7^a$    &$ 0.39^b$ &$50$& \\
        & 2  &(6,6)    & $72^a$  & $0.15^b$& $15.0^a$    &$ 0.44^b$ &$50$& \\\hline
Disk\,2 & 1  &(1,1)    &$46.1 \pm 0.1$ &$  2.2$   &$24.1\pm 0.3$  &$56.34 \pm 0.65$ &$94$ &$0.10\pm0.10^c$\\
        & 2  &(1,1)    &$73.9 \pm 0.1$ &$  3.0$   &$18.4\pm 0.2$  &$58.21 \pm 0.40$ &$94$ &$3.00\pm0.10^c$\\
        & 3  &(1,1)    &$90.2 \pm 0.1$ &$  3.6$   &$16.7\pm 0.1$  &$64.73 \pm 0.33$ &$94$ &$0.10\pm0.10^c$\\
        & 1  &(2,2)    &$44.9 \pm 0.1$ &$  1.8$   &$22.6\pm 0.4$  &$42.52 \pm 0.58$ &$106$&$0.10 \pm0.01 $\\
        & 2  &(2,2)    &$74.3 \pm 0.2$ &$  3.2$   &$17.2\pm 0.3$  &$57.55 \pm 1.43$ &$106$&$0.15 \pm0.01 $\\
        & 3  &(2,2)    &$94.1 \pm 0.3$ &$  2.1$   &$18.6\pm 0.5$  &$41.28 \pm 1.31$ &$106$&$0.10 \pm0.01 $\\
        & 1  &(3,3)    &$48.8 \pm 0.1$ &$  3.4$   &$28.7\pm 0.3$  &$103.6\pm 0.8$   &$98$ &$0.10\pm0.10^c$\\
        & 2  &(3,3)    &$74.5 \pm 0.1$ &$ 14$     &$14.4\pm 0.1$  &$216.2\pm 0.7$   &$98$ &$0.10\pm0.10^c$\\
        & 3  &(3,3)    &$95.8 \pm 0.1$ &$  2.7$   &$18.6\pm 0.2$  &$52.42 \pm 0.51$ &$98$ &$0.10\pm0.10^c$\\
        & 1  &(4,4)    &$50.1 \pm 0.6$ &$  0.8$   &$32.5\pm 1.3$  &$26.23 \pm 0.97$ &$74$ &  \\
        & 2  &(4,4)    &$74.0 \pm 0.1$ &$  3.1$   &$15.0\pm 0.2$  &$49.35 \pm 1.03$ &$74$ &  \\
        & 3  &(4,4)    &$96.2 \pm 0.6$ &$  0.48$   &$20.0\pm 1.2$  &$10.24 \pm 0.57$&$74$ &  \\
        & 1  &(5,5)    &$53.5 \pm 0.1$ &$  0.40$   &$35.2\pm 0.5$  &$14.90 \pm 0.20$&$57$ &  \\
        & 2  &(5,5)    &$73.0 \pm 0.1$ &$  2.0$   &$15.9\pm 0.1$  &$33.44 \pm 0.28$ &$57$ &  \\
        & 3  &(5,5)    &$97.3 \pm 0.6$ &$  0.18$   &$16.3\pm 1.4$  &$3.10  \pm 0.24$&$57$ &  \\
        & 1  &(6,6)    &$56.5 \pm 0.3$ &$  0.69$   &$15.9\pm 0.8$  &$11.80 \pm 0.53$&$51$ &  \\
        & 2  &(6,6)    &$75.7 \pm 0.1$ &$  2.4$   &$15.2\pm 0.2$  &$38.76 \pm 0.48$ &$51$ &  \\
        & 3  &(6,6)    &$96^a$         &$  0.15^b$ &$18.6^a$       &$0.51^b$        &$51$ &  \\\hline
\end{longtable}
\begin{list}{}{}
\item $^a$ no detected. Value taken as reference from the (3,3) transition.  
\item $^b$ upper limits.  
\item $^c$ method NH$_3$ did not converge properly.  
\item $^d$ the fit is not good, due to the no convergence of the method. 
\item $^e \int T_{{\rm MB}}dv$.
\item $^f$ In each source, different velocity components define different clouds.
\end{list}
}  

\clearpage
{\small
\begin{longtable}{cccccccc}
\caption{Gaussian fits and column densities for SiO, C$^{34}$S, HNCO \label{SiO-CS-HNCO}}\\
\hline\hline
Source    & Species   & Velocity Center&$\Delta_v$ &$T_{\rm A}$&$\int T_{\rm A}dv^c$&rms& N\\
          &           & LSR [\kms]     &    [\kms] &  [K]     & [K \kms]        &[mK]&  [cm$^{-2}$]  \\\hline\hline
\endfirsthead
\caption{continued.}\\
\hline\hline
Source    & Species   & Velocity Center&$\Delta_v$ &$T_{\rm A}$&$\int T_{\rm A}dv^c$&rms& N\\
          &           & LSR [\kms]     &    [\kms] &  [K]     & [K \kms]        &[mK]&  [cm$^{-2}$]  \\\hline\hline
\hline
\endhead
\hline
\endfoot
Halo\,1  &SiO (2-1)  & $ 88.0\pm0.1$ &$33.5\pm1.0$ &$0.096$&$ 3.41\pm0.01$&$4.5$ &$5.73\pm0.01\times 10^{12}$\\
          &          & $117.8\pm0.1$ &$17.9\pm0.1$ &$0.993$&$18.96\pm0.08$&$4.5$ &$31.82\pm0.14\times 10^{12}$\\
          &          & $133.5\pm0.3$ &$20.7\pm0.7$ &$0.174$&$ 3.83\pm0.02$&$4.5$ &$6.43\pm0.03\times 10^{12}$\\   
    &$^{29}$SiO(2-1) & $119.7\pm0.6^a$&$17.2\pm1.3$&$0.085$&$ 1.55\pm0.11$&$6.3$ &$2.66\pm0.19\times 10^{12}$\\
    &$^{30}$SiO (2-1)& $120.0\pm0.8^a$&$17.4\pm1.9$&$0.063$&$ 1.16\pm0.11$&$5.9$ &$2.04\pm0.19\times 10^{12}$\\
          &CS (2-1)  & $ 84.8\pm0.5$ &$27.5\pm1.2$ &$0.211$&$ 6.2 \pm1.8$ &$0.2$ &\\
          &          & $116.5\pm0.1$ &$17.1\pm0.1$ &$1.627$&$29.7 \pm0.2$ &$0.2$ &\\
          &          & $130.4\pm0.2$ &$18.2\pm0.6$ &$0.364$&$ 7.06 \pm0.09$ &$0.2$ &\\
          &CS (3-2)  & $ 83.1\pm4.1$ &$23.3\pm4.1$ &$0.075$&$ 1.86\pm0.96$&$9.6$ &\\       
          &          & $116.5\pm4.1$ &$15.4\pm4.1$ &$1.517$&$24.80\pm0.96$&$9.6$ &\\      
          &          & $130.7\pm4.1$ &$11.2\pm4.1$ &$0.279$&$ 3.34\pm0.96$&$9.6$ &\\
   &C$^{34}$S (2-1)  & $-$           &$ - $        &$ - $   & $-$            &  $-$   &$-$\\   
	  &	     & $117. \pm0.3$ &$18.8\pm0.8$ &$0.138$&$ 2.75\pm0.10$&$5.6$&$9.51\pm0.34\times 10^{12}$\\ 
	  &	     & $-$           &$ - $        &$ - $   & $-$            &  $-$ \\          
          &HNCO      & $ 83.2\pm2.0$ &$15.2\pm4.0$ &$0.023$&$ 0.37\pm0.09$&$5.2$ &$10.8\pm2.6\times 10^{12}$\\
          &          & $117.2\pm0.4$ &$21.7\pm1.1$ &$0.119$&$ 2.75\pm0.11$&$5.2$ &$81.0\pm3.4\times 10^{12}$\\
          &          & -             & - & - &- &- &\\
     &C$^{18}$O(1-0) &$ 89.0\pm 0.9$ &$26.8\pm1.9$ &$0.057$&$1.62\pm0.10$ &$4.7$ &$1.11\pm0.07\times10^{15}$\\
          &          &$115.9\pm 0.2$ &$18.8\pm0.6$ &$0.108$&$2.17\pm0.04$ &$4.7$ &$1.49\pm0.03\times10^{15}$\\
          &          &$140.7\pm 1.4$ &$16.6\pm3.8$ &$0.022$&$0.38\pm0.07$ &$4.7$ &$0.26\pm0.05\times10^{15}$\\
     &$^{13}$CO(1-0) &$ 85.0\pm0.4$  &$37.6\pm1.4$ &$0.344$&$13.77\pm0.4$ &$12.0$&$0.94\pm0.03\times10^{15}$\\
          &          &$114.2\pm 0.2$ &$13.3\pm0.4$ &$0.628$&$8.92\pm0.5$  &$12.0$&$0.61\pm0.03\times10^{15}$\\
          &          &$133.7\pm 1.4$ &$27.2\pm3.1$ &$0.167$&$4.84\pm0.6$  &$12.0$&$0.33\pm0.04\times10^{15}$\\ 
          &          &$169.5\pm 2.3$ &$23.6\pm6.0$ &$0.049$&$1.23\pm0.26$ &$12.0$&$0.08\pm0.02\times10^{15}$\\\hline
Halo\,2   &SiO (2-1) & $-84.7\pm1.1$ &$25.5\pm0.8$ &$0.046$&$ 1.25\pm0.01$&$4.0$ &$2.10\pm0.01\times 10^{12}$\\
          &          & $-58.5\pm1.0$ &$22.6\pm2.8$ &$0.045$&$ 1.07\pm0.11$&$4.0$ &$1.80\pm0.19\times 10^{12}$\\
    &$^{29}$SiO(2-1) & $-85^b$ &$25^b$ &$- $&$<0.26$&$6.8$ &$<0.453\times 10^{12}$\\
          &          & $-58^b$ &$23^b$ &$- $&$<0.26$&$6.8$ &$<0.453\times 10^{12}$\\
    &$^{30}$SiO (2-1)& $-58.6\pm2.1^a $ &$18.9\pm4.8 $ &$0.020$&$0.40\pm0.07$&$3.4$&$0.71\pm0.13\times 10^{12}$\\
          &CS (2-1)  & $-80.2\pm6.1$ &$25.8\pm6.1$ &$0.188$&$ 5.15\pm0.30$&$6.2$ &\\
          &          & $-51.9\pm6.1$ &$22.9\pm6.1$ &$0.192$&$ 4.69\pm0.30$&$6.2$ &\\
          &CS (3-2)  & $-77.6\pm1.2$ &$22.2\pm2.7$ &$0.091$&$ 2.16\pm0.30$&$6.5$ &\\
          &          & $-51.8\pm1.5$ &$22.3\pm2.5$ &$0.078$&$ 1.85\pm0.25$&$6.5$ &\\
   &C$^{34}$S (2-1)  & $-81.7\pm13.5$&$47.8\pm23.2$&$0.006$&$ 0.47\pm0.24$&$3.8$ &$1.63\pm0.84\times 10^{12}$\\
          &          & $-35.0\pm14.9$&$35.6\pm21.9$&$0.009$&$ 0.24\pm0.25$&$3.8$ &$0.83\pm0.85\times 10^{12}$\\
          &HNCO      &$-70.4\pm3.9^a$&$30.6\pm8.2$ &$0.032$&$ 1.05\pm0.26$&$5.1$ &$31.0\pm7.7\times 10^{12}$\\\hline
%\\
Halo\,3   &SiO (2-1) & $-62.9\pm1.2$ &$35.9\pm3.4$ &$0.194$&$ 7.43\pm0.52$&$8.4$ &$12.5\pm0.9\times 10^{12}$\\
    &$^{29}$SiO(2-1) & $-56.6\pm3.7$ &$23.1\pm8.6$ &$0.015$&$ 0.37\pm0.11$&$4.8$ &$ 0.64\pm0.18\times 10^{12}$\\
    &$^{30}$SiO (2-1)& $-51.0\pm8.2$ &$34.3\pm15.5$&$0.020$&$ 0.73\pm0.31$&$3.3$ &$ 1.29\pm0.55\times 10^{12}$\\
          &CS (2-1)  & $-64.7\pm0.3$ &$31.4\pm0.8$ &$0.670$&$22.42\pm0.46$&$8.2$ &\\
          &          & $-13.9\pm1.1$ &$33.0\pm2.6$ &$0.186$&$ 6.52\pm0.44$&$8.2$ &\\
          &CS (3-2)  & $-63.6\pm0.3$ &$18.6\pm1.1$ &$0.513$&$10.13\pm0.57$&$14.4$&\\
          &          & $-21.6\pm4.5$ &$42.7\pm9.5$ &$0.077$&$ 3.50\pm0.60$&$14.4$&\\
   &C$^{34}$S (2-1)  & $-63.1\pm1.2$ &$28.8\pm3.6$ &$0.051$&$ 1.58\pm0.15$&$6.5$&$5.46\pm0.52\times 10^{12}$\\
          &          & $-12.6\pm2.4$ &$21.4\pm4.8$ &$0.023$&$ 0.52\pm0.11$&$6.5$&$1.81\pm0.39\times 10^{12}$\\
          &HNCO      & $-67.3\pm0.8$ &$35.3\pm2.0$ &$0.083$&$ 3.10\pm0.15$&$6.0$&$91.5\pm4.3\times 10^{12}$\\
          &          & $-10.6\pm0.9$ &$21.7\pm2.2$ &$0.055$&$ 1.28\pm0.11$&$6.0$&$37.7\pm3.3\times 10^{12}$\\\hline
%\\ 
Halo\,4   &SiO (2-1) & $196.1\pm0.4$ &$20.6\pm0.9$ &$0.123$&$ 2.69\pm0.11$&$3.4$ &$4.51\pm0.19\times 10^{12}$\\
    &$^{29}$SiO(2-1) & $196^b$ &$21^b $ &$-$&$<0.14$&$3.9$ &$<0.242\times 10^{12}$\\
    &$^{30}$SiO (2-1)& $196^b$ &$21^b $ &$-$&$<0.14$&$3.9$ &$<0.250\times 10^{12}$\\
          &CS (2-1)  & $197.1\pm0.3$ &$21.9\pm0.6$ &$0.380$&$ 8.88\pm0.21$&$6.5$ &\\
          &CS (3-2)  & $196.7\pm0.2$ &$22.6\pm0.4$ &$0.183$&$ 4.40\pm0.07$&$5.3$ &\\
   &C$^{34}$S (2-1)  & $195.2\pm2.2$ &$24.8\pm4.8$ &$0.022$&$ 0.57\pm0.10$&$4.9$ & $ 1.96\pm0.34\times 10^{12}$\\
          &HNCO      & $196.9\pm0.2$ &$19.0\pm0.5$ &$0.178$&$ 3.59\pm0.08$&$4.7$ &$106\pm2\times 10^{12}$\\
          &C$^{18}$O & $206.1\pm1.2$ &$ 7.6\pm4.0$ &$0.081$&$ 0.66\pm0.18$&$10.0$&$  5.2\pm2.7\times 10^{15}$ \\
          &$^{13}$CO & $202.3\pm0.6$ &$25.1\pm1.4$ &$0.705$&$18.8\pm0.9$&$25.0$&$  1.28\pm0.06\times 10^{16}$\\\hline
%\\
Halo\,5   &SiO (2-1) & $-62.9\pm0.1$ &$16.3\pm0.3$ &$0.254$&$ 4.39\pm0.07$&$4.6$ &$7.37\pm0.13\times 10^{12}$\\
    &$^{29}$SiO(2-1) & $-64.8\pm0.8$ &$11.9\pm1.4$ &$0.032$&$ 0.41\pm0.05$&$3.3$&$0.70\pm0.09\times 10^{12}$\\
    &$^{30}$SiO (2-1)& $-61.4\pm2.4$ &$22.1\pm7.5$ &$0.917$&$ 0.40\pm0.10$&   &$0.71\pm0.17\times 10^{12}$\\
          &CS (2-1)  & $-62.7\pm0.1$ &$16.7\pm0.1$ &$0.646$&$11.52\pm0.04$&$2.9$ &\\
          &CS (3-2)  & $-62.7\pm0.2$ &$14.4\pm0.4$ &$0.528$&$ 8.12\pm0.18$&$6.4$ &\\
   &C$^{34}$S (2-1)  & $-62.3\pm0.3$ &$13.3\pm0.6$ &$0.074$&$ 1.05\pm0.04$&$3.0$ &$3.62\pm0.14\times 10^{12}$\\
          &HNCO      & $-64.2\pm0.2$ &$14.6\pm0.4$ &$0.164$&$ 2.54\pm0.07$&$4.3$ &$75.0\pm1.9\times 10^{12}$\\\hline
%\\ 	       	     	       	   	    	      
Disk\,X1-1&SiO (2-1) & $178.0\pm0.3$ &$41.3\pm0.7$ &$0.390$ &$17.1\pm0.2$&$9.1$ &$28.75\pm0.39\times 10^{12}$\\
    &$^{29}$SiO(2-1) & $179.4\pm2.5$ &$43.4\pm5.0$ &$0.031$ &$ 1.43\pm0.16$&$5.4$ &$2.46\pm0.27\times 10^{12}$\\
    &$^{30}$SiO (2-1)& $188.0\pm1.0$ &$24.6\pm2.6$ &$0.039$ &$ 1.00\pm0.08$&$3.3$ &$1.77\pm0.13\times 10^{12}$\\
          &CS (2-1)  & $176.2\pm0.3$ &$36.4\pm0.7$ &$0.718$ &$27.81\pm0.41$&$6.3$ &\\
          &CS (3-2)  & $174.3\pm0.1$ &$31.1\pm0.3$ &$0.591$ &$19.60\pm0.16$&$7.7$ &\\
   &C$^{34}$S (2-1)  & $177.5\pm1.3$ &$35.1\pm2.8$ &$0.040$ &$ 1.51\pm0.11$&$5.1$ &$5.21\pm0.38\times 10^{12}$\\
          &HNCO      & $187.6\pm0.6$ &$16.4\pm1.8$ &$0.170$ &$ 2.96\pm0.24$&$9.3$ &$87.4\pm7.2\times 10^{12}$\\
          &C$^{18}$O & $177.9\pm6.4$ &$49.5\pm13.0$&$0.040$ &$ 2.12\pm0.46$&$13.3$&$1.45\pm0.31\times10^{15}$\\
          &$^{13}$CO & $175.6\pm2.6$ &$56.4\pm6.1$ &$0.603$ &$36.2\pm3.2$  &$21.2$&$2.47\pm0.22\times 10^{16}$\\\hline
%\\	   
Disk\,X2-1&SiO (2-1) &  $99.7\pm0.3$ &$33.2\pm0.8$ &$0.282$&$ 9.96\pm0.21$&$9.1$ &$16.72\pm0.36\times 10^{12}$\\
    &$^{29}$SiO(2-1) &  $95.2\pm3.3$ &$57.7\pm19.1$&$0.021$&$ 1.26\pm0.25$&$5.4$ &$2.17\pm0.43\times 10^{12}$\\
    &$^{30}$SiO (2-1)&  $94.1\pm3.3$ &$41.4\pm9.8$ &$0.013$&$ 0.56\pm0.10$&$3.3$ &$0.99\pm0.17\times 10^{12}$\\
          &CS (2-1)  &  $98.1\pm0.2$ &$30.7\pm0.4$ &$0.953$&$31.10\pm0.37$&$6.3$ &\\
          &CS (3-2)  &  $97.1\pm0.1$ &$26.0\pm0.3$ &$0.580$&$16.08\pm0.16$&$7.7$ &\\
   &C$^{34}$S (2-1)  &  $99.7\pm0.6$ &$21.0\pm1.8$ &$0.064$&$ 1.43\pm0.09$&$5.1$ &$4.94\pm0.33\times 10^{12}$\\
          &HNCO      &  $96.3\pm0.5$ &$30.9\pm1.0$ &$0.293$&$ 9.61\pm0.29$&$9.3$ &$283.5\pm8.6\times 10^{12}$\\
          &C$^{18}$O &  $94.5\pm0.8$ &$28.9\pm1.9$ &$0.194$&$ 5.96\pm0.34$&$13.3$&$4.10\pm0.24 \times 10^{15}$ \\
          &$^{13}$CO &  $95.3\pm0.5$ &$32.1\pm1.3$ &$2.056$&$70.3 \pm2.5$ &$21.2$&$4.80\pm 0.17\times 10^{16}$\\ \hline
%\\		   		   			   		   
Disk\,X1-2&SiO (2-1) &  $69.1\pm0.1$ &$17.4\pm0.3$ &$0.522$&$ 9.70\pm0.12$&$5.7$ &$16.28\pm0.20\times 10^{12}$\\
    &$^{29}$SiO(2-1) &  $68.1\pm2.4$ &$18.0\pm5.2$ &$0.020$&$ 0.39\pm0.10$&$5.5$ &$0.67\pm0.18\times 10^{12}$\\
    &$^{30}$SiO (2-1)&  $72.5\pm3.3$ &$44.5\pm7.8$ &$0.018$&$ 0.83\pm0.13$&$4.7$ &$1.47\pm0.22\times 10^{12}$\\
          &CS (2-1)  &  $70.0\pm0.2$ &$16.1\pm0.5$ &$1.349$&$23.05\pm0.66$&$4.7$ &\\
          &CS (3-2)  &  $70.6\pm0.2$ &$12.9\pm0.4$ &$1.320$&$18.16\pm0.45$&$7.3$ &\\
   &C$^{34}$S (2-1)  &  $71.1\pm0.4$ &$14.8\pm1.0$ &$0.106$&$ 1.67\pm0.09$&$4.7$ &$ 5.77\pm0.32\times 10^{12}$\\
          &HNCO      &  $68.7\pm0.4$ &$18.3\pm0.8$ &$0.151$&$ 2.94\pm0.12$&$7.0$ &$86.7 \pm3.4\times 10^{12}$\\
          &C$^{18}$O &  $68.3\pm2.8$ &$16.7\pm5.6$ &$0.058$&$ 1.04\pm0.32$&$5.6$ &$ 0.71\pm0.22\times 10^{15}$ \\
          &$^{13}$CO &  $68.0\pm2.2$ &$16.6\pm5.4$ &$0.567$&$10.0 \pm2.9$ &$8.0$ &$ 0.68\pm0.20\times 10^{16}$\\ \hline
%\\
Disk\,X2-2&SiO (2-1) & $-43.5\pm2.0$ &$29.3\pm4.5$ &$0.037$&$ 1.15\pm0.16$&$5.7$ &$1.92\pm0.26\times 10^{12}$\\
    &$^{29}$SiO(2-1) & $-43^b$ &$29^b$ &$- $&$<0.23$&$5.5$ &$<0.402\times 10^{12}$\\
    &$^{30}$SiO (2-1)& $-51.8\pm2.8$ &$28.0\pm4.9$ &$0.018$&$ 0.54\pm0.10$&$4.7$ &$0.96\pm0.17\times 10^{12}$\\
          &CS (2-1)  & $-43.8\pm0.3$ &$19.6\pm0.7$ &$1.057$&$22.09\pm0.71$&$4.7$ &\\
          &CS (3-2)  & $-43.4\pm0.4$ &$21.6\pm1.0$ &$0.648$&$14.88\pm0.57$&$7.3$ &\\
   &C$^{34}$S (2-1)  & $-45.0\pm0.5$ &$21.0\pm1.5$ &$0.093$&$ 2.07\pm0.11$&$4.7$ &$ 7.15\pm0.39\times 10^{12}$\\
          &HNCO      & $-47.1\pm0.7$ &$21.7\pm1.7$ &$0.087$&$ 2.01\pm0.13$&$7.0$ &$59.2\pm3.9\times 10^{12}$\\
          &C$^{18}$O & $-46.1\pm0.5$ &$18.2\pm1.3$ &$0.301$&$ 5.81\pm0.35$&$5.6$ &$ 3.99\pm0.24\times 10^{15}$ \\
          &$^{13}$CO & $-44.0\pm0.4$ &$17.1\pm1.1$ &$3.312$&$60.2 \pm3.0$&$8.0$ &$ 4.11\pm0.20\times 10^{16}$\\ \hline
%\\	   			   		   
Disk\,1   &SiO (2-1) & $54.1\pm0.6$ &$11.5\pm1.8$ &$0.040$&$0.49\pm0.07$&$3.2$ &$0.82\pm0.12\times 10^{12}$\\
          &          & $73.6\pm0.5$ &$18.0\pm1.4$ &$0.062$&$1.18\pm0.07$&$3.2$ &$1.98\pm0.12\times 10^{12}$\\
    &$^{29}$SiO(2-1) & $54^b$ &$12$ &$- $&$<0.06$&$2.2$ &$<0.105\times 10^{12}$\\
          &          & $74^b$ &$18$ &$- $&$<0.07$&$2.2$ &$<0.129\times 10^{12}$\\
    &$^{30}$SiO (2-1)& $54^b$ &$12$ &$- $&$<0.06$&$2.2$ &$<0.106\times 10^{12}$\\
          &          & $74^b$ &$18$ &$- $&$<0.07$&$2.2$ &$<0.130\times 10^{12}$\\
          &CS (2-1)  & $56.8\pm0.3$ &$14.5\pm0.6$ &$0.087$&$1.35\pm0.05$&$2.2$ &\\
          &          & $74.9\pm0.1$ &$17.2\pm0.2$ &$0.327$&$5.99\pm0.07$&$2.2$ &\\
          &CS (3-2)  & $56.3\pm0.3$ &$ 9.8\pm0.8$ &$0.036$&$0.38\pm0.03$&$2.1$ &\\
          &          & $74.8\pm0.1$ &$14.8\pm0.2$ &$0.177$&$2.78\pm0.02$&$2.1$ &\\
   &C$^{34}$S (2-1)  & $50.8\pm4.5$ &$12.8\pm7.3$ &$0.003$&$0.05\pm0.03$&$1.9$ &$0.16\pm0.10\times 10^{12}$\\
          &          & $73.8\pm0.9$ &$15.2\pm2.2$ &$0.016$&$0.26\pm0.03$&$1.9$ &$0.09\pm0.11\times 10^{12}$\\
          &HNCO      & $55.3\pm2.8$ &$16.5\pm6.3$ &$0.024$&$0.43\pm0.15$&$4.2$ &$12.6\pm4.5\times 10^{12}$\\
          &          & $75.6\pm0.7$ &$15.7\pm1.6$ &$0.100$&$1.66\pm0.16$&$4.2$ &$49.1\pm4.7\times 10^{12}$\\
          &C$^{18}$O & $55.5\pm0.8$ &$11.5\pm1.6$ &$0.075$&$0.92\pm0.12$&$9.3$ &$0.63\pm0.09\times 10^{15}$ \\
          &          & $78.6\pm0.9$ &$11.6\pm2.3$ &$0.067$&$0.83\pm0.13$&$9.3$ &$0.57\pm0.09\times 10^{15}$ \\
          &$^{13}$CO & $57.3\pm0.3$ &$12.6\pm0.8$ &$0.649$&$9.89\pm0.48$&$10.1$&$0.59\pm0.03\times 10^{16}$ \\
          &          & $77.6\pm0.3$ &$14.4\pm0.8$ &$0.647$&$9.89\pm0.48$&$10.1$&$0.68\pm0.03\times 10^{16}$ \\\hline
%
%\\	   
Disk\,2   &SiO (2-1) & $52.3\pm0.7$ &$18.3\pm1.8$ &$0.519$&$10.12\pm0.92$& $7.9$ &$17.0\pm1.5\times 10^{12}$\\
          &          & $76.9\pm0.3$ &$19.0\pm0.8$ &$1.233$&$24.94\pm0.91$& $7.9$ &$41.9\pm1.5\times 10^{12}$\\
    &$^{29}$SiO(2-1) &$69.0\pm1.0^a$ &$26.2\pm2.3$&$0.240$&$ 6.67\pm0.48$& $5.0$ &$11.5\pm0.8\times 10^{12}$\\ 
    &$^{30}$SiO (2-1)&$69.2\pm1.8^a$ &$26.2\pm4.4$&$0.184$&$ 5.13\pm0.73$& $5.4$ &$ 9.0\pm1.3\times 10^{12}$\\ 
          &CS (2-1)  & $14.2\pm6.1$ &$33.0\pm6.1$ &$0.461$&$16.2\pm2.5$ &$12.9$ &\\
          &          & $55.9\pm6.1$ &$21.2\pm6.1$ &$1.544$&$34.9\pm2.5$ &$12.9$ &\\
          &          & $79.0\pm6.1$ &$17.7\pm6.1$ &$1.922$&$36.1\pm2.5$ &$12.9$ &\\
          &CS (3-2)  & $14.7\pm5.0$ &$26.9\pm13.4$&$0.317$&$ 9.1\pm3.5$ &$14.8$ &\\
          &          & $52.7\pm1.4$ &$16.4\pm3.5$ &$1.110$&$19.3\pm3.7$ &$14.8$ &\\
          &          & $76.9\pm0.9$ &$17.5\pm2.7$ &$1.640$&$30.6\pm3.8$ &$14.8$ &\\
   &C$^{34}$S (2-1)  & -      &  -      & -  &  -   & -  &\\
          &          & $33.9\pm2.6$ &$37.5\pm5.9$ &$0.145$&$ 5.79\pm0.82$&$2.9$ &$20.0\pm2.8\times 10^{12}$\\
          &          & $71.6\pm0.5$ &$23.1\pm1.2$ &$0.533$&$13.12\pm0.75$&$2.9$ &$45.4\pm2.6\times 10^{12}$\\
          &HNCO      & -      &  -      & -  &  -   & -  &\\
          &          & $67.1\pm0.1$ &$25.1\pm0.2$ &$8.868$&$236.8 \pm1.5$&$11.8$ &$6982\pm44\times 10^{12}$\\
          &          &   -      &  -      & -  &  -   & -  &\\
          &C$^{18}$O &  -    &  -      & -  &  -   & -  & \\
          &          & $68.0\pm0.8$ &$23.4\pm2.0$ &$0.945$&$23.5\pm1.6$&$14.1$&$16.2\pm1.1\times 10^{15}$\\
          &          &  -    &  -      & -  &  -   & -  & \\
          &$^{13}$CO &- &  -      & -  &  -   & -  &\\
          &          & $69.1\pm0.6$&$28.5\pm1.5$ &$5.932$ &$179.9\pm7.6$&$22.2$ &$12.3\pm0.5\times 10^{16}$\\
          &          & &  -      & -  &  -   & -  &\\\hline
\end{longtable}
\begin{list}{}{}
\item $^a$ only possible to fit one velocity components.
\item $^b$ not detected, value for reference taken by the main isotope.
\item $^c$formal errors of a Gaussian fit. The calibration errors may be larger. 
\end{list}
} 
\end{appendix}
\twocolumn
\begin{appendix}
\section{Complementary Figures}
This Appendix presents the LVG diagrams of the metastable inversion transitions of para-NH$_3$. Most of the sources show two kinetic temperature components:
 one warm which is plotted in a red line; and one cool which is plotted in a blue line. In the cases where only the warm kinetic temperature component 
was present, the result is showed with a red line. We show the n(H$_2$) derived for each component from the CS data, which was used as a fixed parameter 
in the RADEX program. The shaded regions correspond to the error associated to each para-NH$_3$ line. When a line was not detected we plot their $3-\sigma$ 
level in a dashed line. The error associated to the kinetic temperature and column density was estimated computing the $\chi^2$ of the line intensities over 
the grid used for the LVG model. We impose $\Delta \chi^2=\chi^2-\chi^2_{\mathrm{min}}=1$ which translate in the $68.3$\% confidence level projected for each 
parameter axis which is showed in a black elipse.

\clearpage
\begin{figure}
\vbox{
\includegraphics[width=0.35\textwidth, angle=90]{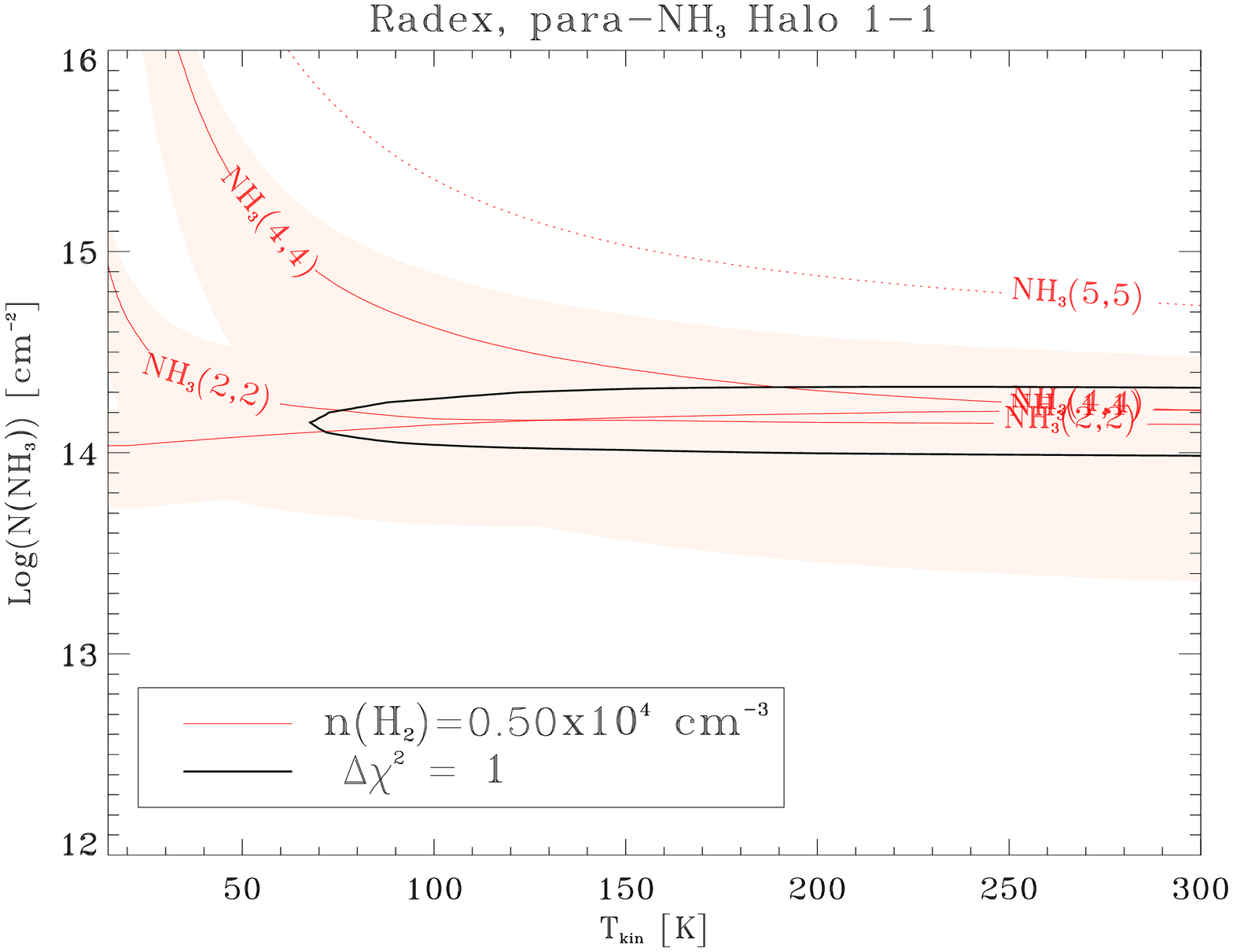}
\includegraphics[width=0.35\textwidth, angle=90]{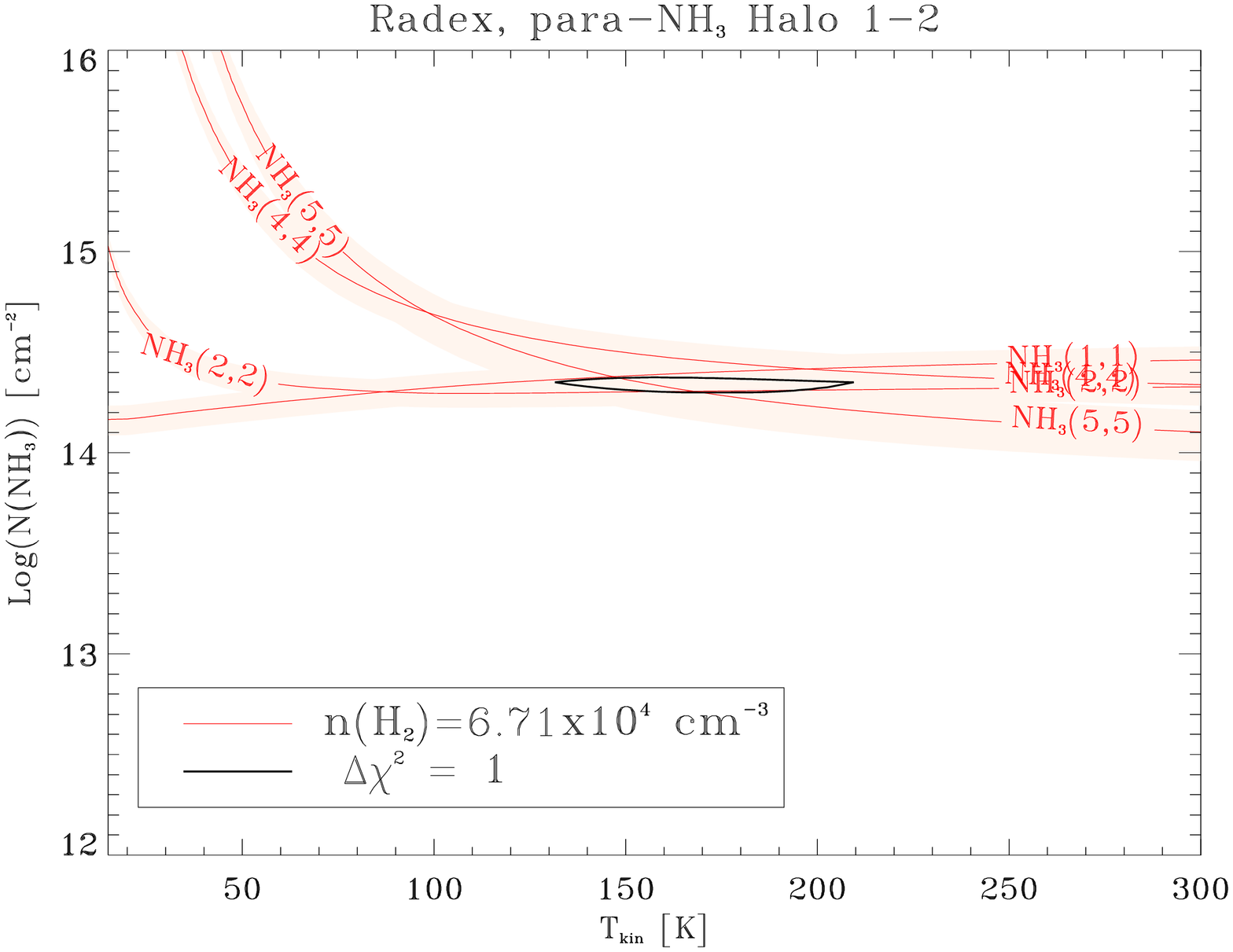}
\includegraphics[width=0.35\textwidth, angle=90]{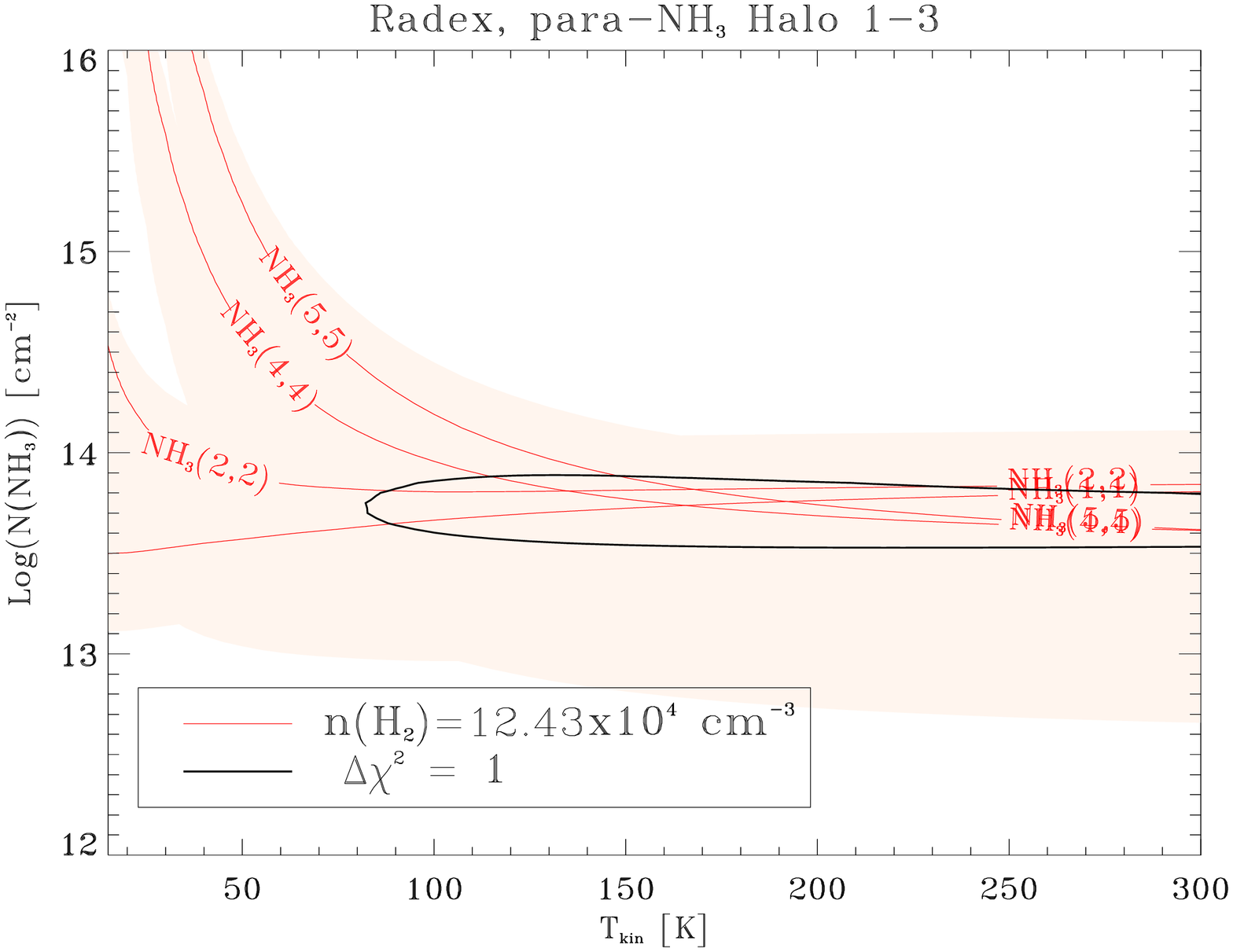}
}
\caption{LVG diagrams of NH$_3$ for each velocity component of Halo\,1.  Top: $87.7\, {\rm km\,s}^{-1}$ . Middle: $117.2 \,{\rm km\,s}^{-1}$.  Bottom: $138.6\, {\rm km\,s}^{-1}$.}
\label{modeloHalo1}
\end{figure}

\begin{figure}
\includegraphics[width=0.35\textwidth, angle=90]{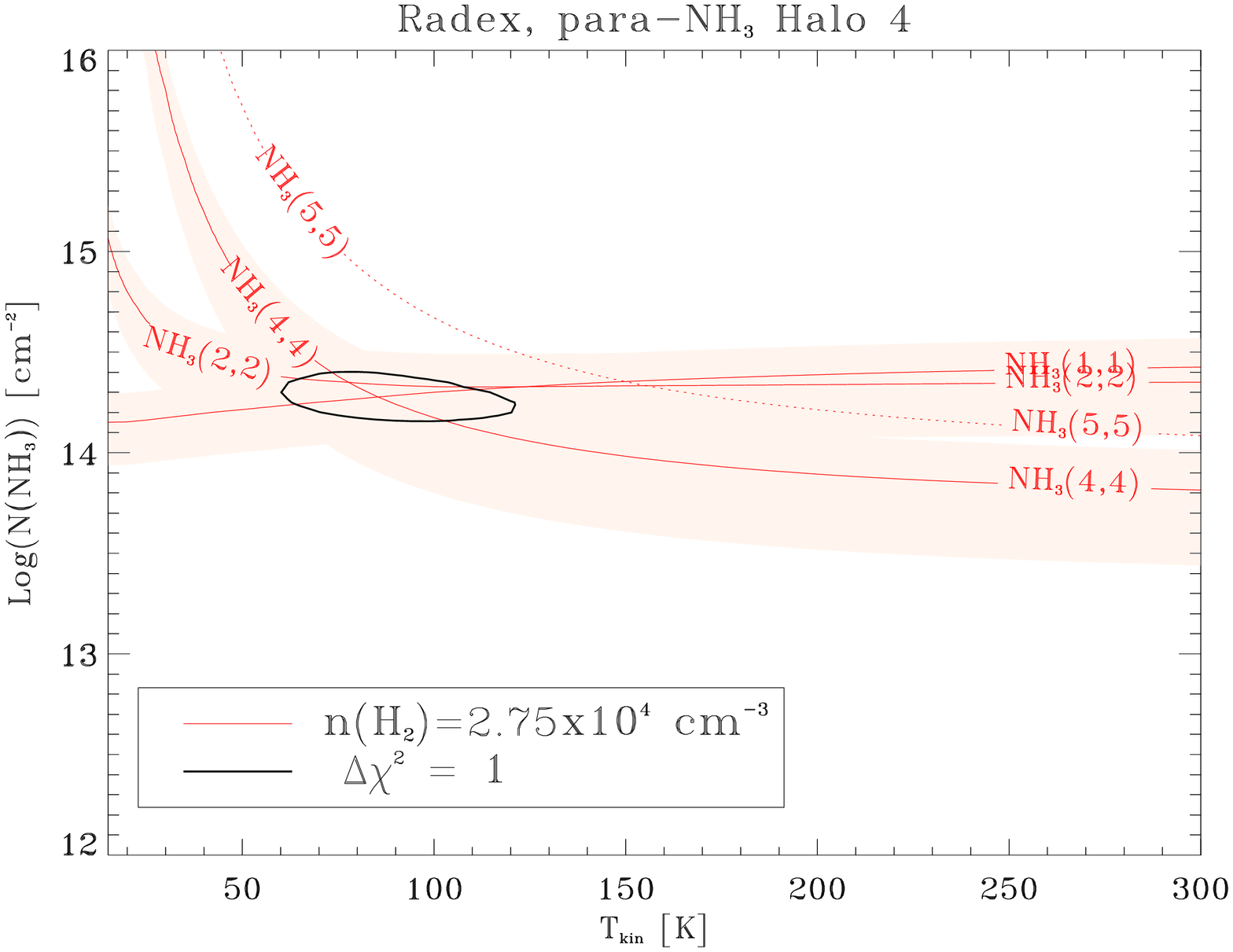}
\caption{LVG diagram of NH$_3$ for Halo\,4.}
\label{modeloHalo4}
\end{figure}

\begin{figure}
\includegraphics[width=0.35\textwidth, angle=90]{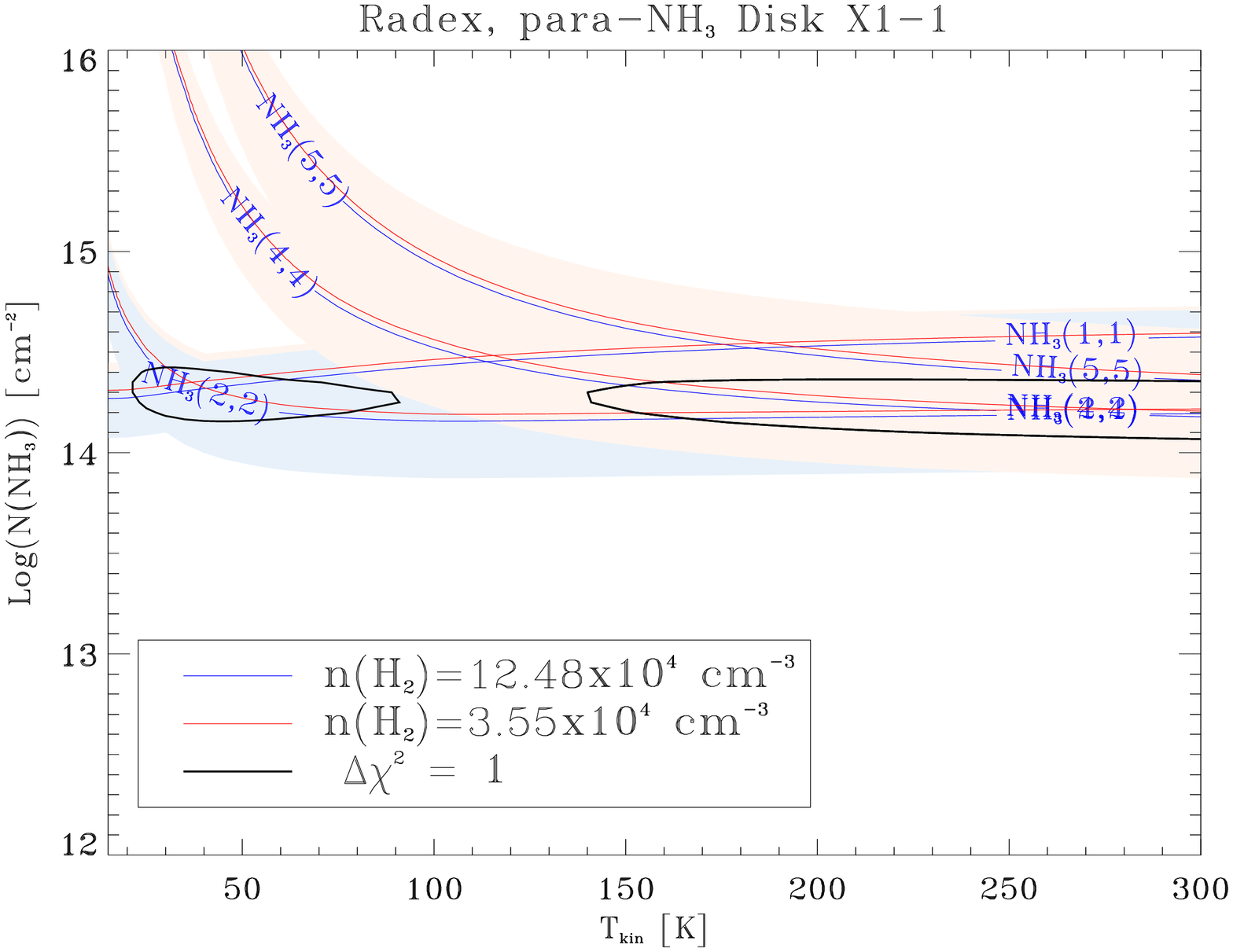}
\caption{LVG diagram of NH$_3$ for Disk\,X1-1. The blue lines correspond to the low temperature regimen, and the red lines to the high temperature regimen.}
\label{modelo13complexX1}
\end{figure}

\begin{figure}
\includegraphics[width=0.35\textwidth, angle=90]{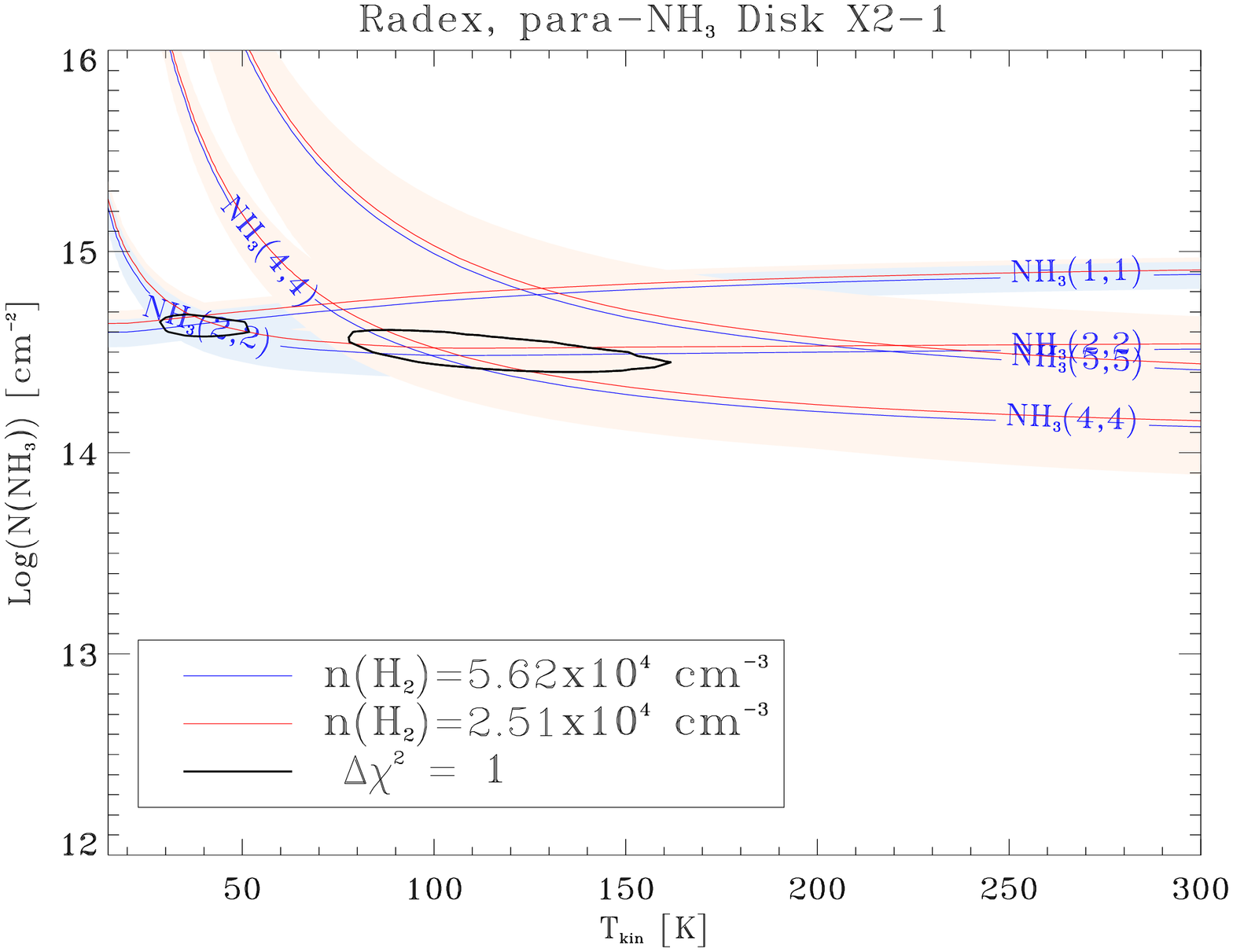}
\caption{LVG diagram of NH$_3$ for Disk\,X2-1. The blue lines correspond to the low temperature regimen, and the red lines to the high temperature regimen.}
\label{modelo13complexX2}
\end{figure}

\begin{figure}
\includegraphics[width=0.35\textwidth, angle=90]{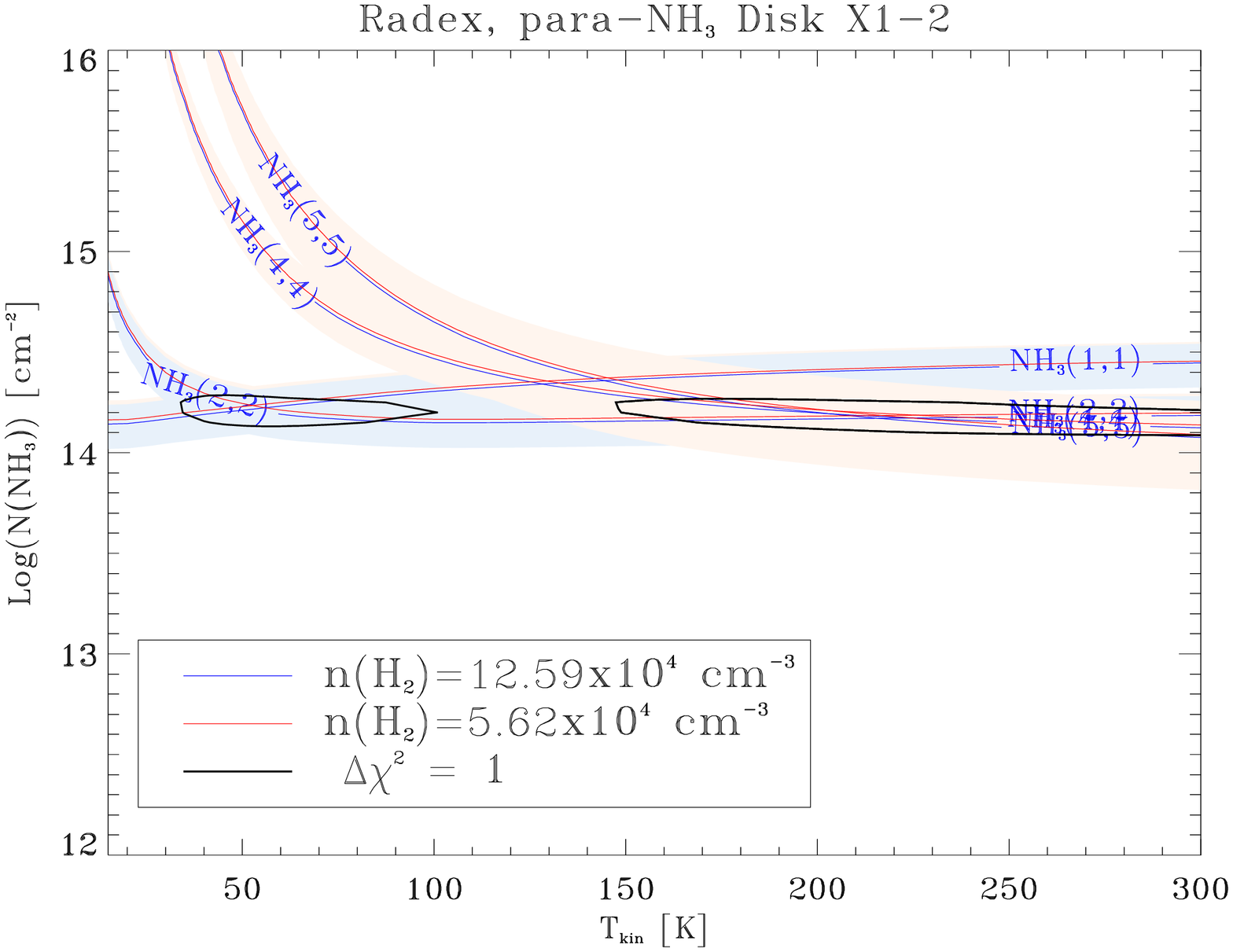}
\caption{LVG diagram of NH$_3$ for  Disk\,X1-2. The blue lines correspond to the low temperature regimen, and the red lines to the high temperature regimen.}
\label{modeloSgrC1X1}
\end{figure}

\begin{figure}
\includegraphics[width=0.35\textwidth, angle=90]{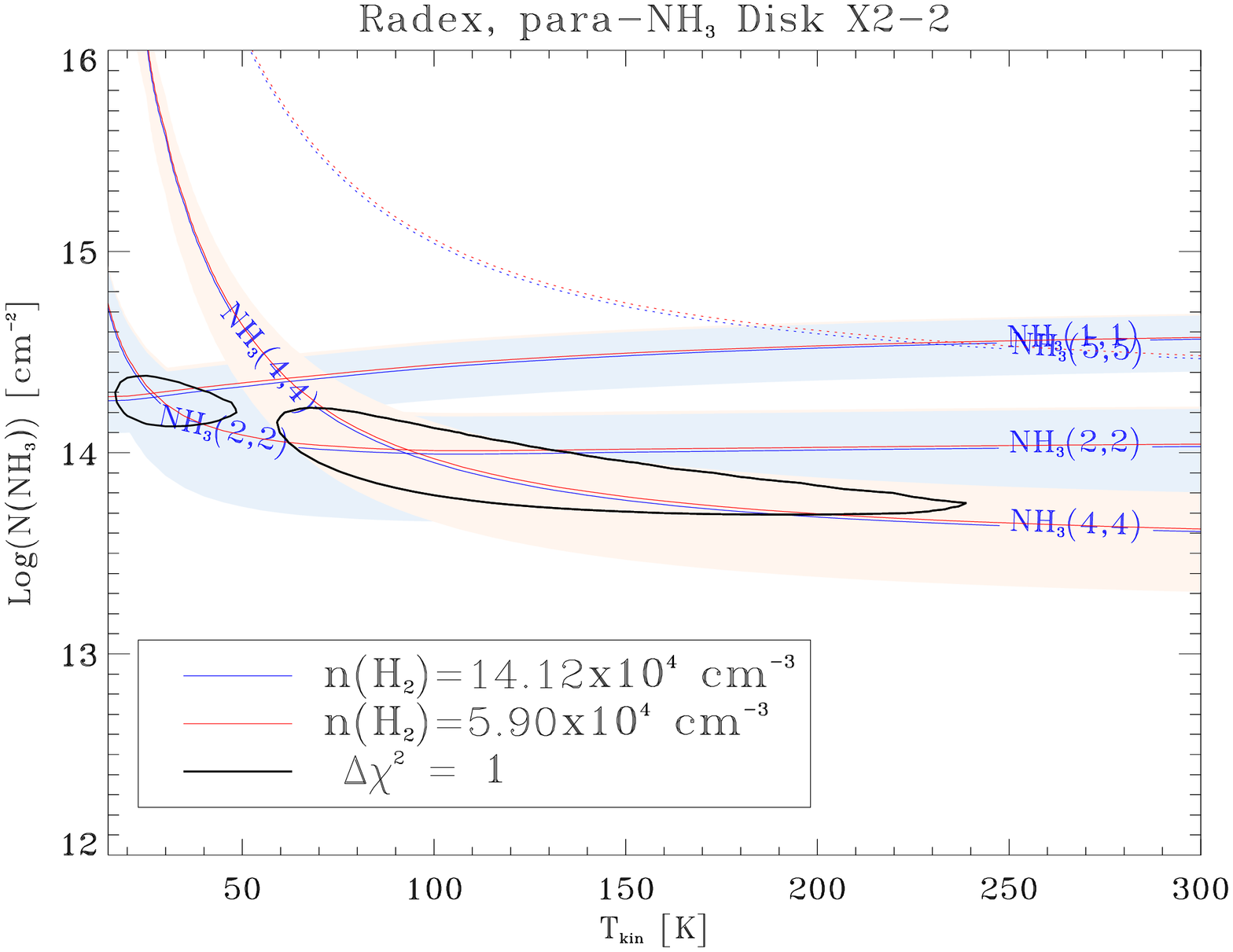}
\caption{LVG diagram of NH$_3$ for Disk\,X2-2. The blue lines correspond to the low temperature regimen, and the red lines to the high temperature regimen.}
\label{modeloSgrC1X2}
\end{figure}

%\clearpage

\begin{figure}
\vbox{
\includegraphics[width=0.35\textwidth, angle=90]{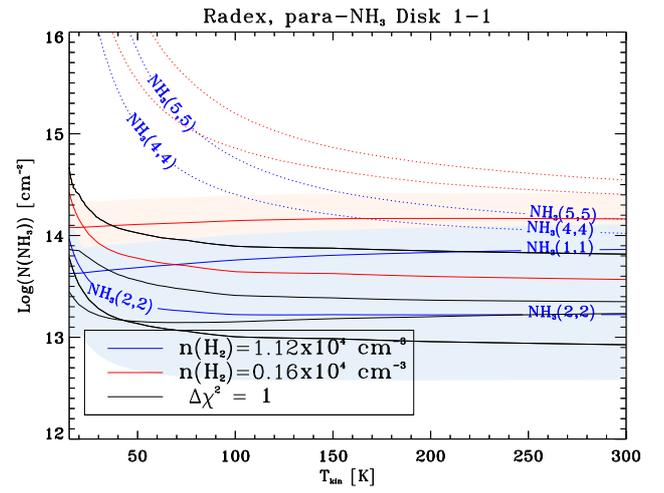}
\includegraphics[width=0.35\textwidth, angle=90]{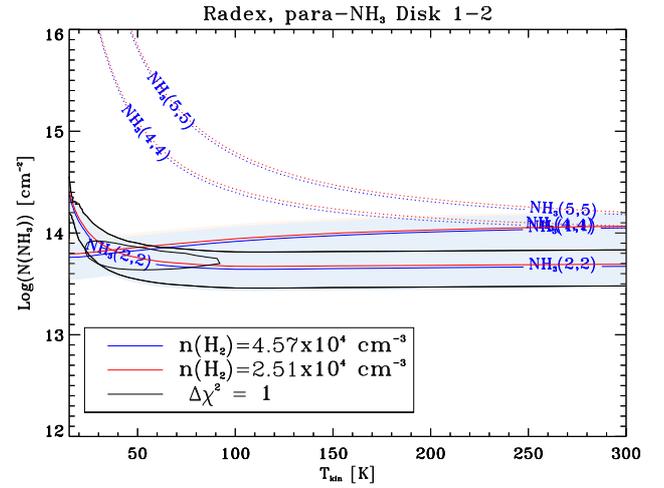}
}
\caption{LVG diagrams of NH$_3$ for each velocity component of Disk\,1. Top: $56.5\, {\rm km\,s}^{-1}$. Bottom: $72.4 \,{\rm km\,s}^{-1}$. The blue lines correspond to the low temperature regimen, and the red lines to the high temperature regimen.}
\label{modelocontrolD}
\end{figure}

\begin{figure}
\vbox{
\includegraphics[width=0.35\textwidth, angle=90]{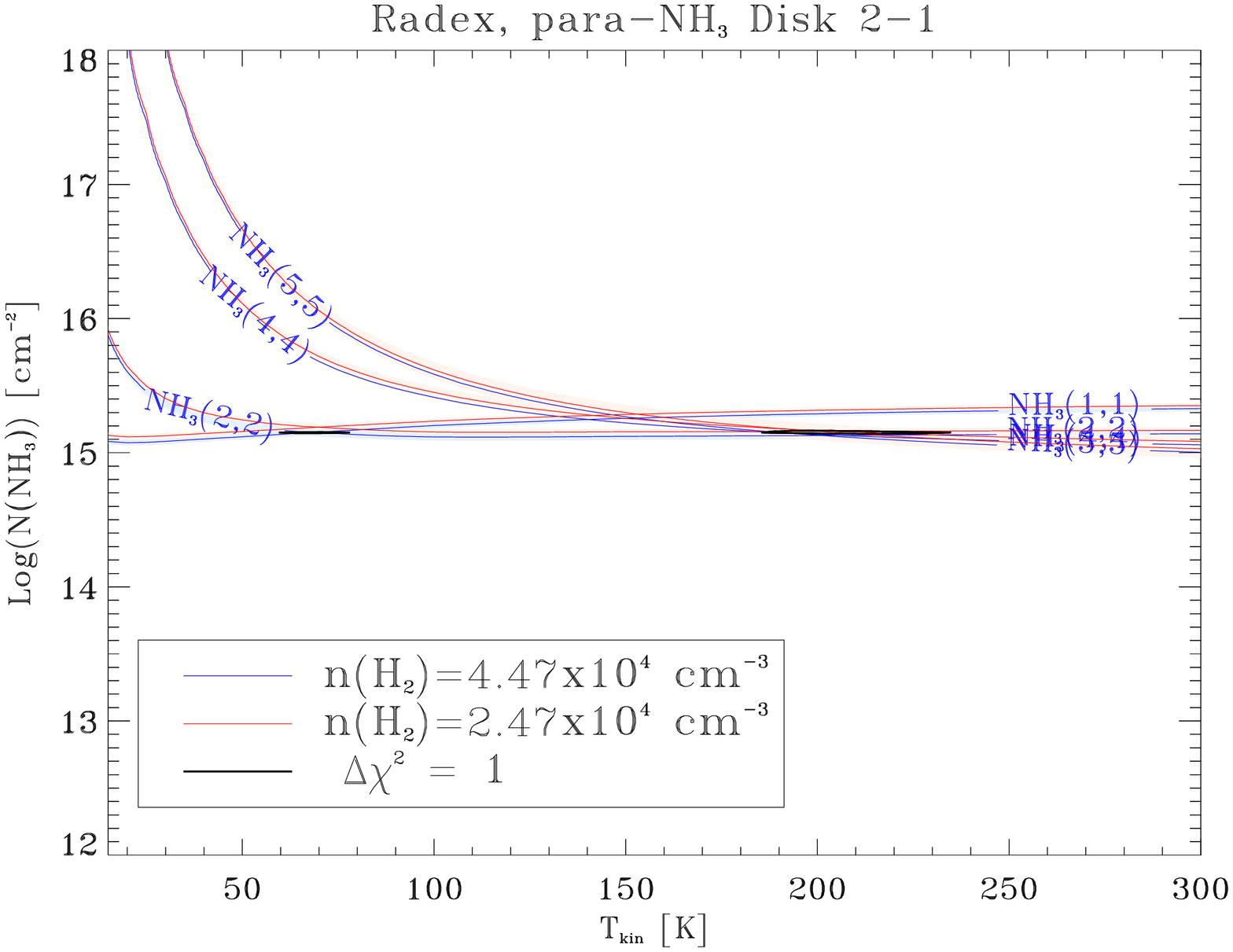}
\includegraphics[width=0.35\textwidth, angle=90]{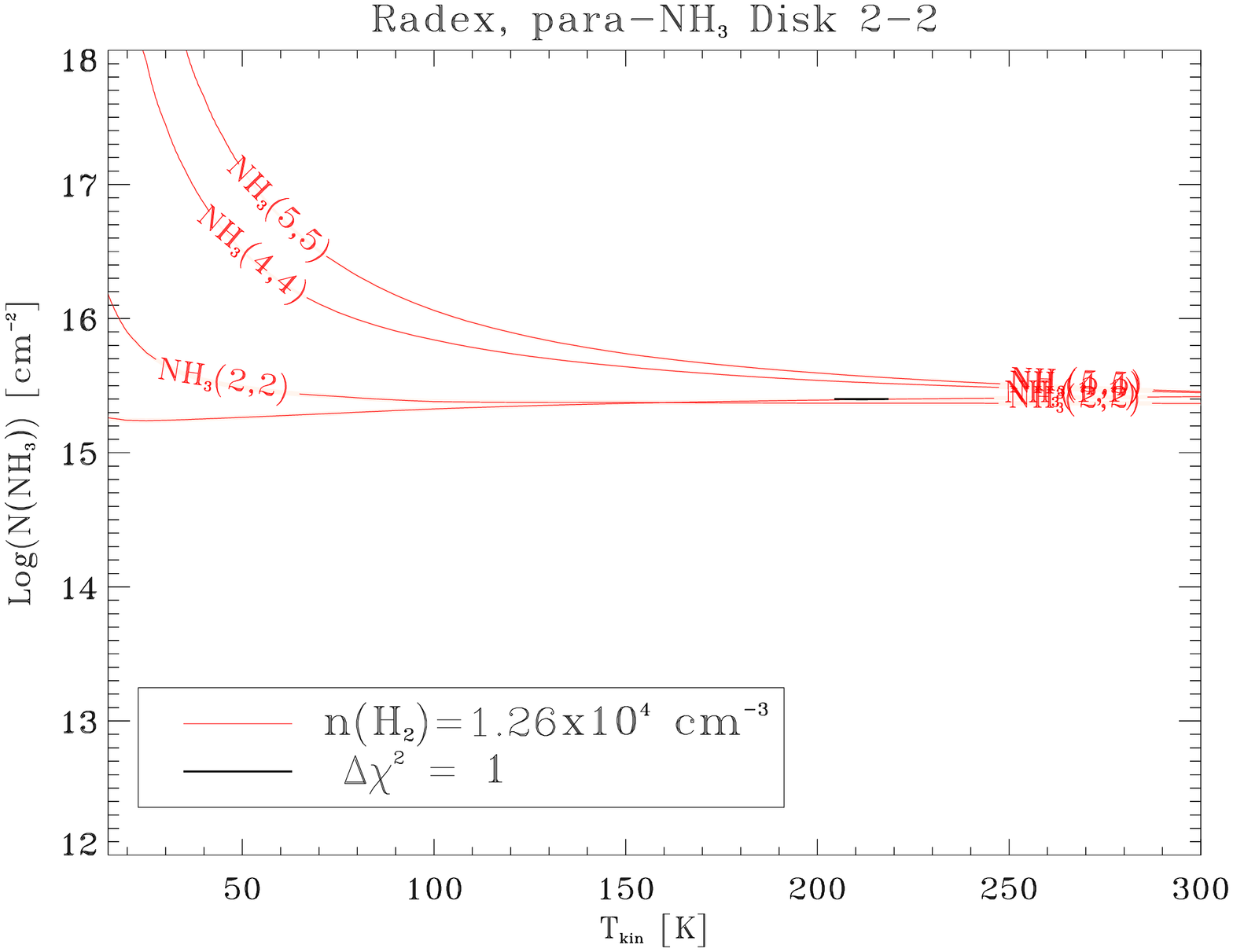}
\includegraphics[width=0.35\textwidth, angle=90]{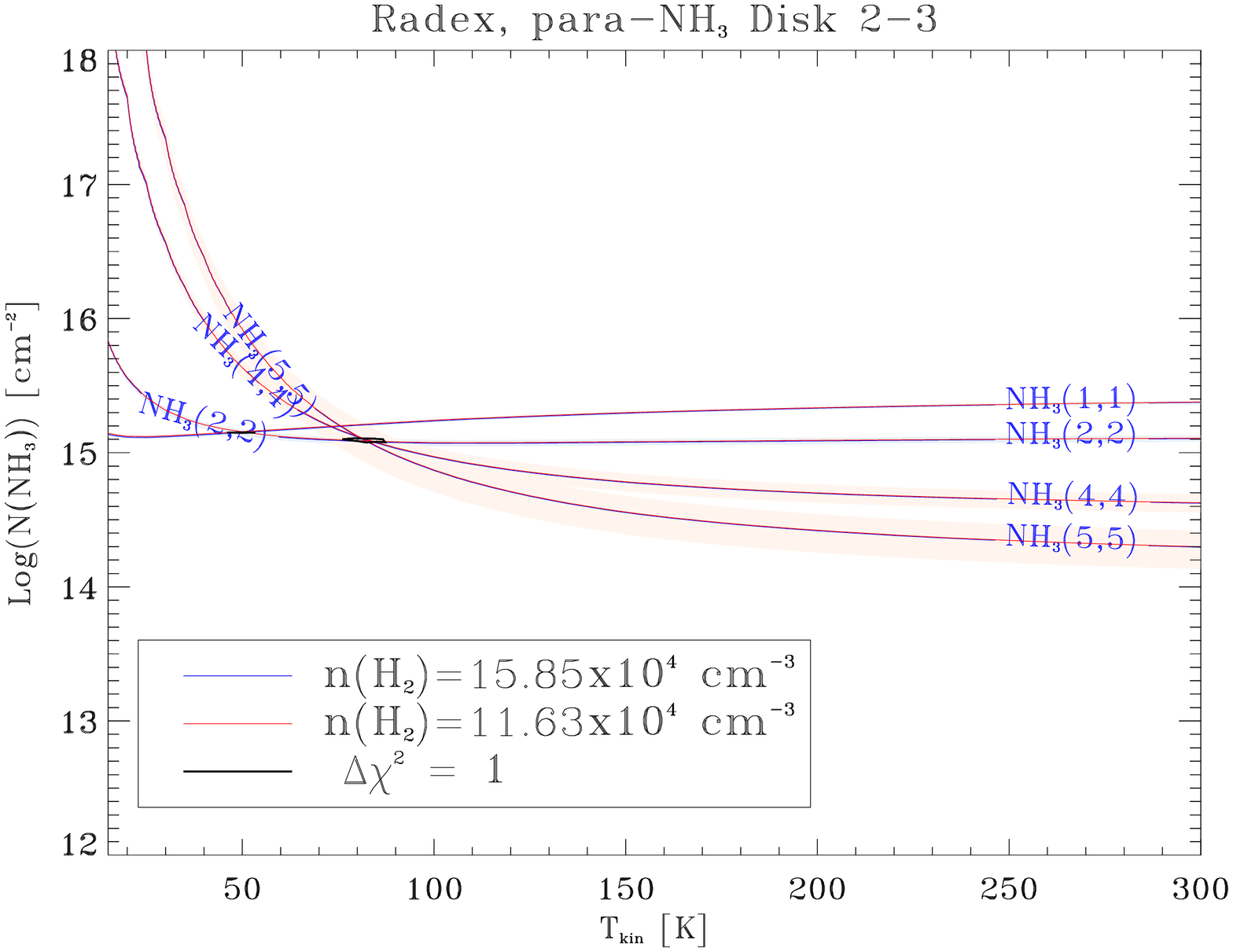}
}
\caption{LVG diagrams of NH$_3$ for each velocity component of
Disk\,2. Top: $ 48.8\, {\rm km\,s}^{-1}$. Middle: $74.5 \,{\rm km\,s}^{-1}$.  Bottom: $95.8\, {\rm km\,s}^{-1}$. The blue lines correspond to the low temperature regimen, and the red lines to the high temperature regimen.}
\label{modeloSgrB2}
\end{figure}
\clearpage
%_________________________________________________________________________________________________
\begin{figure}
\vbox{
\includegraphics[width=0.4\textwidth, angle=90]{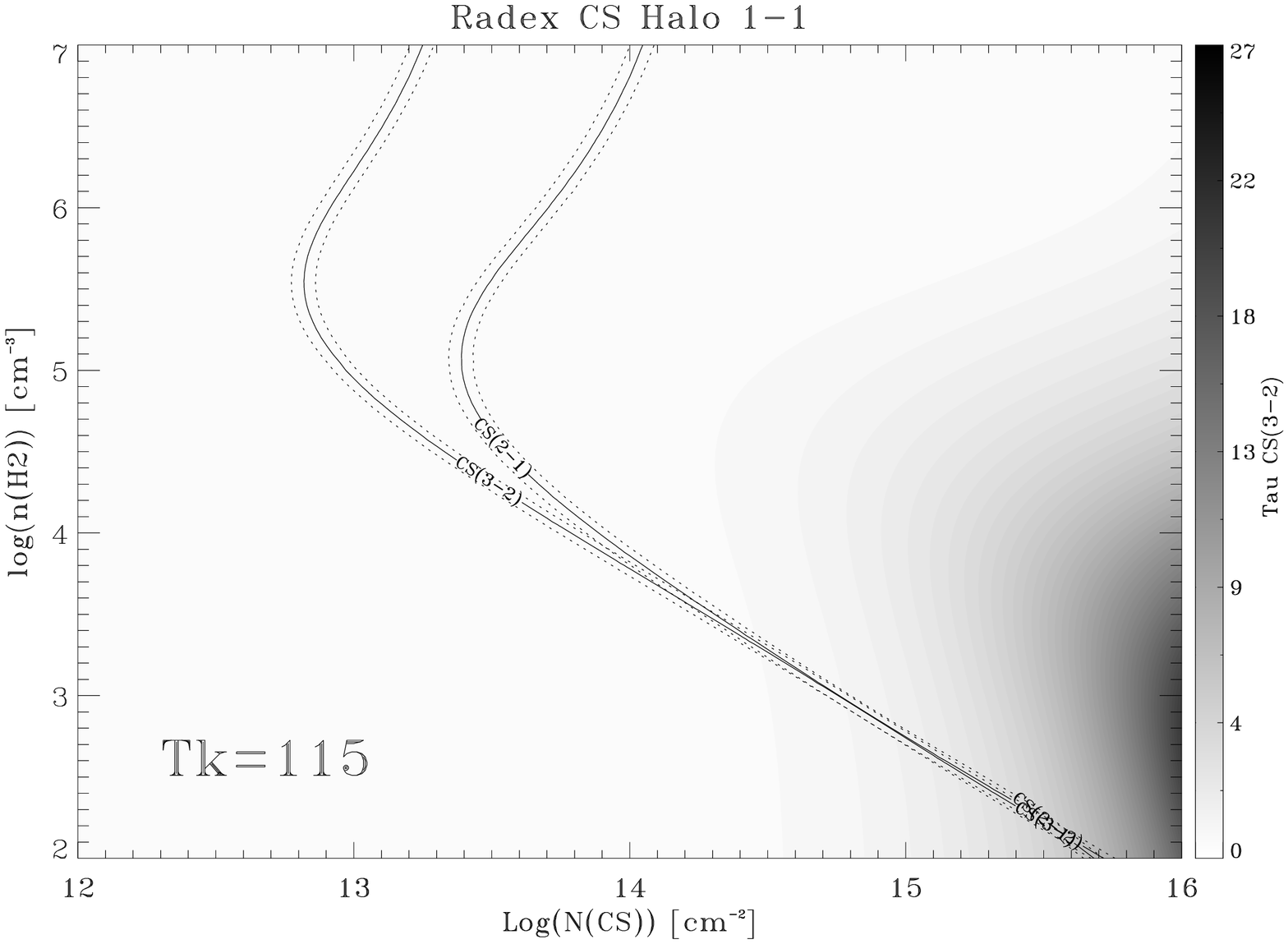}
\includegraphics[width=0.4\textwidth, angle=90]{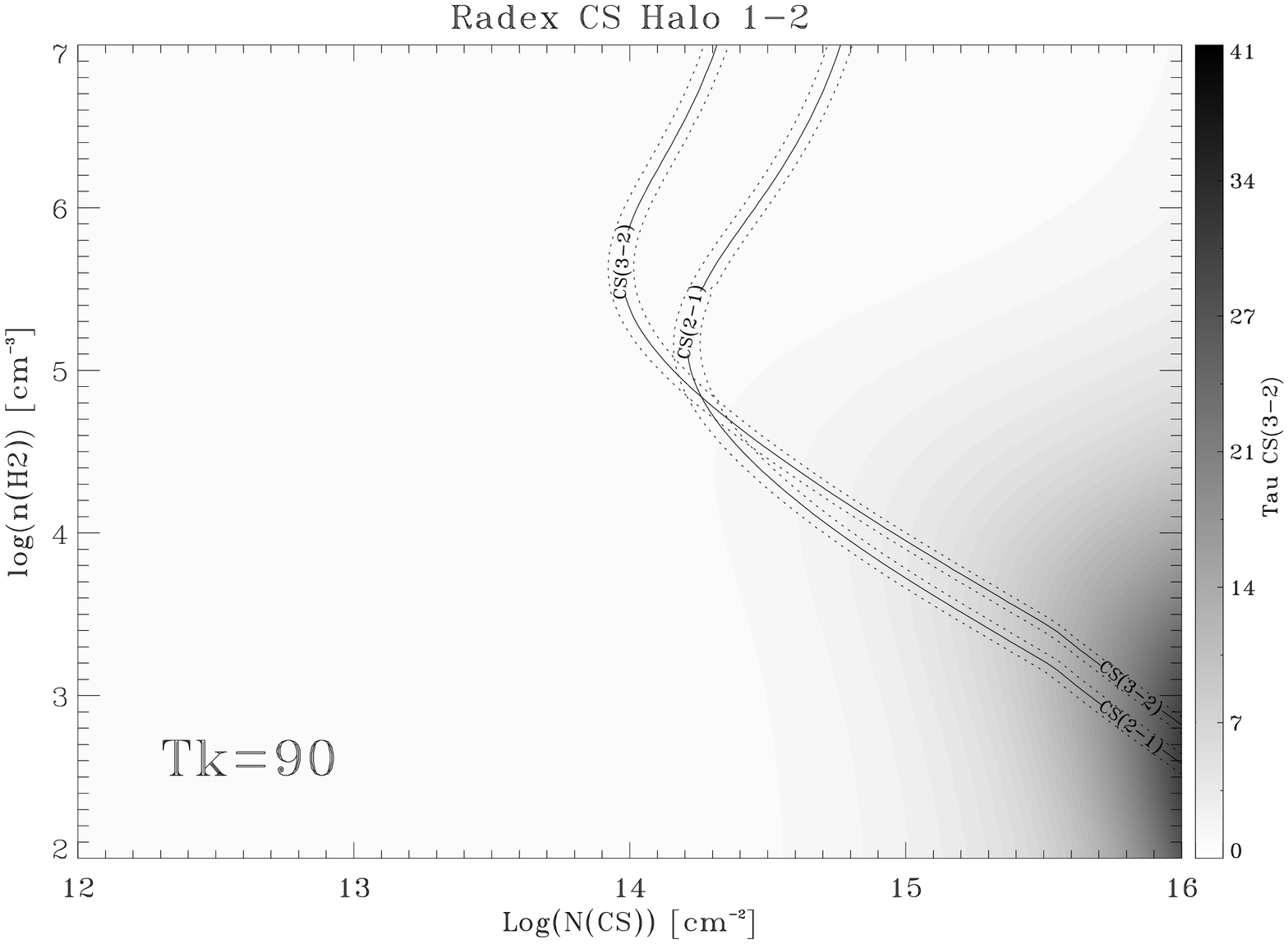}
\includegraphics[width=0.4\textwidth, angle=90]{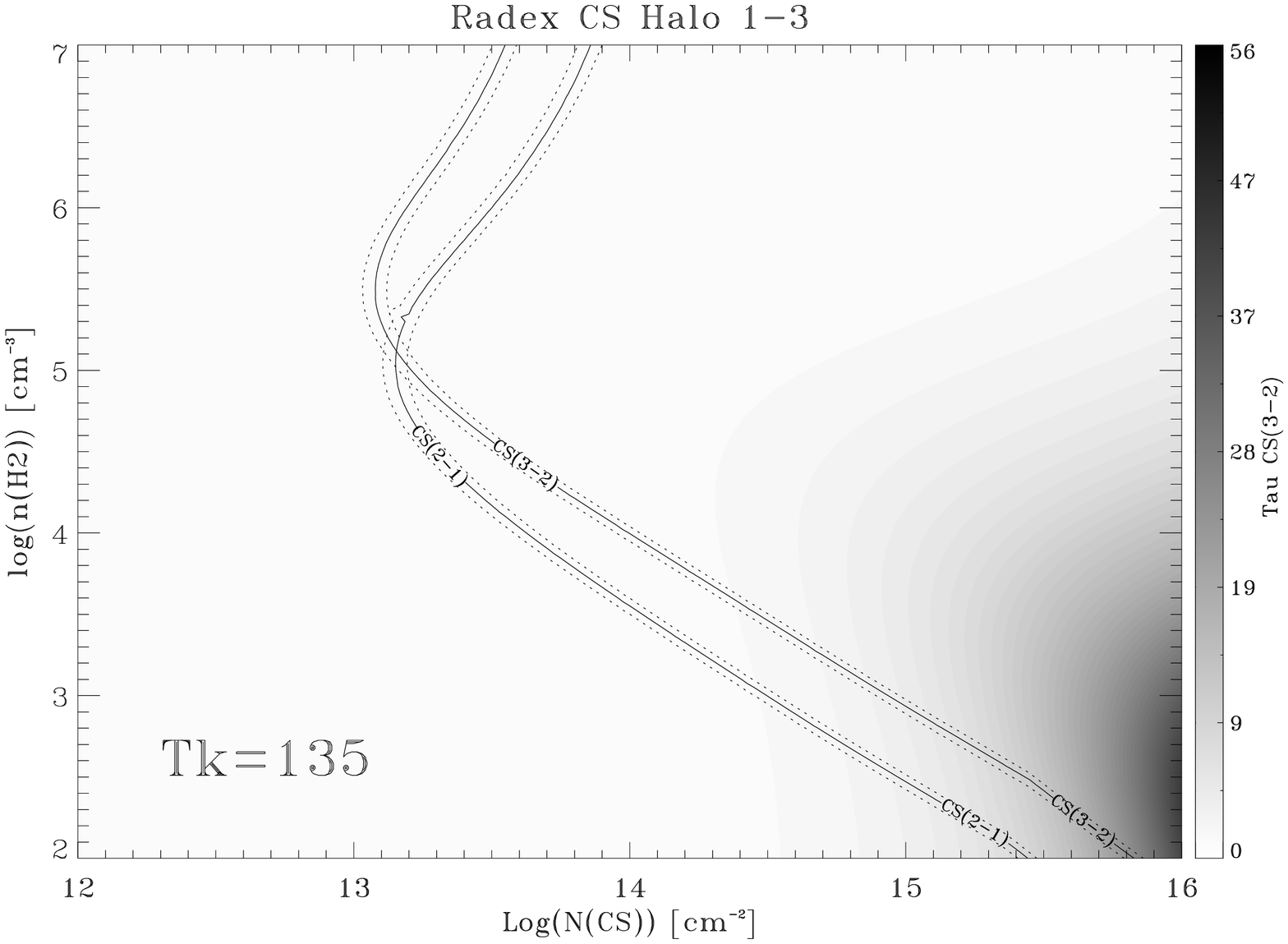}
}
\caption{LVG diagrams of CS for each velocity component (defined by the J=2-1 CS) of Halo\,1 (see Table \ref{SiO-CS-HNCO}) . 
Top: 83.8 km\,s$^{-1}$. Middle: $117.4\,{\rm km\,s}^{-1}$.  Bottom: $136.9\, {\rm km\,s}^{-1}$. The $T_{{\rm kin}}$ is indicated in the lower left corner in each plot.}
\label{modeloCSHalo1}
\end{figure}

\begin{figure}
\vbox{
\includegraphics[width=0.4\textwidth, angle=90]{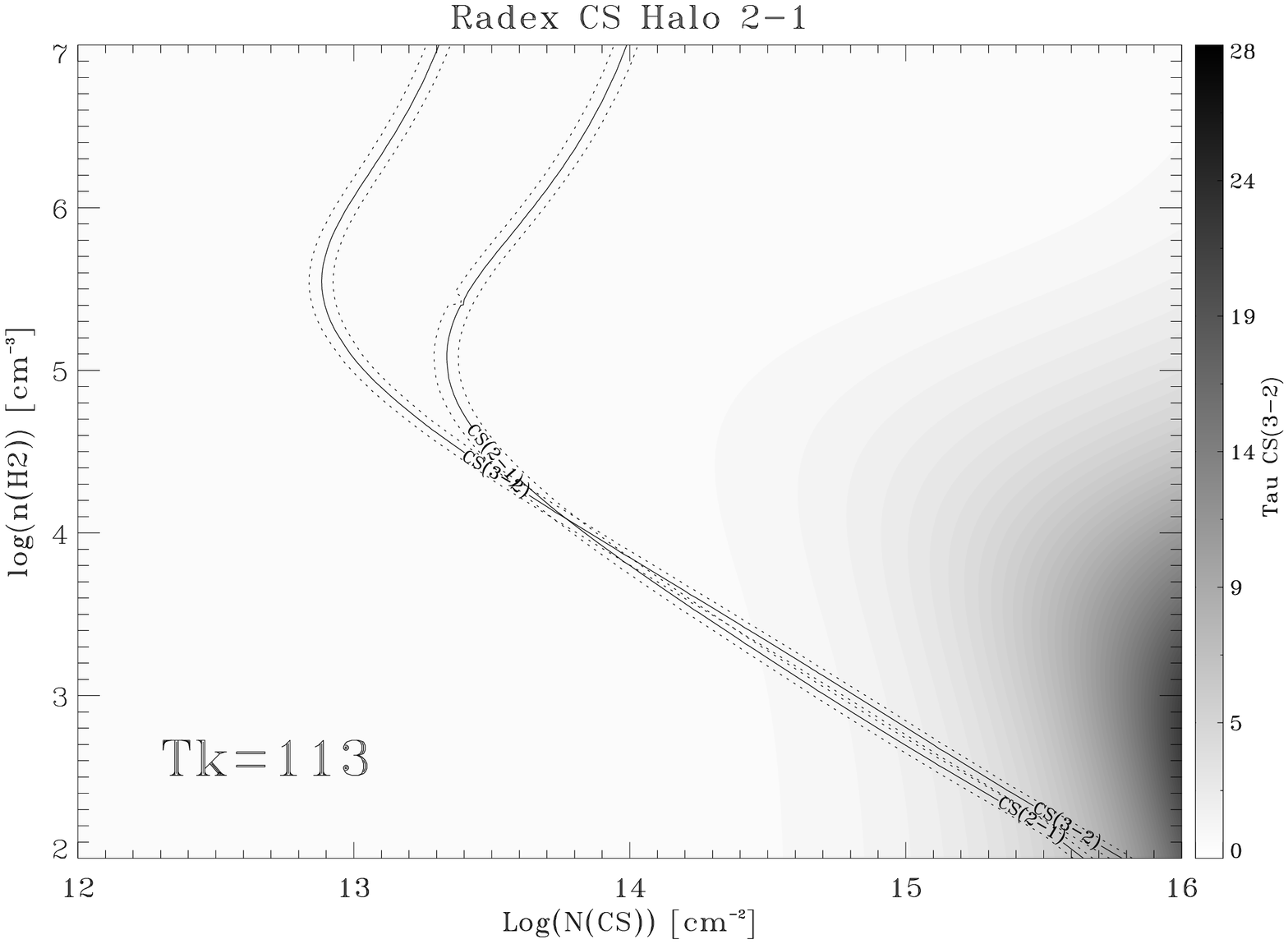}
\includegraphics[width=0.4\textwidth, angle=90]{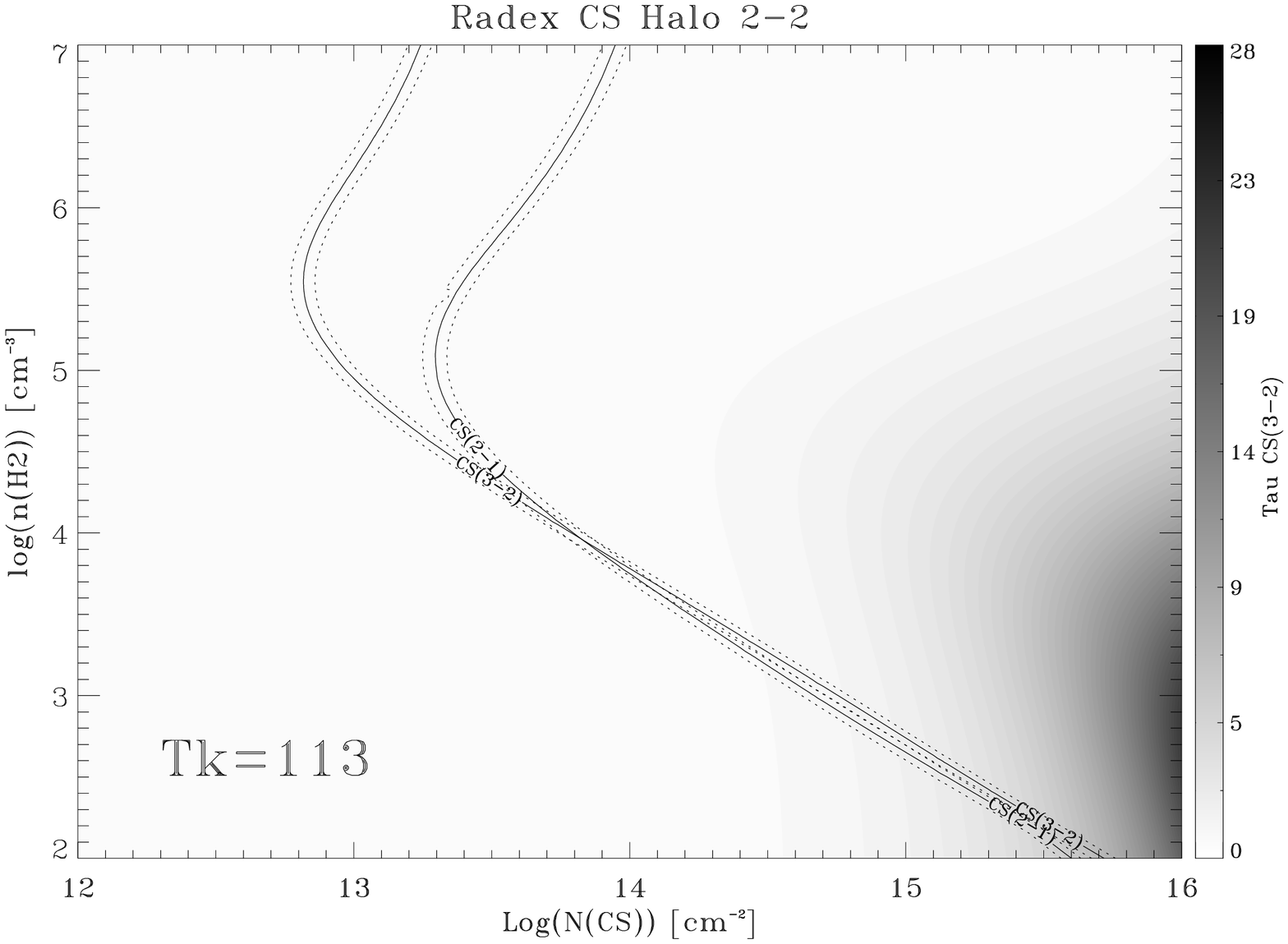}
}
\caption{LVG diagrams of CS for each velocity component of Halo\,2. Top:$-80\, {\rm km\,s}^{-1}$.  Bottom: $-51.9\, {\rm km\,s}^{-1}$.}
\label{modeloCSHalo2}
\end{figure}

\begin{figure}
\vbox{
\includegraphics[width=0.4\textwidth, angle=90]{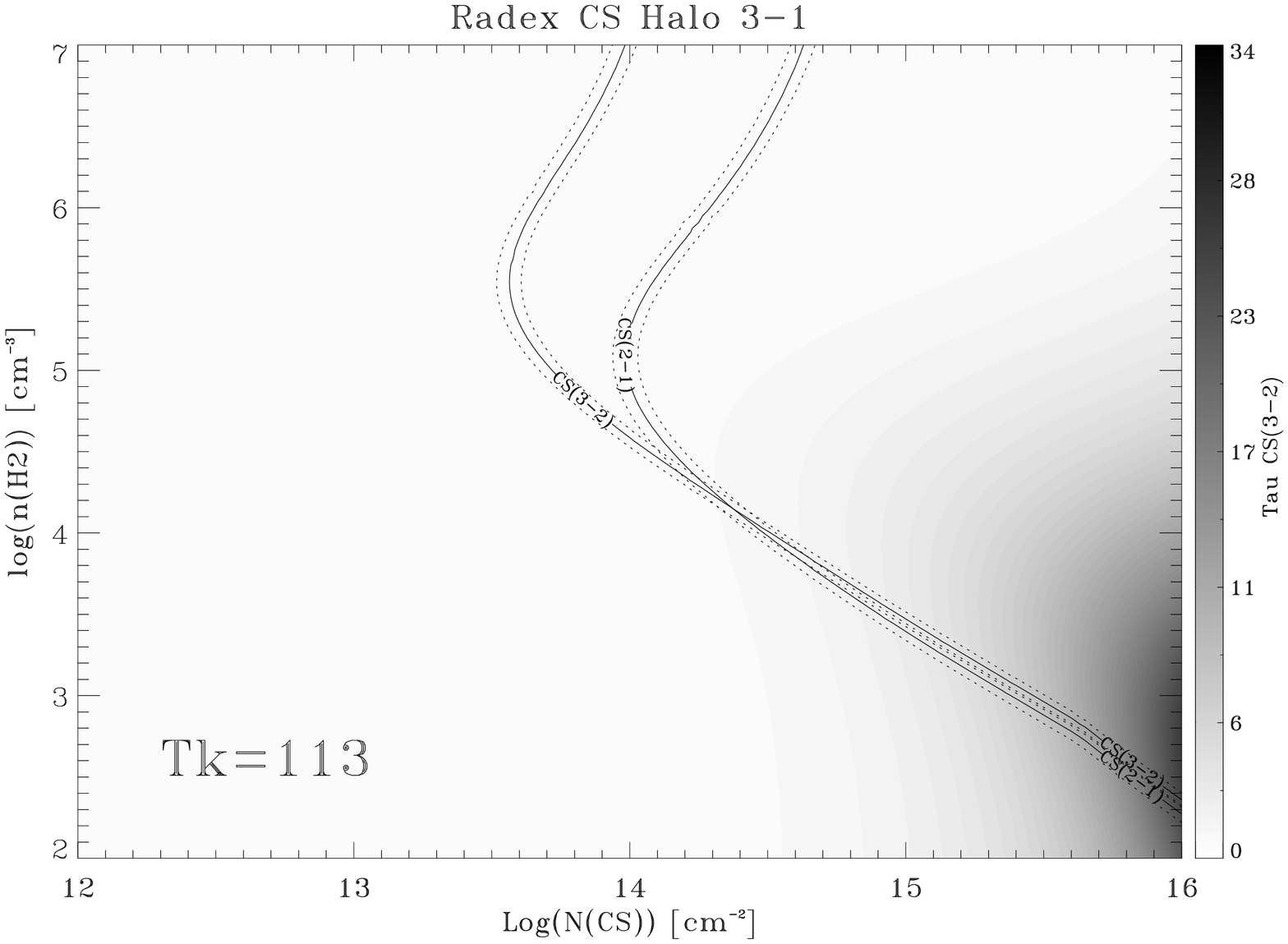}
\includegraphics[width=0.4\textwidth, angle=90]{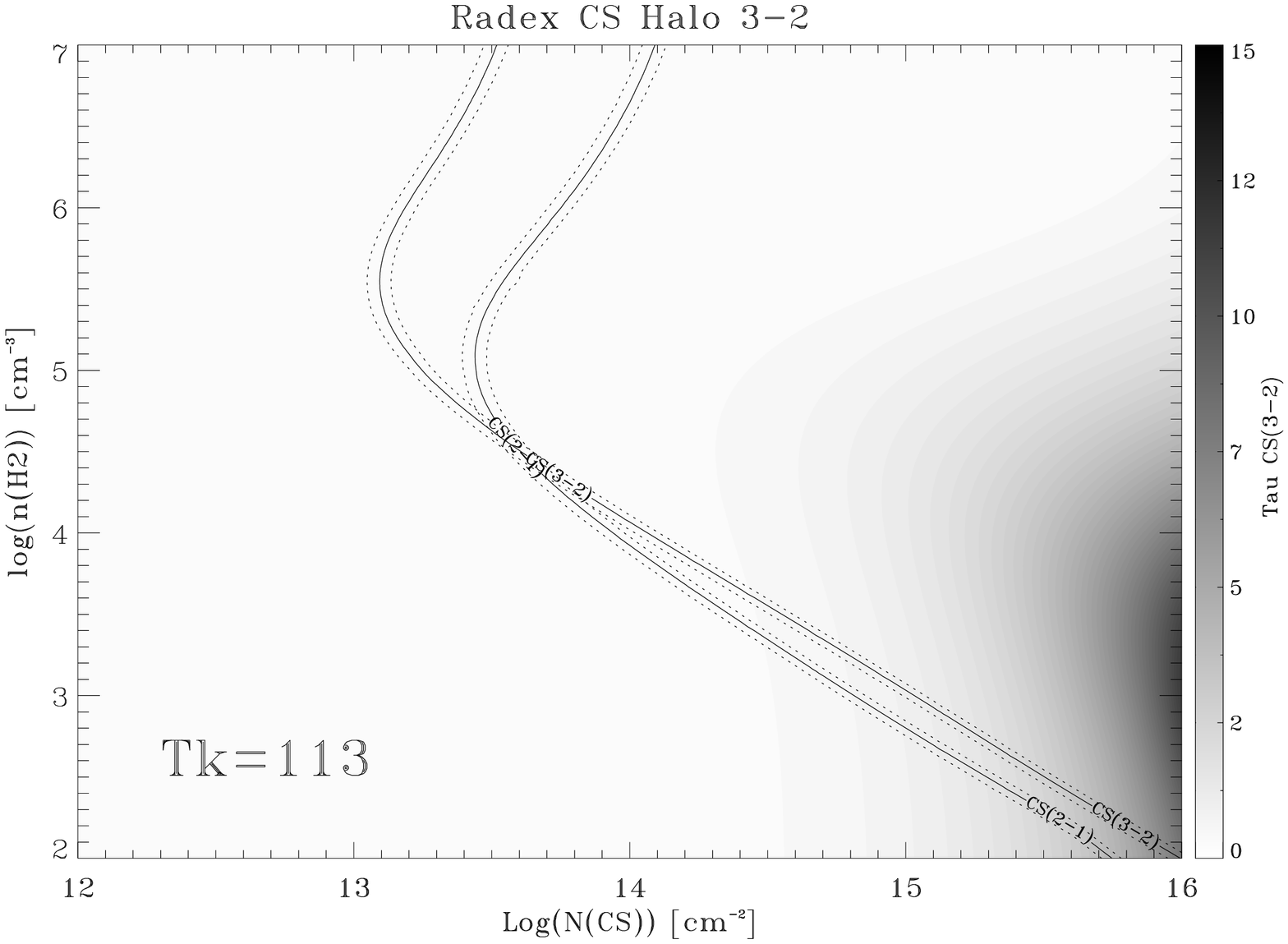}
}
\caption{LVG diagrams of CS for each velocity component of Halo\,3. Top: $-64.7 \,{\rm km\,s}^{-1}$.  Bottom: $-13.9\, {\rm km\,s}^{-1}$.}
\label{modeloCSHalo3}
\end{figure}

\begin{figure}
\includegraphics[width=0.4\textwidth, angle=90]{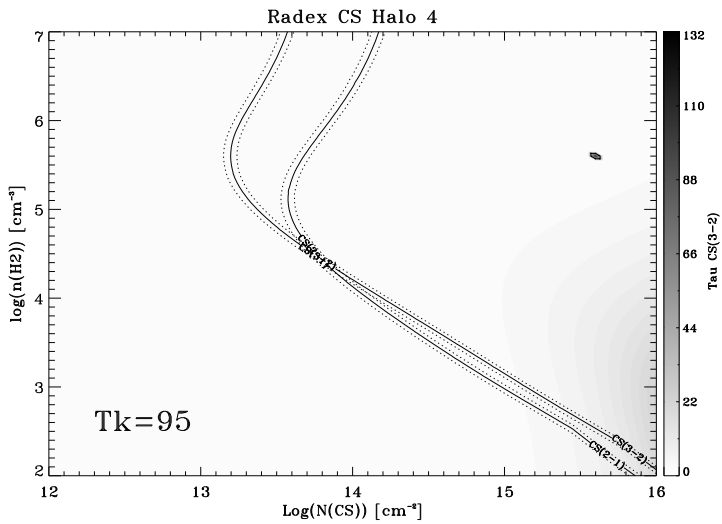}
\caption{LVG diagrams of CS for Halo\,4.}
\label{modeloCSHalo4}
\end{figure}

\begin{figure}
\includegraphics[width=0.4\textwidth, angle=90]{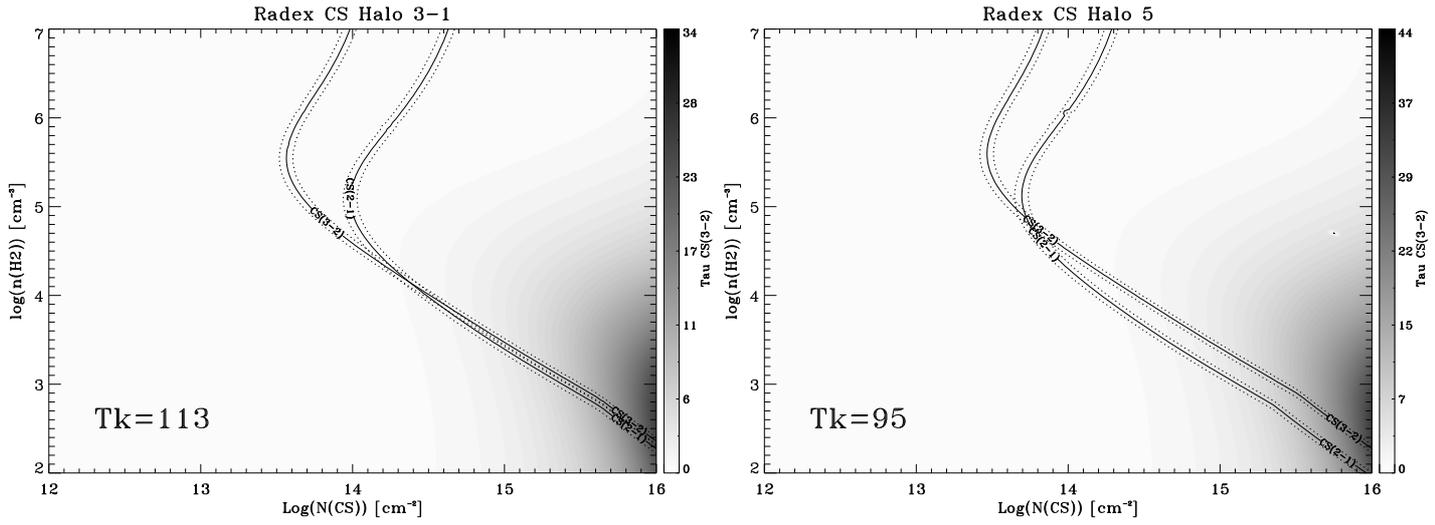}
\caption{LVG diagrams of CS for Halo\,5.}
\label{modeloCSHalo5}
\end{figure}
\clearpage

\begin{figure*}
\hbox{
\includegraphics[width=0.4\textwidth, angle=90]{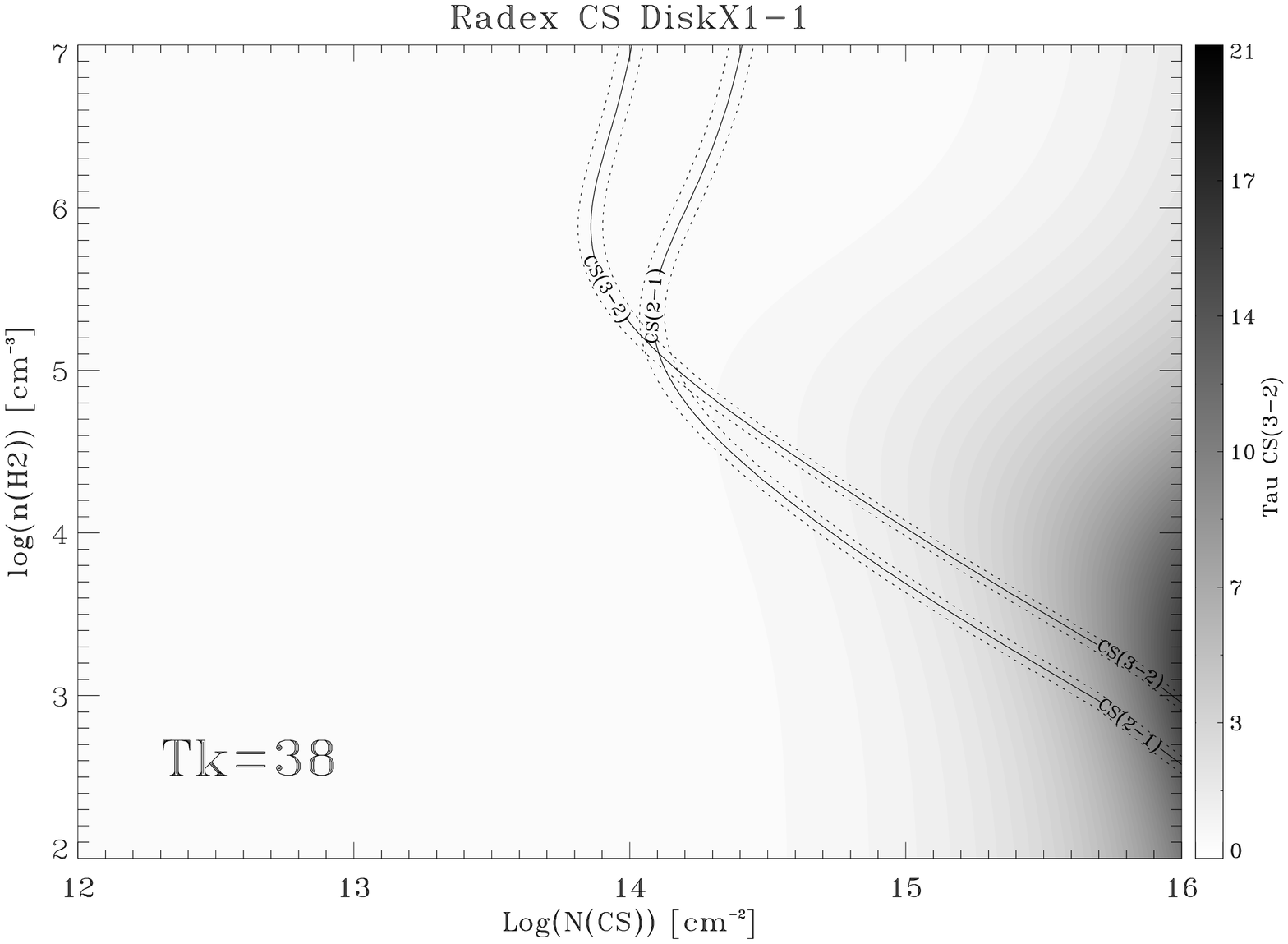}
\hspace{-0.8cm}
\includegraphics[width=0.4\textwidth, angle=90]{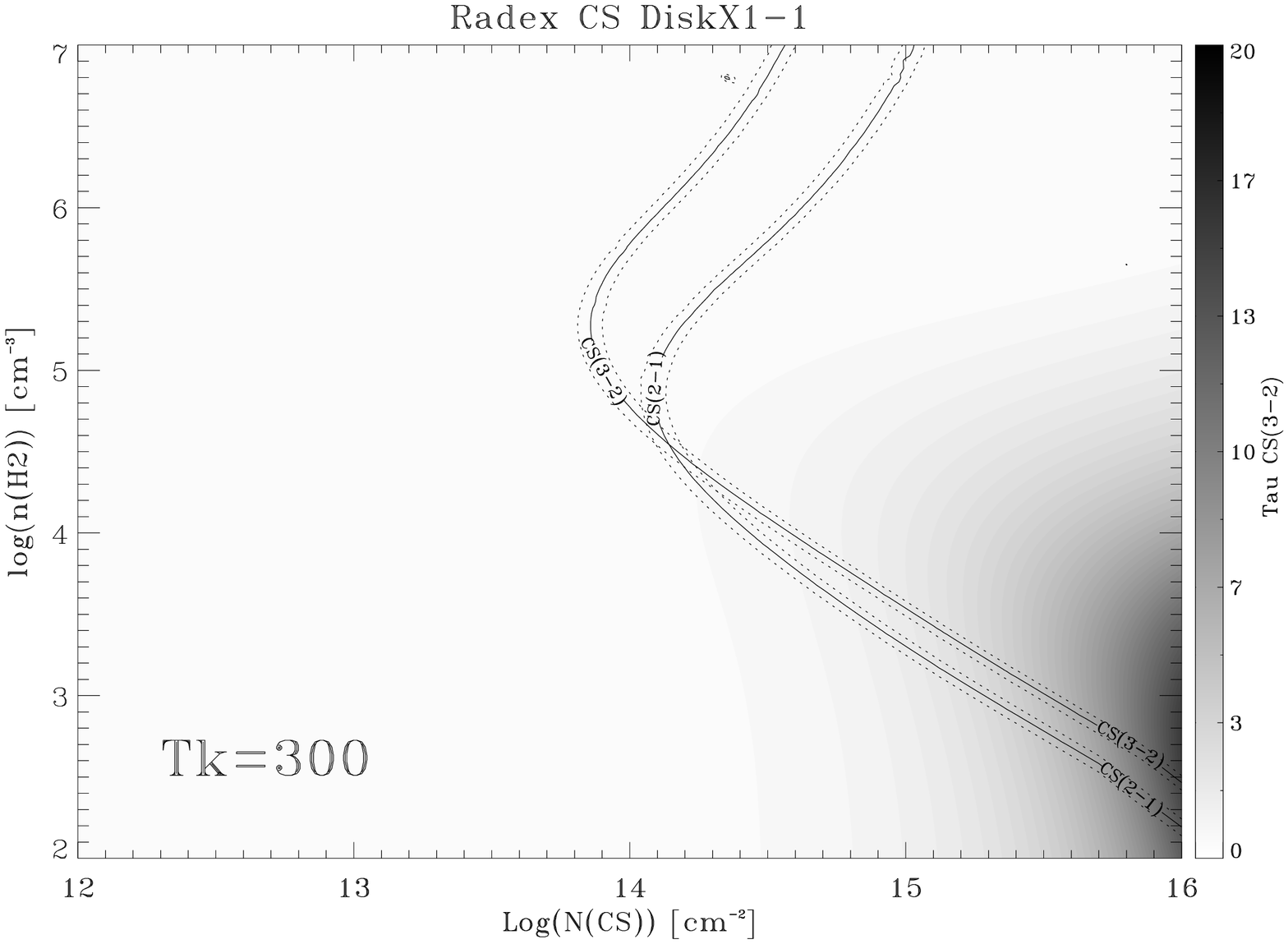}
}
\caption{LVG  diagrams of CS for each kinetic temperature regime of Disk\,X1-1. }
\label{modeloCS13complexX1}
\end{figure*}

\begin{figure*}
\hbox{
\includegraphics[width=0.4\textwidth, angle=90]{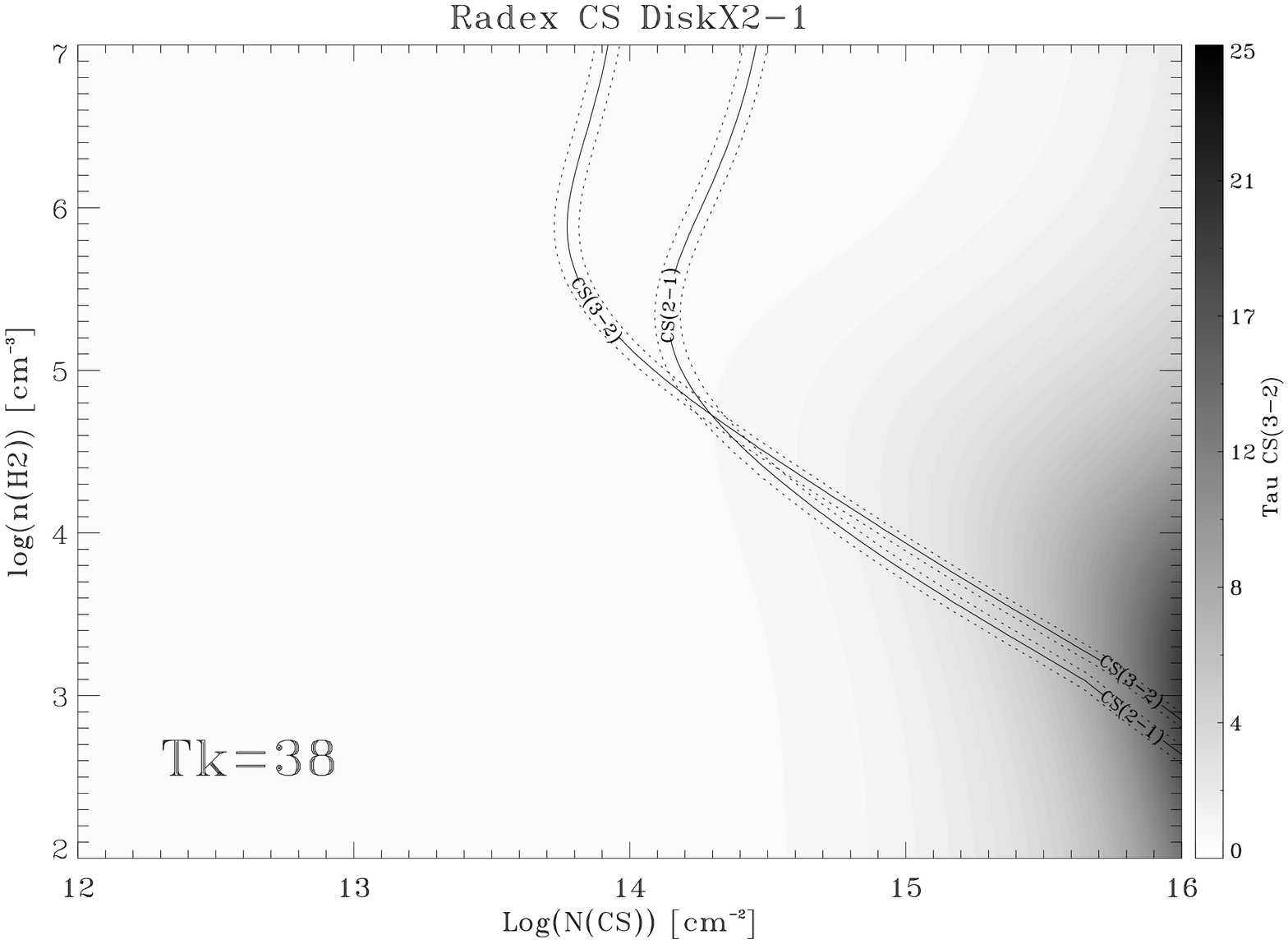}
\hspace{-0.8cm}
\includegraphics[width=0.4\textwidth, angle=90]{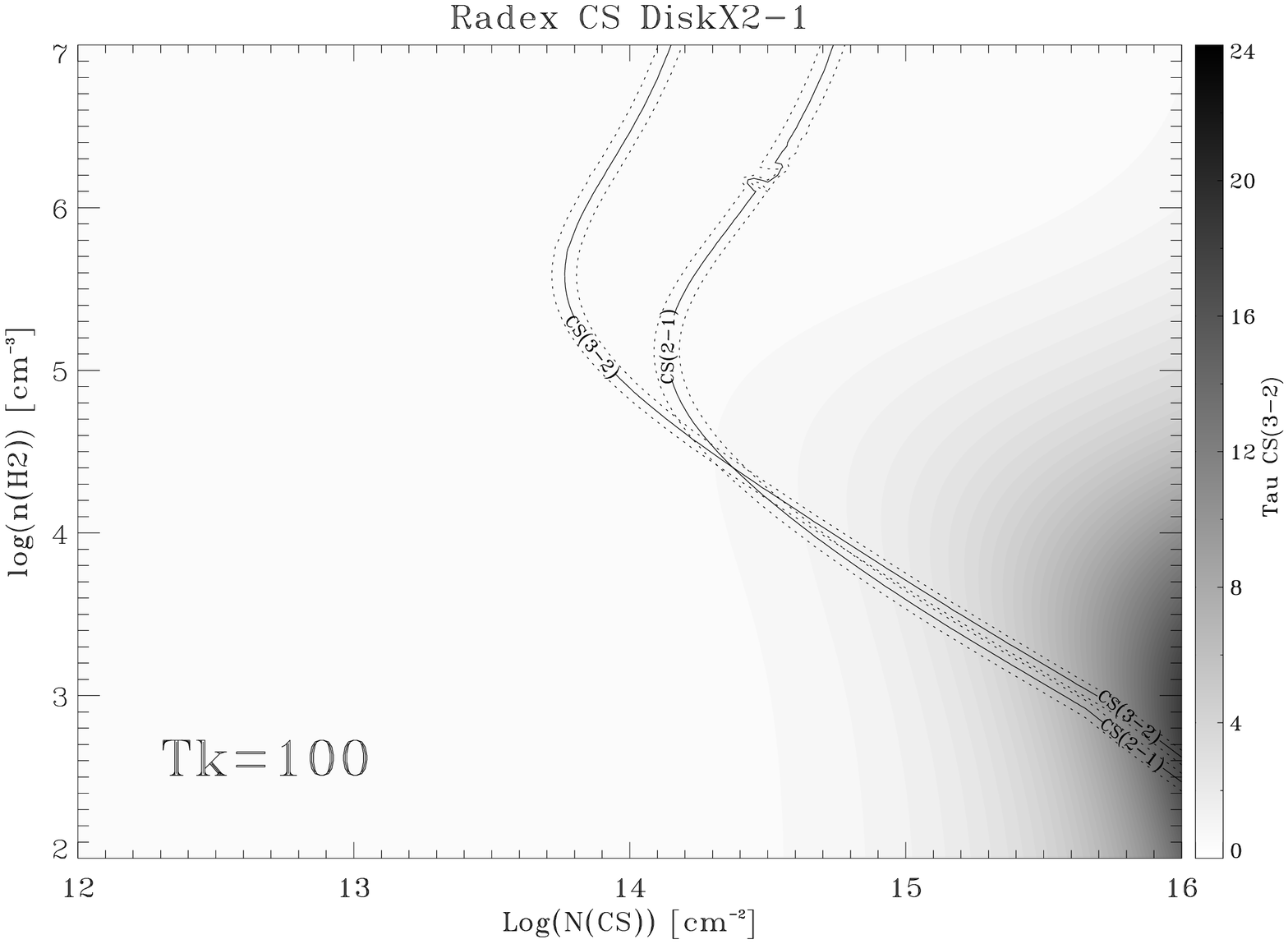}
}
\caption{LVG diagrams of CS for each kinetic temperature regime of Disk\,X2-1}
\label{modeloCS13complexX2}
\end{figure*}

\begin{figure*}
\hbox{
\includegraphics[width=0.4\textwidth, angle=90]{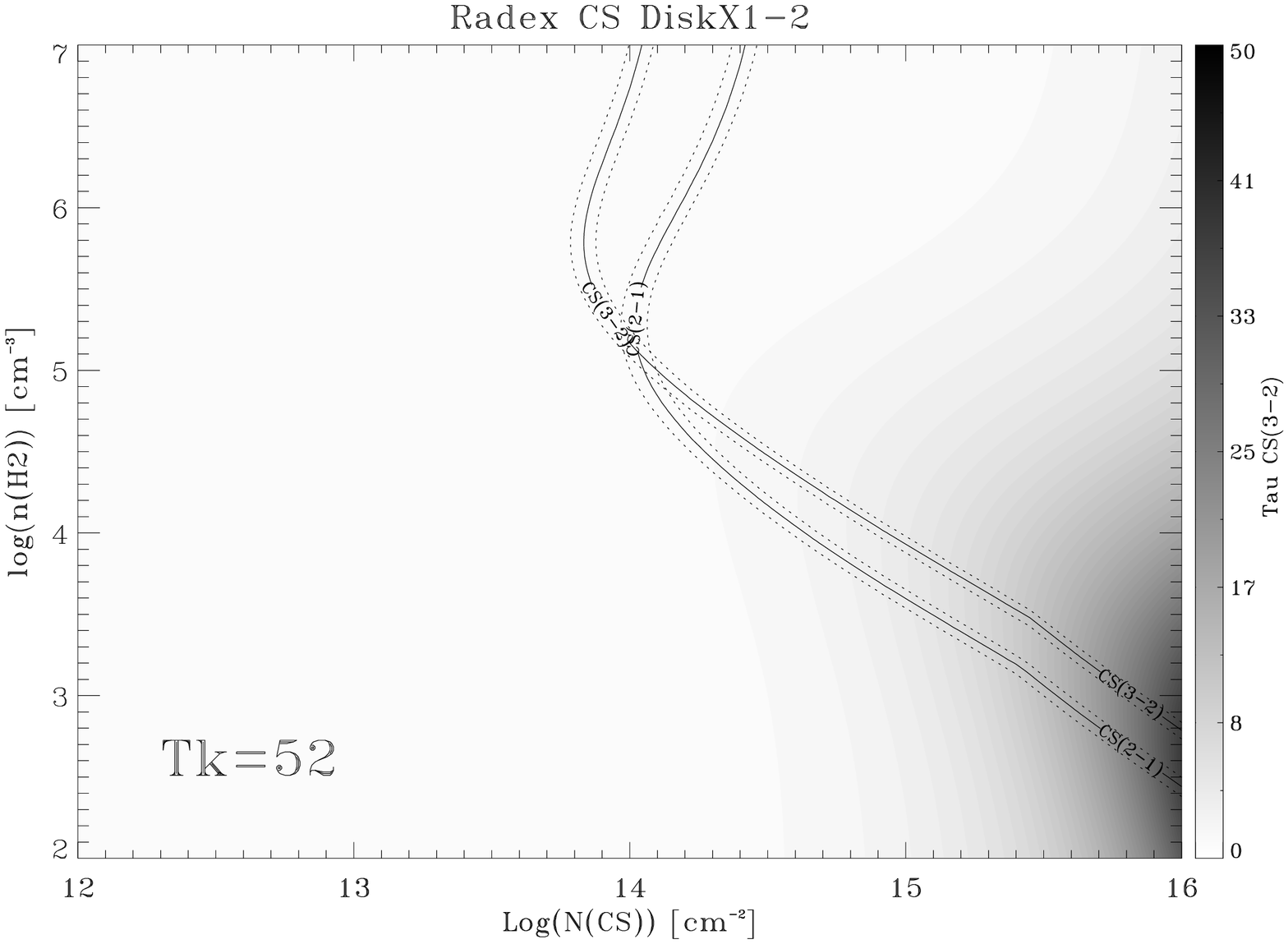}
\hspace{-0.8cm}
\includegraphics[width=0.4\textwidth, angle=90]{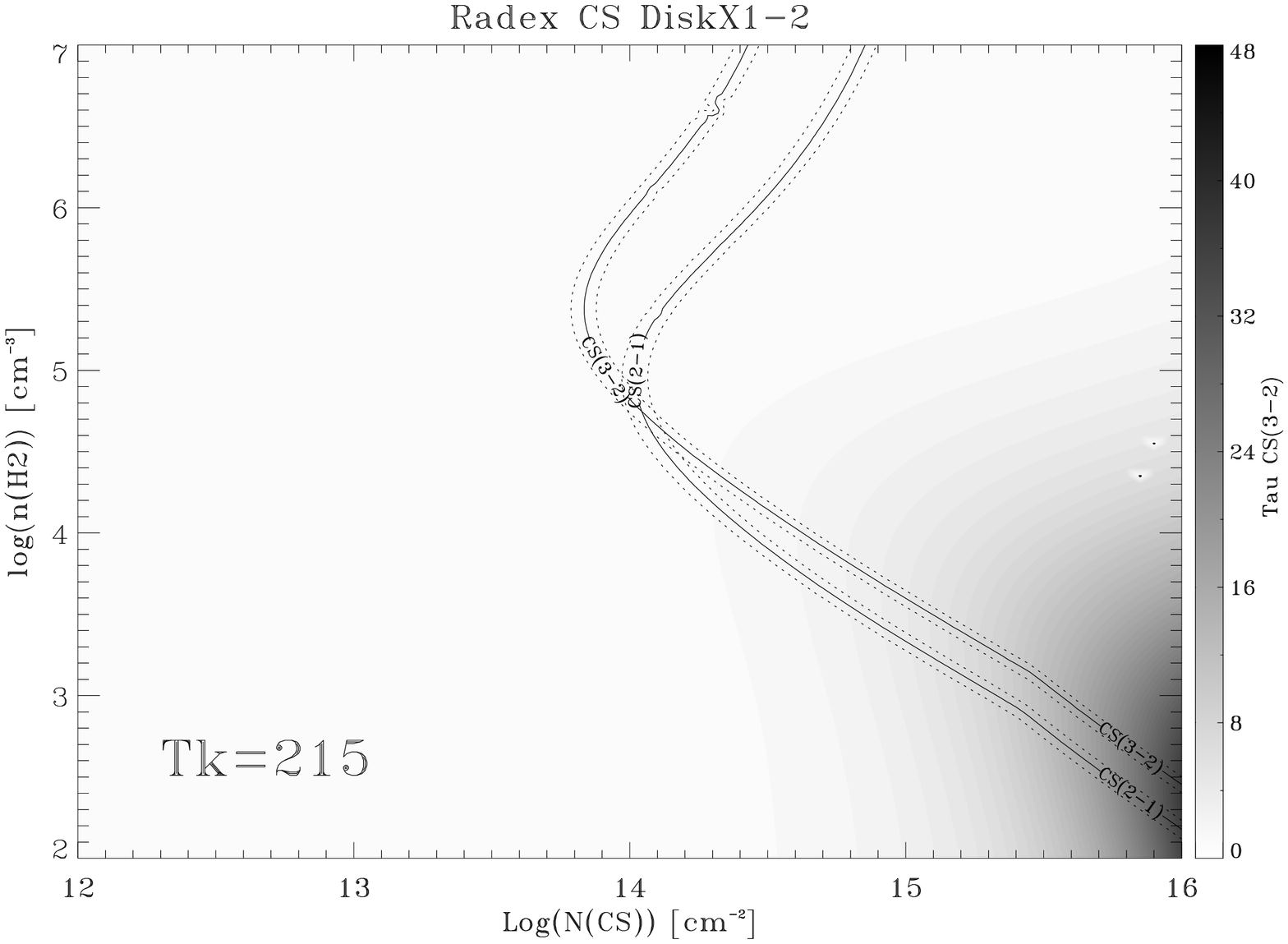}
}
\caption{LVG diagrams of CS for each kinetic temperature regime of Disk\,X1-2}
\label{modeloCSSgrC1X1}
\end{figure*}

\begin{figure*}
\hbox{
\includegraphics[width=0.4\textwidth, angle=90]{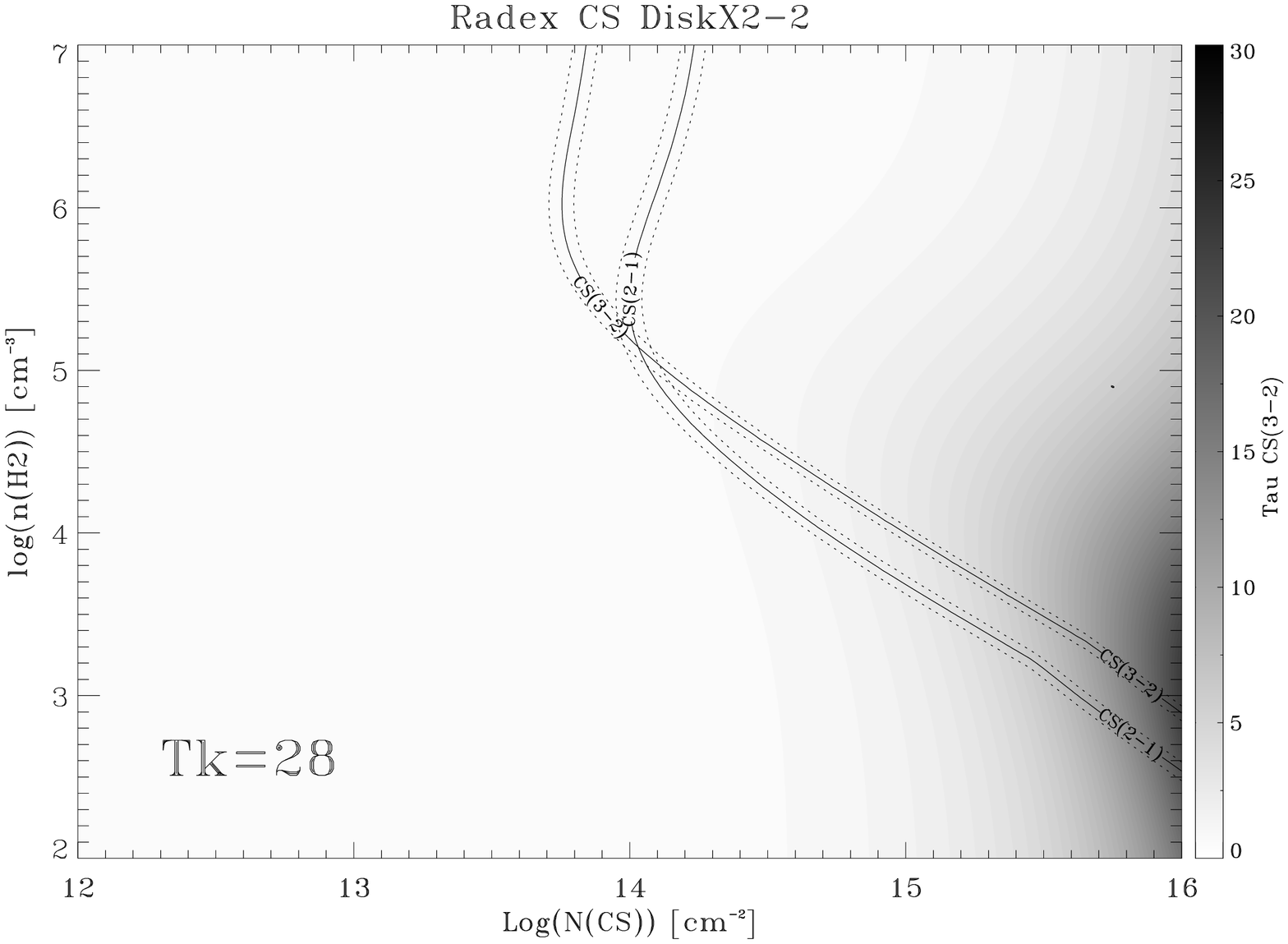}
\hspace{-0.8cm}
\includegraphics[width=0.4\textwidth, angle=90]{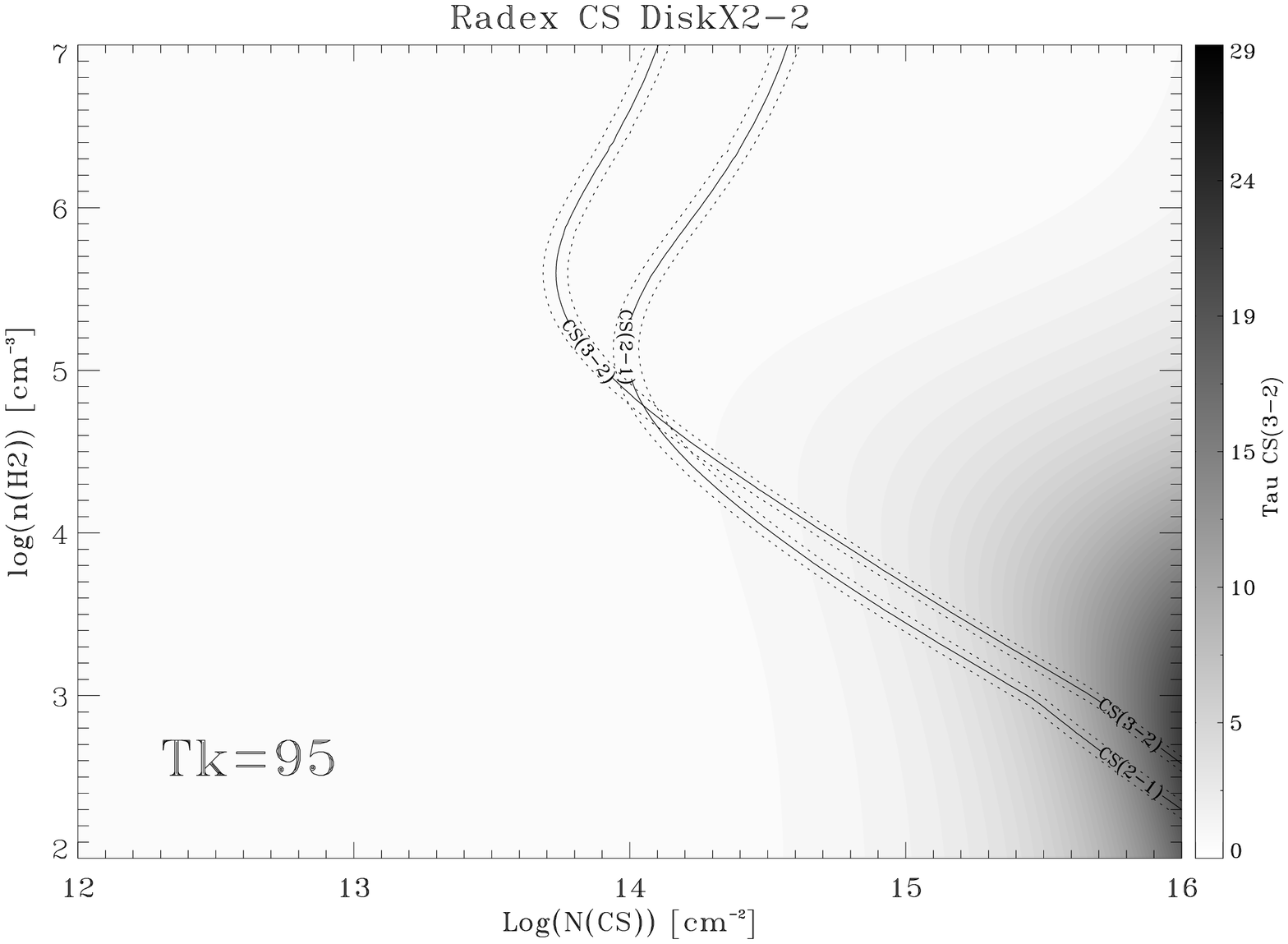}
}
\caption{LVG diagrams of CS for each kinetic temperature regime of Disk\,X2-2}
\label{modeloCSSgrC1X2}
\end{figure*}

\begin{figure*}
\vbox{
\hbox{
\includegraphics[width=0.4\textwidth, angle=90]{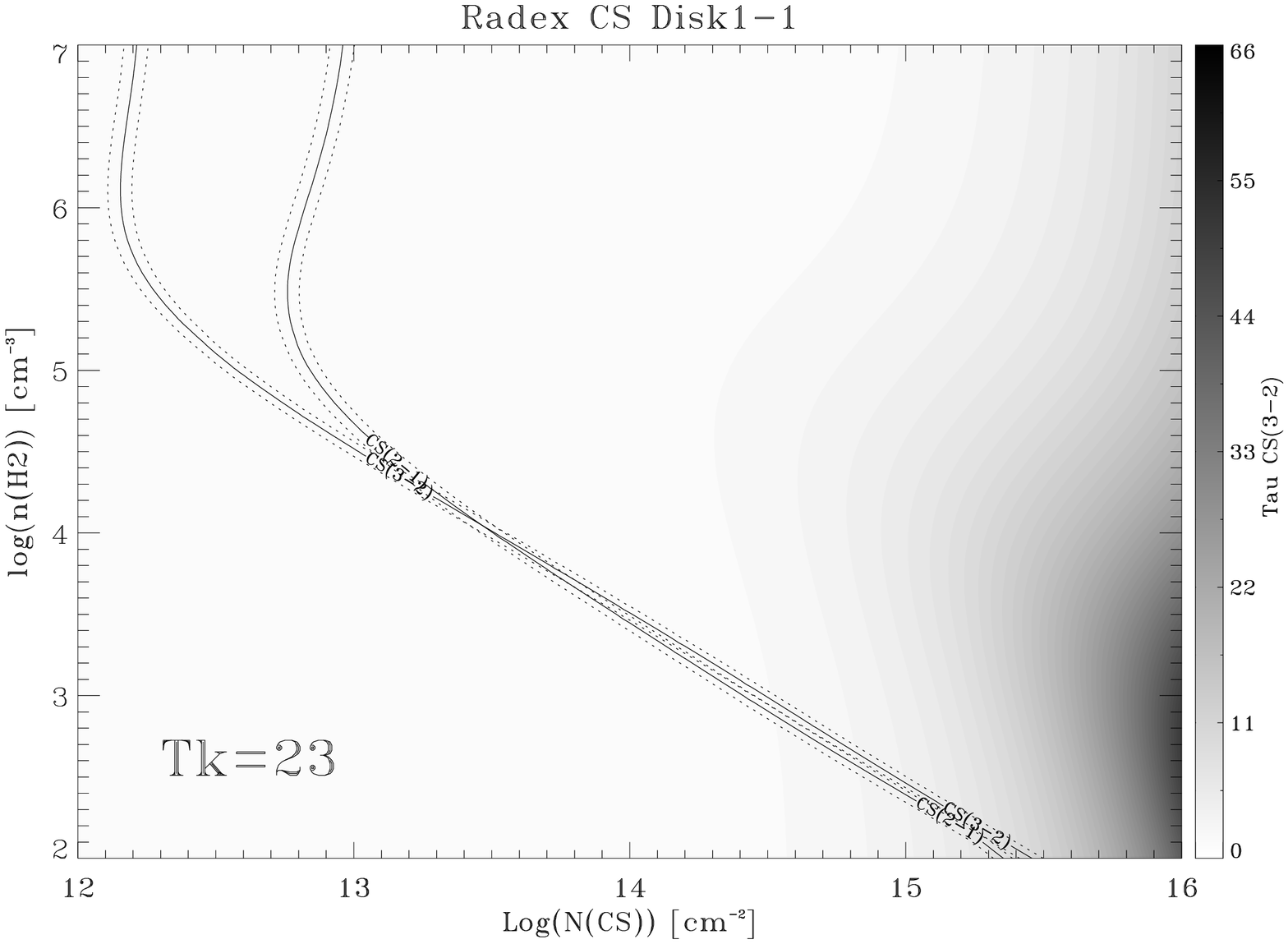}
\hspace{-0.8cm}
\includegraphics[width=0.4\textwidth, angle=90]{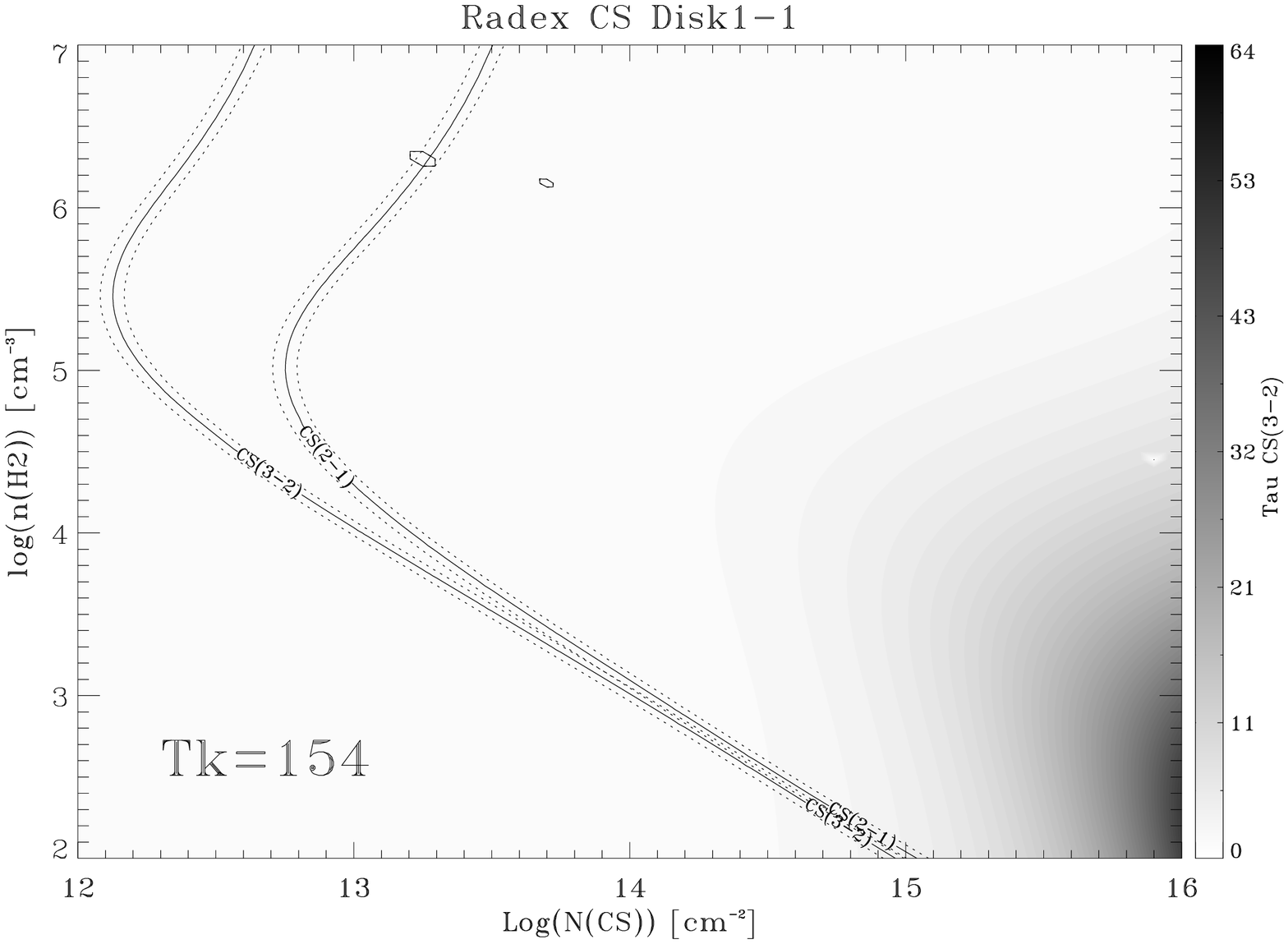}
}
\hbox{
\includegraphics[width=0.4\textwidth, angle=90]{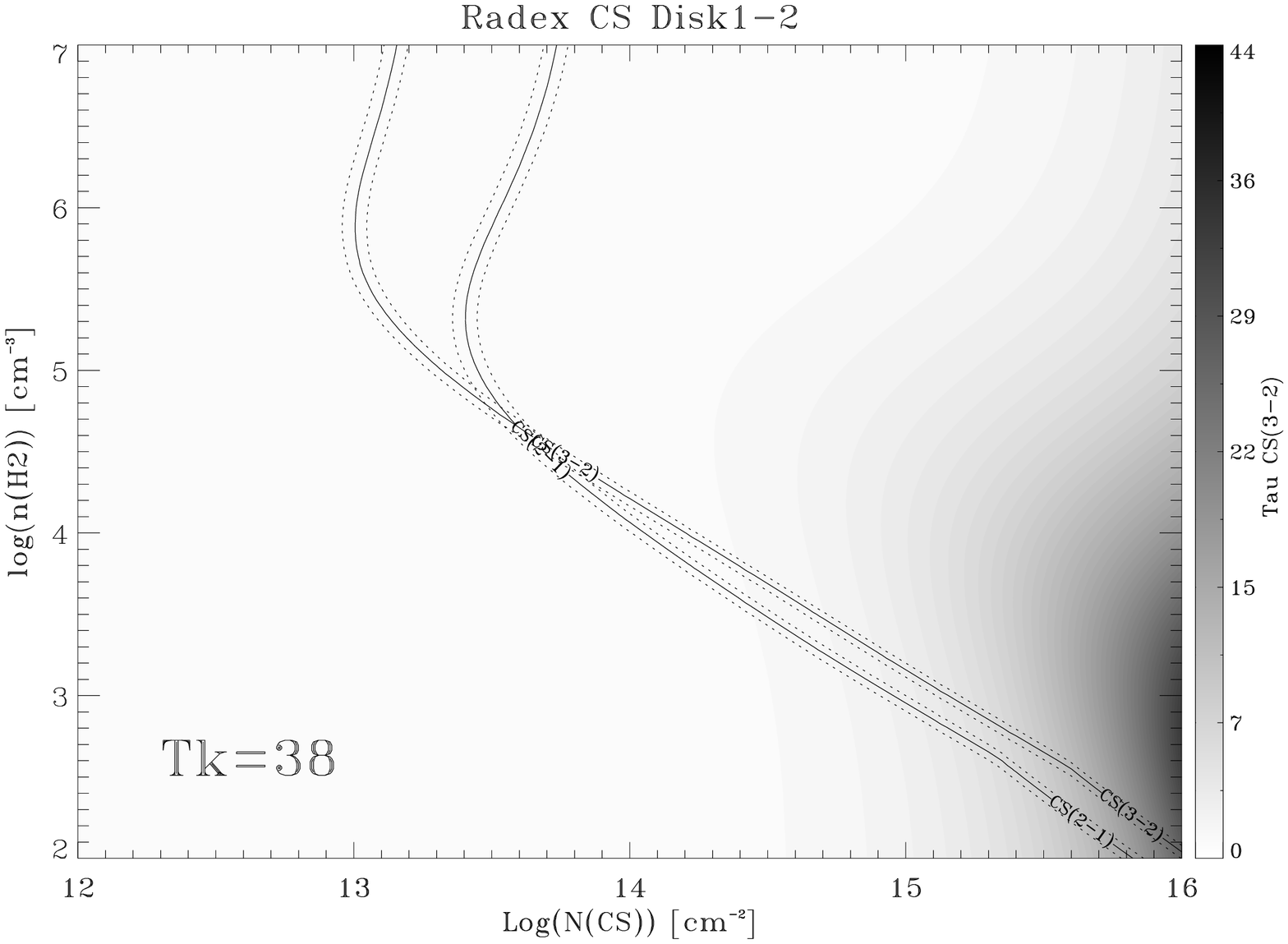}
\hspace{-0.8cm}
\includegraphics[width=0.4\textwidth, angle=90]{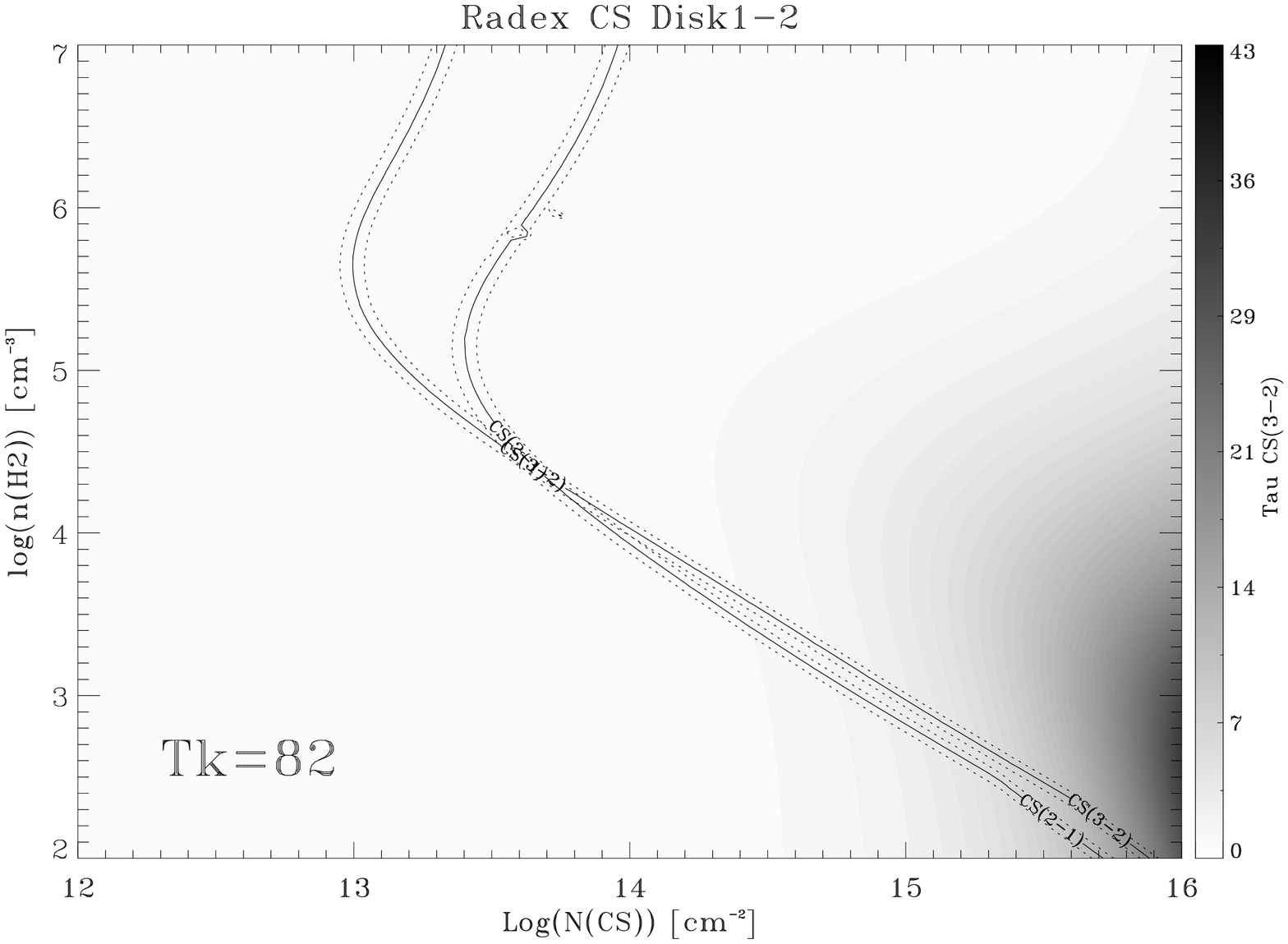}
}
}
\caption{LVG diagrams of CS for each kinetic temperature regime, for each velocity component of Disk\,1. Top: $ 56.8\,{\rm km\,s}^{-1}$. Bottom: $74.9\, {\rm km\,s}^{-1}$.}\label{modeloCScontrolD}
\end{figure*}

\begin{figure*}
\vbox{
\hbox{
\includegraphics[width=0.4\textwidth, angle=90]{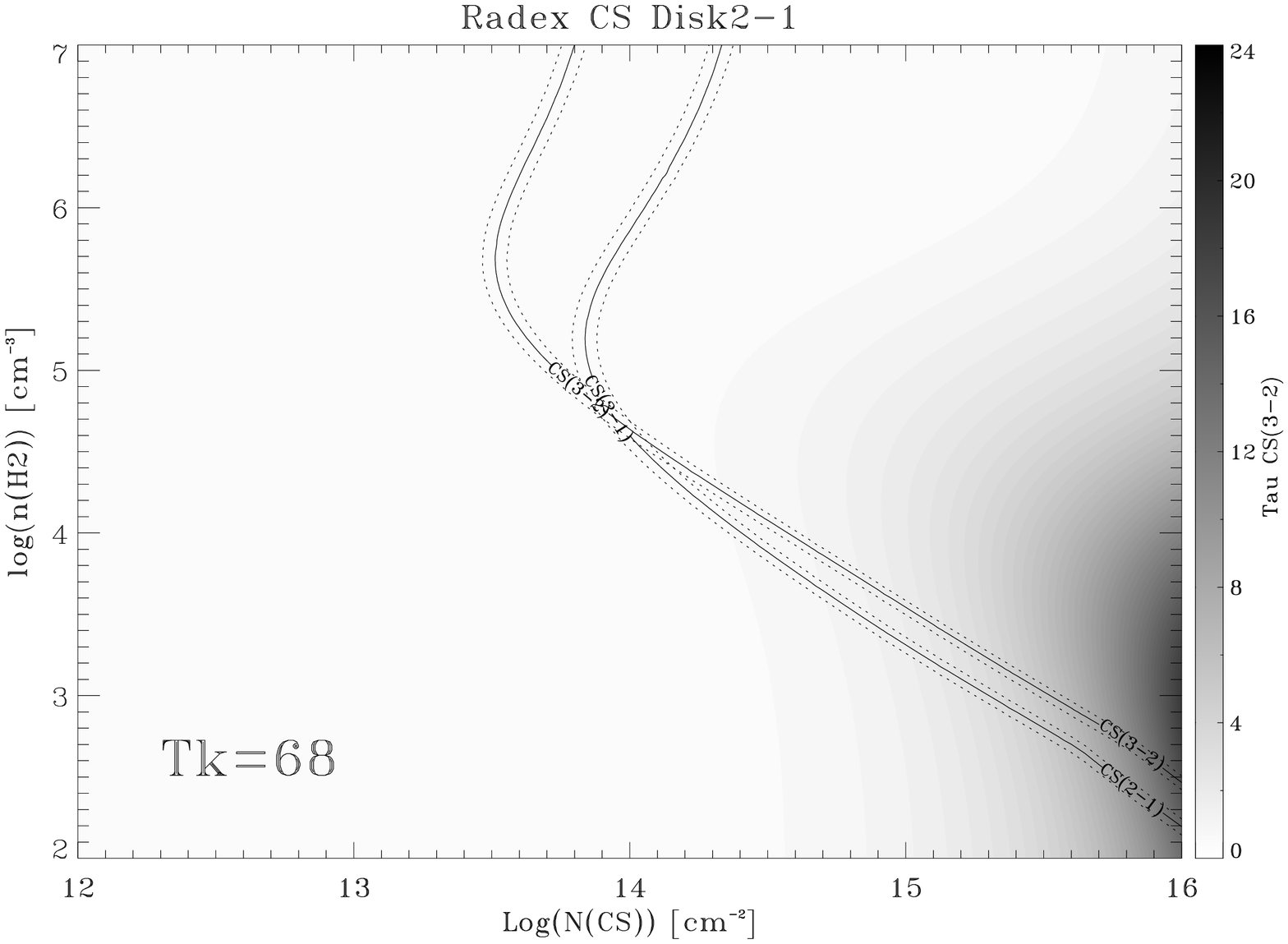}
\hspace{-0.8cm}
\includegraphics[width=0.4\textwidth, angle=90]{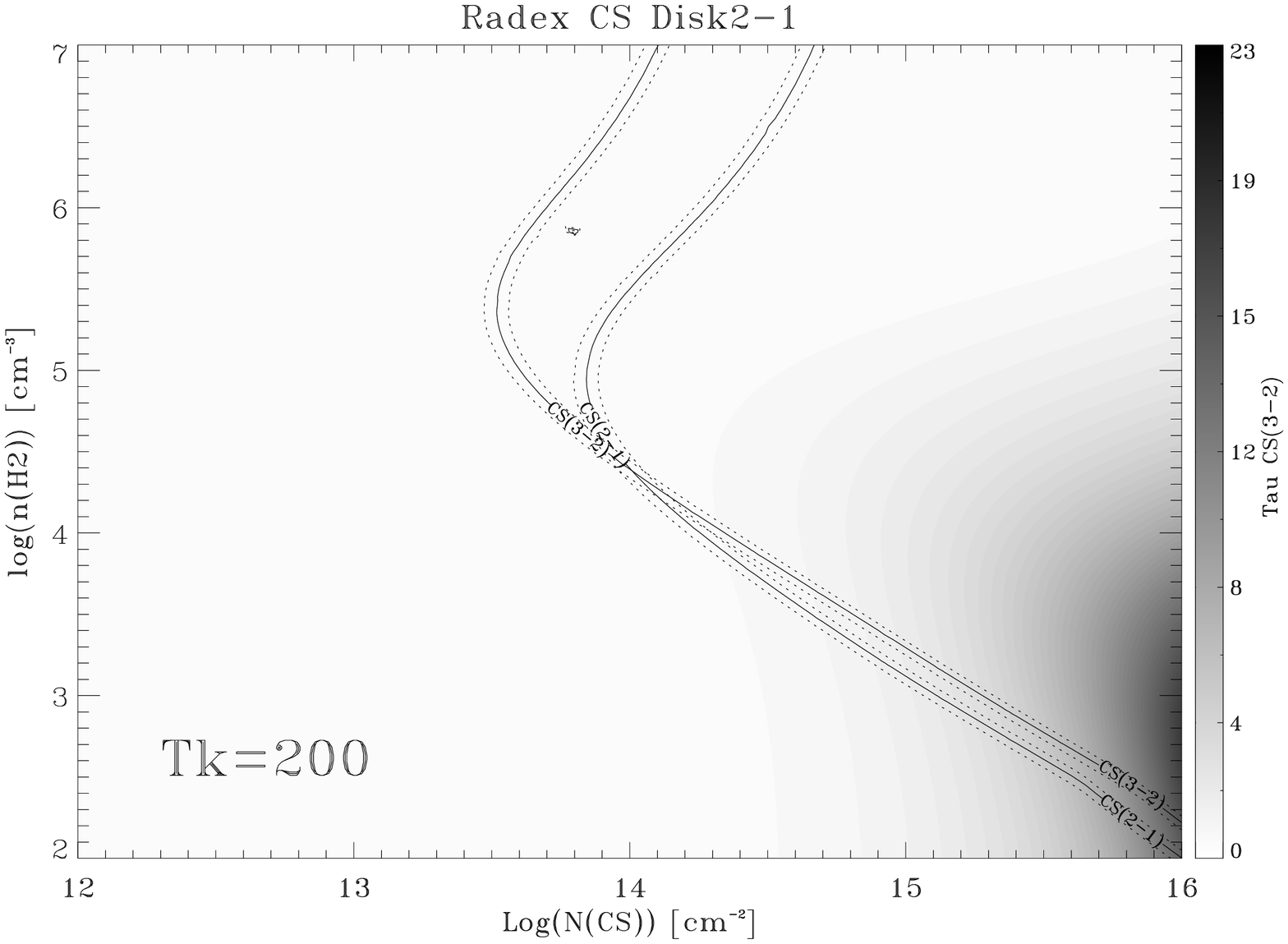}
}
\hbox{
\includegraphics[width=0.4\textwidth, angle=90]{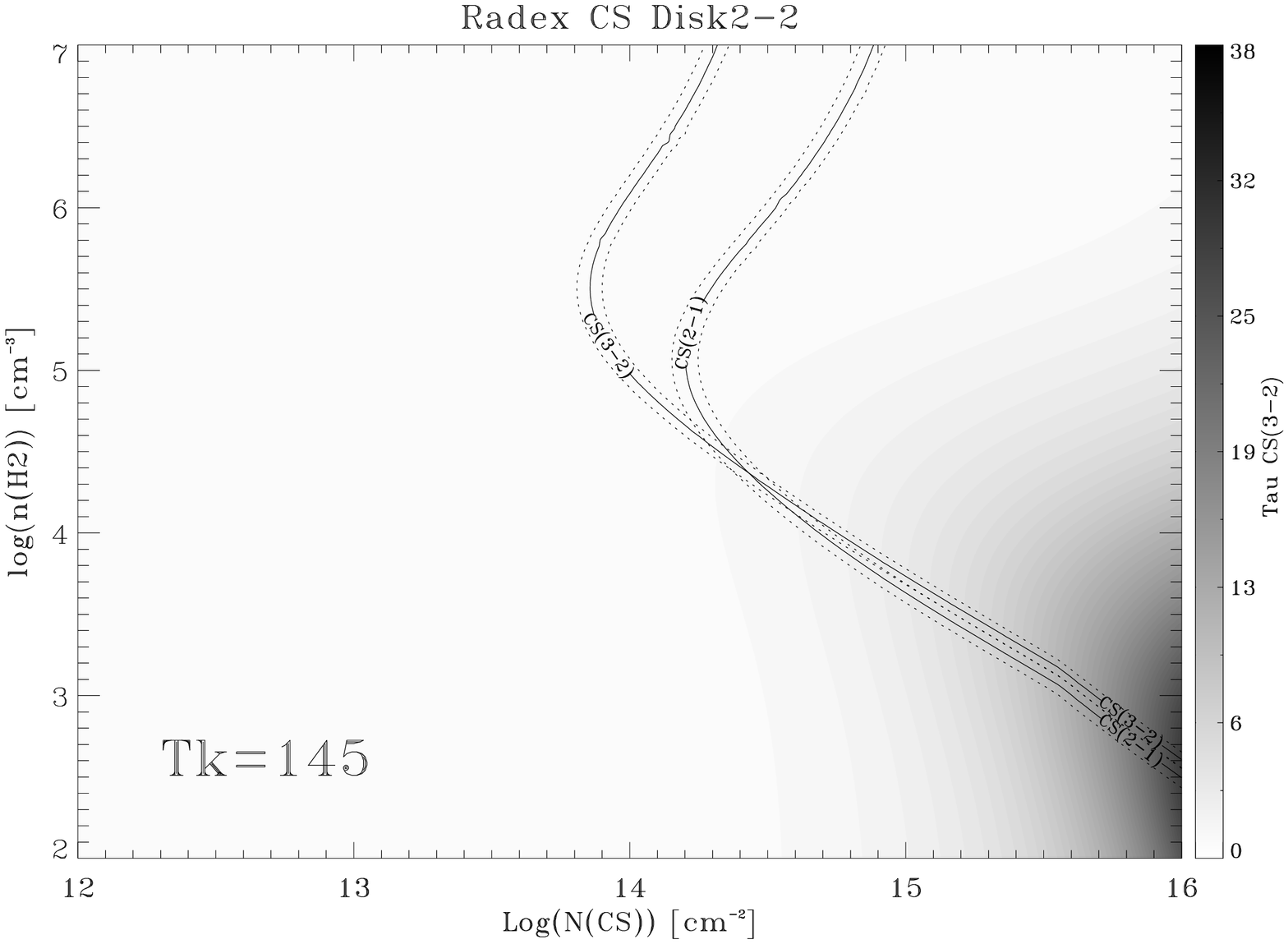}
\hspace{-0.8cm}
}
\hbox{
\includegraphics[width=0.4\textwidth, angle=90]{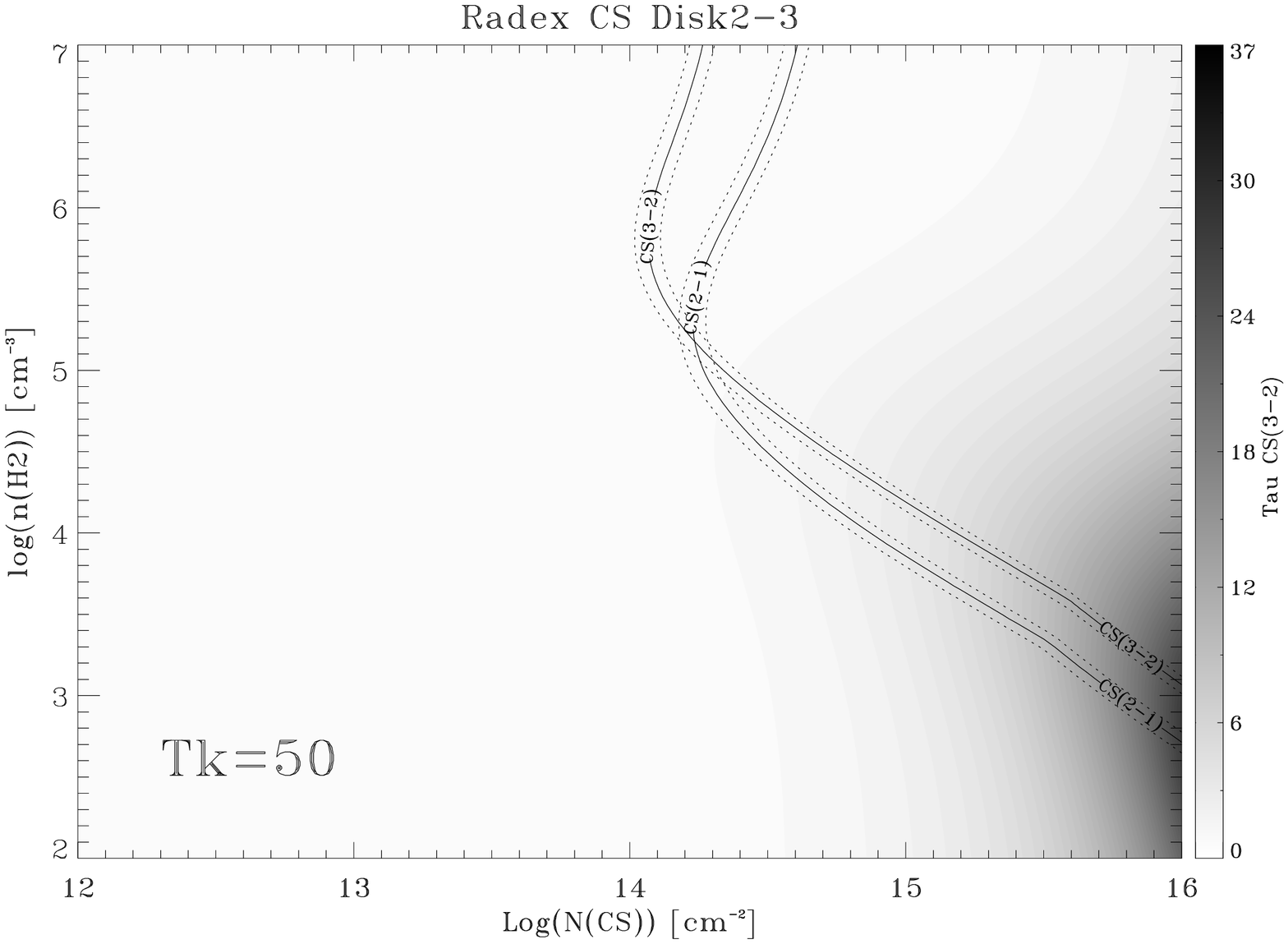}
\hspace{-0.8cm}
\includegraphics[width=0.4\textwidth, angle=90]{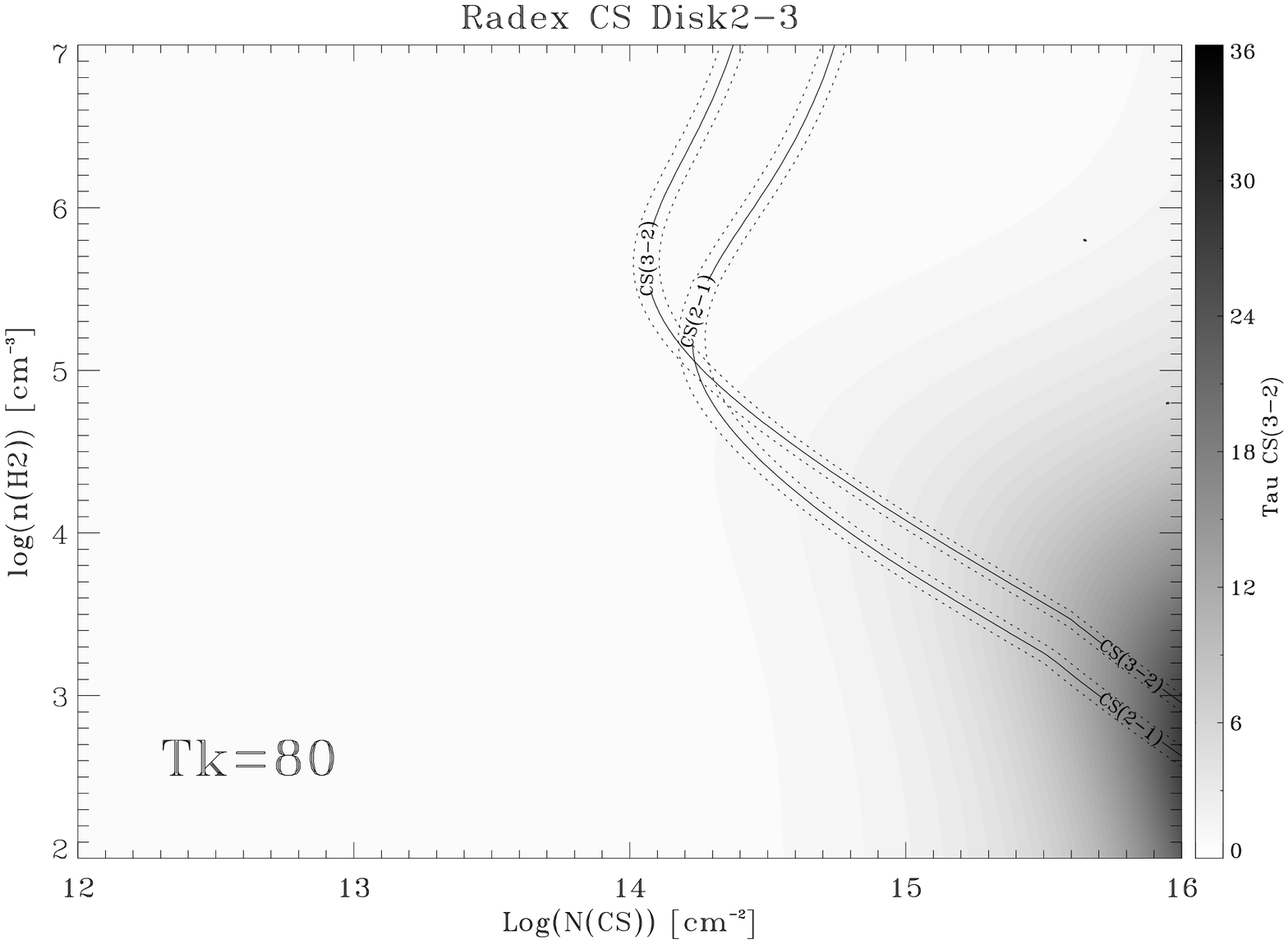}
}
}
\caption{LVG diagrams of CS for each kinetic temperature regime, for each velocity component of
Disk\,2. Top: $14.2\, {\rm km\,s}^{-1}$. Middle: $55.9\, {\rm km\,s}^{-1}$.  Bottom: $79.0\,{\rm km\,s}^{-1}$.}
\label{modeloCSSgrB2}
\end{figure*}
\end{appendix}
%___________________________________________________________________________________________________
\end{document}